\documentclass[nolinenumbers]{aastex631}

\shorttitle{Evidence for non-minimally coupled DM}
\shortauthors{G. Gandolfi et al.}

\graphicspath{{./}{figures/}}

\usepackage{amsmath,latexsym,amssymb}
\usepackage{graphicx}
\usepackage{array}

\begin{document}

\title{Empirical Evidence of Non-Minimally Coupled Dark Matter \\ in the Dynamics of Local Spiral Galaxies?}

\author[0000-0003-3248-5666]{Giovanni Gandolfi}\affiliation{SISSA, Via Bonomea 265, 34136 Trieste, Italy}\affiliation{IFPU - Institute for fundamental physics of the Universe, Via Beirut 2, 34014 Trieste, Italy}

\author[0000-0002-4882-1735]{Andrea Lapi}
\affiliation{SISSA, Via Bonomea 265, 34136 Trieste, Italy}\affiliation{IFPU - Institute for fundamental physics of the Universe, Via Beirut 2, 34014 Trieste, Italy}\affiliation{IRA-INAF, Via Gobetti 101, 40129 Bologna, Italy}
\affiliation{INFN-Sezione di Trieste, via Valerio 2, 34127 Trieste,  Italy}

\author[0000-0002-7632-7443]{Stefano Liberati}\affiliation{SISSA, Via Bonomea 265, 34136 Trieste, Italy}\affiliation{IFPU - Institute for fundamental physics of the Universe, Via Beirut 2, 34014 Trieste, Italy}\affiliation{INFN-Sezione di Trieste, via Valerio 2, 34127 Trieste,  Italy}

\begin{abstract}
We look for empirical evidence of a non-minimal coupling (NMC) between dark matter (DM) and gravity in the dynamics of local spiral galaxies. In particular, we consider a theoretically motivated NMC that may arise dynamically from the collective behavior of the coarse-grained DM field (e.g., via Bose–Einstein condensation) with averaging/coherence length $L$. In the Newtonian limit, this NMC amounts to modify the Poisson equation by a term $L^{2} \nabla^{2} \rho$ proportional to the Laplacian of the DM density itself. We show that such a term, when acting as a perturbation over the standard Navarro--Frenk--White (NFW) profile of cold DM particles, can substantially alter the dynamical properties of galaxies, in terms of their total radial acceleration within the disk and rotation velocity. Specifically, we find that this NMC model can properly fit the stacked rotation curves of local spiral galaxies with different velocities at the optical radius, including dwarfs and low-surface brightness systems, at a level of precision comparable to, and in some instances even better than, the phenomenological Burkert profile. Finally we show that, extrapolating down to smaller masses the scaling of $L$ vs. halo mass found from the above rotation curve analysis, the NMC model can adequately reproduce the radial acceleration relation (or RAR) in shape and normalization down to the dwarf spheroidal galaxy range, a task which constitutes a serious challenge for alternative DM models even inclusive of baryonic effects.
\end{abstract}


\section{Introduction} \label{sec:intro}

The analysis of spiral galaxy rotation curves (RCs) has empirically highlighted since the late 1970s a discrepancy between the amount of luminous matter and the mass budget required to explain the overall kinematic properties of such systems (see \citealt{1978ApJ...225L.107R}; \citealt{1978PhDT.......195B}). The common lore traces back the missing mass to an unseen component called dark matter (DM), constituted by cold (i.e. non-relativistic) and weakly interacting massive particles. Despite such a cold DM paradigm has proven to be relatively successful on cosmological scales, it struggles to fully describe the observed phenomenology on galactic scales, especially in DM-dominated dwarfs. In this respect, two crucial issues will be our focus here.

The first is the well-known cusp-core controversy about the inner shape of the DM density profile in galaxies. Analysis of observed galaxy RCs in the standard Newtonian framework seems to favor an inner core, while gravity-only simulations in the standard cold DM framework produce the universal Navarro-Frenk-White (NFW; see \citealt{Boylan-Kolchin:2003xvl}; \citealt{2006aglu.confE..30N}; \citealt{2010AdAst2010E...5D}; \citealt{Navarro:2016bfs}) profile with a cuspy inner behavior. The second, somewhat related, point concerns the so called radial acceleration relation (RAR), linking the (total) radial acceleration $g_{\mathrm{tot}}$ inferred from galaxy RCs with different masses/velocities and that associated to the luminous matter distribution $g_{\mathrm{bar}}$ mainly probed by photometric observations. The RAR is a remarkably tight relationship (see e.g. \citealt{Lelli:2017vgz}; \citealt{2018A&A...615A...3L}), that subsumes/generalizes many well-known dynamical laws of galaxies (see \citealt{Lelli:2017vgz}), and has been extensively studied
(e.g. \citealt{Keller:2016gmw}; \citealt{Burrage:2016yjm}; \citealt{Lelli:2017sul};  \citealt{Chae:2017bhk}; \citealt{Li:2018tdo}; \citealt{DiPaolo:2018mae}; \citealt{Green:2019cqm}; \citealt{Tian:2020qjd}; \citealt{rodrigues_marra_2020}).

From the theoretical point of view, a plethora of physical effects have been invoked to clear the above issues.
As for the cusp-core problem, it has been advocated that dynamical friction (\citealt{El-Zant:2001jai}; \citealt{2006ApJ...649..591T}; \citealt{Romano-Diaz:2008bab}; \citealt{Goerdt:2008pw}; \citealt{El-Zant:2016byp}) or feedback effects from stars and active galactic nuclei (see \citealt{2012MNRAS.422.1231G}; \citealt{2013MNRAS.429.3068T}; \citealt{Pontzen:2014lma}; \citealt{2017MNRAS.472.2153P}; \citealt{2020MNRAS.491.4523F}; \citealt{2020MNRAS.499.2912F}) during the galaxy formation process can induce violent fluctuations in the inner gravitational potential and/or transfer of energy and angular momentum from the baryons to DM, so erasing the central cusp. Another class of solutions conceives DM as constituted by non-standard particle candidates, thus leading to abandon the cold DM hypothesis in favor of more exotic alternatives (see the review by \citealt{2019A&ARv..27....2S} and references therein). As for the RAR, it has been claimed to emerge naturally from the self-similarity of cold DM halos (e.g., \citealt{Navarro:2016bfs}) or to be explained by properly accounting for the effects of baryons (e.g., \citealt{DiCintio:2014xia}; \citealt{DiCintio:2015eeq}; \citealt{2016MNRAS.455..476S}; \citealt{2017ApJ...835L..17K}; \citealt{2017PhRvL.118p1103L}; \citealt{2017MNRAS.464.4160D}; \citealt{Navarro:2016bfs}; \citealt{Wheeler_2019}). Note that it has been hinted that cored profiles could have a better chance of correctly reproducing the RAR (\citealt{DiCintio:2015eeq}, \citealt{DiPaolo:2018mae}, \citealt{Li:2020iib}).

On a different ground, the incompleteness of the cold DM model on galactic scales has led to the emergence of numerous theories of modified gravity. Perhaps the most famous framework on galactic scales is the Modified Newtonian Dynamics (MOND), that was originally proposed by \cite{1983ApJ...270..365M} and further investigated in a rich literature (see \citealt{Bekenstein:2004ne}; \citealt{PhysRevD.76.124012}; \citealt{Milgrom:2009an}; \citealt{Famaey:2011kh}). As the name suggests, MOND aims to explain the mass discrepancy in galaxies through a modification of Newtonian gravity (or more generally of the Newton second law) that comes into action at accelerations well below a definite universal threshold; in its original formulation, DM is not included and baryons are the only source of the gravitational field. As for the two aforementioned issues, it has been claimed that MOND (or theories reducing to it in the weak field limit) can properly fit galactic RCs (\citealt{1998ApJ...508..132D}; \citealt{Sanders:2002pf}), and provide a satisfying description of the RAR (e.g., \citealt{2018A&A...615A...3L}). 

In this paper, we entail yet another viewpoint to modify the standard cold DM framework and make it capable of accurately describe the observed galaxy RCs and \emph{at the same time} to faithfully reproduce the RAR. Specifically, we put forward the possibility that DM could be non-minimally coupled to gravity, as conjectured in a series of previous works from our team and collaborators (see \citealt{Bruneton:2008fk}; \citealt{Bertolami:2009ic}; \citealt{Bettoni:2011fs}; \citealt{Bettoni_2014}; \citealt{Bettoni_2015}; \citealt{Ivanov:2019iec}; \citealt{Gandolfi:2021jai}). Introducing such a coupling can retain the success of the cold DM on large cosmological scales while improving its behavior in galactic systems, recovering there a MOND-like (even if not exactly MONDian, since DM is there) dynamics. The word ``non-minimal'' in this context means that DM, or more precisely its gradients, are directly coupled to the Einstein tensor. We caveat that such non-minimal coupling (NMC) is not necessarily a fundamental feature of the DM particles, but rather may dynamically develop when the averaging/coherence length $L$ associated with the fluid description of the DM collective behavior is comparable to the local curvature scale. In the Newtonian limit the NMC here considered implies a modification of the Poisson equation by a term $L^2 \nabla^2 \rho$ proportional to the DM density $\rho$ (as in \citealt{Bettoni_2014}). This apparently simple addition can significantly change the internal dynamics of galaxies with respect to a pure cold DM framework, and in fact it has already proven to alleviate some problems in DM-dominated systems (see \citealt{Gandolfi:2021jai}). Incidentally, note that on large scales non-minimally coupled fluids behave under certain conditions similarly to a repulsive dark energy component, and thus the NMC could have a cosmological relevance too (e.g., \citealt{Bettoni:2011fs}; \citealt{2013arXiv1309.0292B}).

We will show that the NMC term, when acting as a perturbation on a galaxy system characterised by the cuspy NFW profile for the DM, can  substantially alter its dynamical properties. Such a NMC model can thus provide accurate fits to the stacked RCs of spiral galaxies with different velocities at the optical radius, including dwarfs and low-surface brightness systems. Moreover, we will show that the same NMC model can properly account for the RAR as well. The plan of the paper is as follows: in Sect.~\ref{2|theory} we briefly review the theoretical background behind our NMC model; in Sect.~\ref{3|fitRCs} we analyse a sample of stacked RCs of spiral galaxies with the NMC model; in Sect.~\ref{4|RAR} we build empirically-based mock RCs of galaxies with different masses and construct the RAR, showing that it is well reproduced by the NMC model; in Sect.~\ref{5|conclusion} we summarize our findings and outline future perspectives and applications of the NMC framework.  

Throughout this work, we adopt the standard flat $\Lambda$CDM cosmology (\citealt{Planck:2018vyg}) with rounded
parameter values: matter density $\Omega_{M}=0.3$, dark energy density $\Omega_{\Lambda}=0.7$, baryon density $\Omega_{b}=0.05$, and Hubble constant $H_{0}=100 h \mathrm{~km} \mathrm{~s}^{-1} \mathrm{Mpc}^{-1}$ with $h=0.7$. Unless otherwise specified, $G \approx 6.67 \times 10^{-8} \mathrm{~cm}^{3} \mathrm{~g}^{-1} \mathrm{~s}^{-2}$ indicates the standard gravitational (Newton) constant.

\section{A theoretical background for the NMC}\label{2|theory}

In this section, we recall the basic theoretical background behind the NMC hypothesis, referring the reader to \cite{Gandolfi:2021jai} for further details. A very basic NMC model can be built by adding a coupling term $S_{\rm int}$ between DM and gravity in the total Einstein--Hilbert action (in the Jordan frame) with shape:
\begin{equation}
    S_{\text {int }}\left[\tilde{g}_{\mu \nu}, \varphi\right] = \epsilon L^{2} \int \mathrm{d}^{4} x \,\sqrt{-\tilde{g}} \, \widetilde{G}^{\mu \nu}\, \nabla_{\mu}\, \varphi \nabla_{\nu} \varphi~;
\end{equation}
here $\varphi$ is the (real) DM scalar field, $\epsilon = \pm 1$ is the polarity of the coupling, $\widetilde{G}^{\mu \nu}$ is the Einstein tensor, and $L$ is the NMC characteristic length-scale. Note that $L$ may be not a new fundamental length-scale of Nature, but rather can emerge dynamically from some collective behavior of the coarse-grained DM field (e.g., Bose--Einstein condensation). Therefore, such a NMC model does not consist in a modified gravity theory, but simply in a formalization of an emergent behavior of cold DM in galactic environments.  Note that, from a purely theoretical perspective, such form of the NMC is allowed by the Einstein equivalence principle (e.g., \citealt{2015AmJPh..83...39D}). We will keep $\epsilon$ indicated as a bookkeeping parameter, but based on \cite{Gandolfi:2021jai} we will set it to $\epsilon=-1$ (repulsive coupling).

Adopting the fluid approximation for the field $\varphi$ (as in \citealt{2012JCAP...07..027B}) and taking the Newtonian limit, it can be shown that the NMC boils down to a simple modification of the usual Poisson equation  (\citealt{Bettoni_2014}, \citealt{Gandolfi:2021jai})
\begin{equation}
\label{poissmod}
    \nabla^{2} \Phi=4 \pi G\left[\left(\rho+\rho_{\mathrm{bar}}\right)-\epsilon L^{2} \nabla^{2} \rho\right],
\end{equation}
where $\Phi$ is the Newtonian potential, and $\rho_{\mathrm{bar}}$ and $\rho$ are the baryonic and DM densities. In spherical symmetry, Eq.~(\ref{poissmod}) implies that the total gravitational acceleration writes 
\begin{equation}
\label{gnmc}
    g_{\text{tot}}(r) = -\frac{G\,M(< r)}{r^2}+4\pi\, G\, \epsilon L^2\, \frac{{\rm d}\rho}{{\rm d}r}\; ,
\end{equation}
where $M(<r)$ is the total mass enclosed in the radius $r$; the first term is the usual Newtonian acceleration and the second term is the additional contribution from the NMC. Plainly, the related RCs $v_{\rm tot}^2(r)=|g_{\rm tot}(r)|\, r$ of spiral galaxies predicted in this framework will differ from the standard Newtonian case.

In \cite{Gandolfi:2021jai} we have highlighted that Eq.~(\ref{poissmod}) gives rise to some interesting features for strongly DM-dominated systems in self-gravitating equilibria: the NMC can help to develop an inner core in the DM density profile, enforcing a shape closely following the Burkert one out to several core scale radii; DM-dominated halos with NMC are consistent with the core-column density relation (see e.g. \citealt{Salucci:2000ps}, \citealt{10.1111/j.1365-2966.2009.15004.x}, \citealt{burkert2015structure}, \citealt{2013ApJ...770...57B}, \citealt{Burkert:2020laq}), i.e. with the observed universality of the product between the core radius $r_0$ and the core density $\rho_0$. However, the NMC hypothesis needs still to be tested in galaxies with different velocities at the optical radius, where the contribution of the baryonic component to the dynamics can be substantial, which is precisely our aim in the next sections. 

\section{TESTING THE NMC WITH STACKED RC\lowercase{s}}\label{3|fitRCs}

In this section, we will apply the NMC to mass-model stacked RCs of local spiral galaxies with different velocities at the optical radius and related properties. We will then compare our results with fits obtained from the standard Newtonian case for two other classic DM halo shapes, namely the standard NFW profile (see \citealt{1996ApJ...462..563N}) emerging from gravity-only simulations of cold DM, and the phenomenological Burkert profile (see \citealt{1995ApJ...447L..25B}). 

For the analysis, we rely on the samples of stacked RCs collected by \cite{1996MNRAS.281...27P} for normal spirals divided in $11$ average velocity bins, by \cite{Dehghani:2020cvl} for low surface brightness (LSB) spirals divided in $5$ average velocity bins and by \cite{2017MNRAS.465.4703K} for low-luminosity dwarfs. These stacked RCs are built by co-adding high-quality individual RCs of thousands galaxies with similar velocities at the optical radius and related properties, after properly normalizing the velocity and radial variables to reference scales for each galaxy, which are typically the optical radius $r_{\rm opt}$ and the optical circular velocity $v_{\rm opt}\equiv v(r_{\rm opt})$; the interested reader can find details of such a procedure in \cite{2018ApJ...859....2L}. The average properties of our sample of stacked RCs are listed in Table~\ref{binlist}. 

We mass-model the stacked RCs as the sum of a baryonic (disk) component $v_{\rm d}^2(r)=G\,M_{\rm d}(<r)/r$ plus a DM contribution $v_{\rm DM}^2(r)=G\,M_{\rm DM}(<r)/r$, with $M_{\rm d}(<r)$ and $M_{\rm DM}(<r)$ the cumulative disk and DM mass, respectively.
The overall velocity model plainly reads $v_{\mathrm{tot}}^2(r)=v_{\mathrm{d}}^2(r)+v_{\mathrm{DM}}^2(r)$. 
The distribution followed by baryonic matter is modeled as a razor-thin exponential disk (see \citealt{1970ApJ...160..811F}) with exponential surface density
    \begin{align*}
        \Sigma_{\mathrm{d}}(r)=\Sigma_{0} \exp \left(-r / r_{\mathrm{d}}\right)\, ;
    \end{align*}
    here $\Sigma_{0}=M_{\rm d}/2\pi\,r_{\rm d}^2$ is the central value in terms of the total disk mass $M_{\rm d}=M_{\rm d}(<\infty)$ and of the disk scale-length $r_{\mathrm{d}}\approx r_{\mathrm{opt}}/3.2$. The related contribution to the RC is given by (e.g., \citealt{1987gady.book.....B})
    \begin{align}
    \label{expdisk}
       v_{\mathrm{d}}^2(r)=\frac{G\,M_{\rm d}}{r_{\mathrm{d}}}\, 2\,y^{2}\left[I_{0}(y) K_{0}(y)-I_{1}(y) K_{1}(y)\right]\; ,
    \end{align}
    where $y\equiv r/(2\,r_{\mathrm{d}})$, while $I_{0,1}(\cdot)$ and $K_{0,1}(\cdot)$ are modified Bessel functions. Since the fit is performed in a radial range $r\lesssim r_{\rm opt}$ we have checked that any contribution from a gaseous disk (typically more important at larger radii) is negligible and largely unconstrained, so we include only the stellar disk in the mass-modeling. 
    
\subsection{DM models}\label{3.1|theoreticalremarks}

We exploit three different DM models. Two are based on standard Newtonian gravity, but differ in the form of the DM profile shape: NFW or Burkert. The other model is based on the NFW profile but include a perturbative correction to the dynamics via the NMC term of Sect. \ref{2|theory}.

\begin{itemize}
    
    \item \underline{NFW profile} 
    
    The standard NFW profile features the shape (see \citealt{1996ApJ...462..563N}; \citealt{2001MNRAS.321..155L})
    \begin{equation}
    \label{NFW}
        \rho_{\rm DM}(r)=\frac{\delta_{\mathrm{c}}\, \rho_\mathrm{c}\, r_s^3}{r\,(r+r_s)^2}\; ,
    \end{equation}
    where $\delta_{\mathrm{c}}$ is the (dimensionless) characteristic overdensity of the halo, $\rho_\mathrm{c}=3\, H_0^2/8\pi\,G$ is the local critical density, and $r_s$ is the scale radius. The virial mass $M_{\rm v}$ and the concentration $c\equiv r_{\rm v}/r_s$, defined in terms of the virial radius $r_{\rm v}\approx 260\, (M_{\rm v}/10^{12}\, M_\odot)^{1/3}$,  can be used to fully characterize the profile since $\delta_c\, \rho_c = M_{\rm v}\, c^3\, g(c)/4\pi\,r_{\rm v}^3$ with $g(c)\equiv [\ln(1+c)-c/(1+c)]^{-1}$. The corresponding RC writes
    \begin{equation}
    \label{nfwmass}
        v_{\rm DM}^2(r) = \frac{G M_\mathrm{v}}{r_{\rm v}}\,\frac{g(c)}{s}\,\left[\ln (1+c\, s)-\frac{c \, s}{1+c\, s}\right]\; ,
    \end{equation}
    where $s\equiv r/r_\mathrm{v}$. From the above, it is clear that the overall galaxy RC can be specified in terms of three parameters: the halo mass $M_{\rm v}$, the halo concentration $c$ and the disk mass $M_{\rm d}$.
    
    \item \underline{NMC model} 
    
    We include the effect of the NMC as a perturbative correction to the dynamics based on Eq. (\ref{gnmc}), retaining the standard NFW profile for the DM. The perturbative parameter in our analysis is the term $L^2/r_s^2$, which, as we will show with our results, is always small for the range of masses probed in our study. Plugging Eq. (\ref{NFW}) in Eq. (\ref{gnmc}) and after some simple yet tedious algebra we obtain the RC
    \begin{equation}
    \label{nmcmass}
        v_{\rm DM}^2(r) = \frac{G M_\mathrm{v}}{r_{\rm v}}\,\frac{g(c)}{s}\,\left[\ln (1+c\, s)-\frac{c \, s}{1+c\, s}+ \frac{\epsilon L^2}{r_s^2}\,\frac{1+3\,c\,s}{(1+c\,s)^3} \right]\; .
    \end{equation}
     The overall RC model has four free parameters: the halo concentration $c$, the halo virial mass $M_{\mathrm{v}}$, the NMC length-scale $L$, and the disk mass $M_{\mathrm{d}}$. 

    \item \underline{Burkert profile} 
    
    The phenomenological Burkert profile features the shape
    \begin{equation}
    \label{burkert}
        \rho_{\rm DM}(r)=\frac{\rho_{0}\,r_{0}^{3}}{\left(r+r_{0}\right)\, \left(r^{2}+r_{0}^{2}\right)}\, ,
    \end{equation}
    where $r_0$ is the core radius and $\rho_0$ the core density. 
    The RC writes (see \citealt{Salucci:2000ps})
    \begin{equation}
    \label{burkertmass}
       v_{\rm DM}^2(r) = \frac{4\,G M_{0}}{r}\, \left\{\ln \left(1+\frac{r}{r_{0}}\right)-\tan^{-1}\left(\frac{r}{r_{0}}\right)+\frac{1}{2} \ln \left[1+\left(\frac{r}{r_{0}}\right)^{2}\right]\right\},
    \end{equation}
    where $M_{0}=1.6\, \rho_{0}\, r_{0}^{3}$. When using the Burkert profile, we adhere to the customary approach of describing the total RC in terms of three parameters: the core radius $r_0$, the core mass $M_0$, and the ratio $\kappa\equiv v^2_{\rm d}(r_{\rm opt})/v^2_{\rm tot}(r_{\rm opt})$ of the disk to the total velocity at the optical radius.

\end{itemize}
\subsection{Fitting procedure and results}\label{3.2|fitresults}

\begin{figure}[ht]
    \begin{center}
    \includegraphics[width=.495\textwidth]{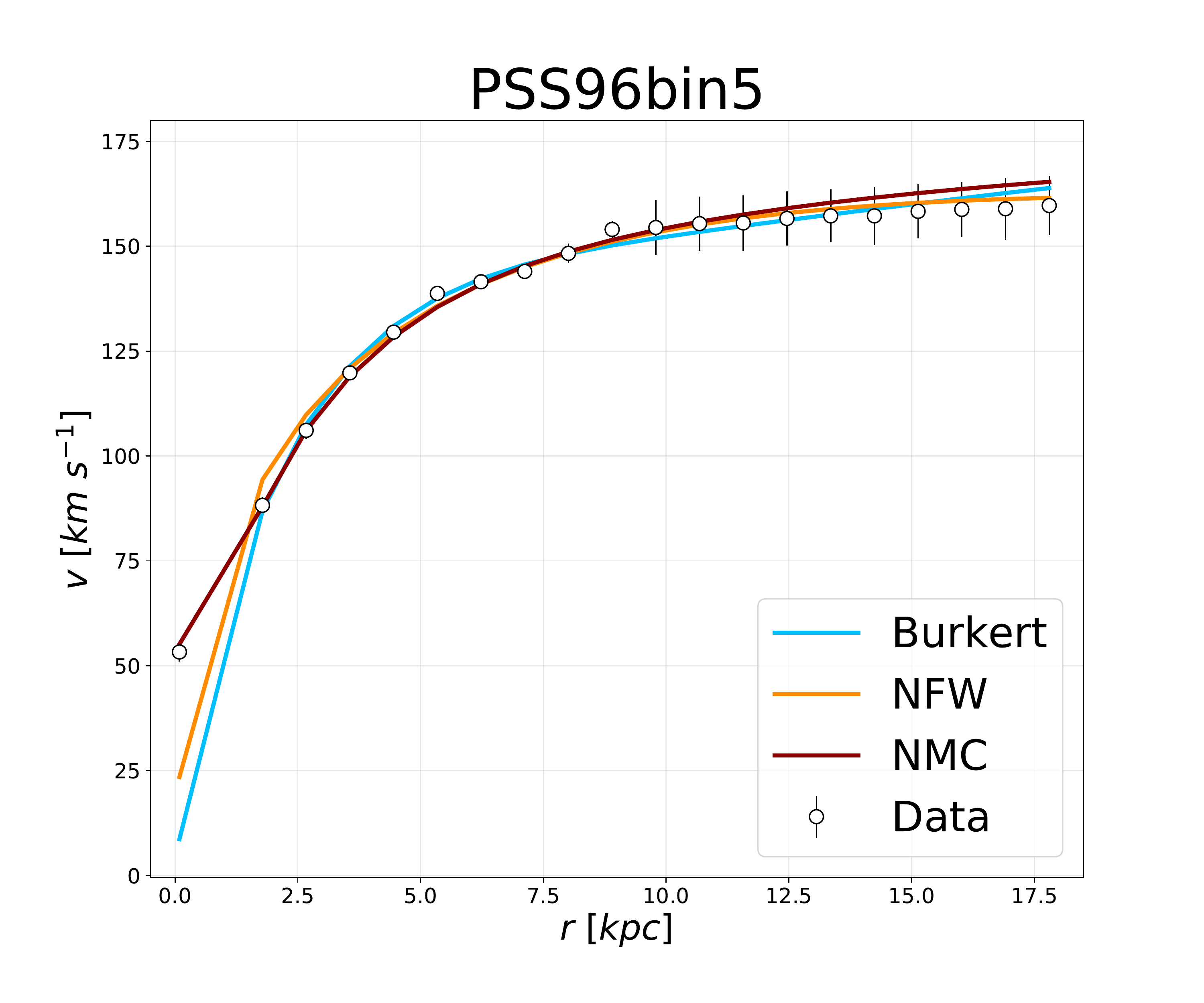}
    \includegraphics[width=.495\textwidth]{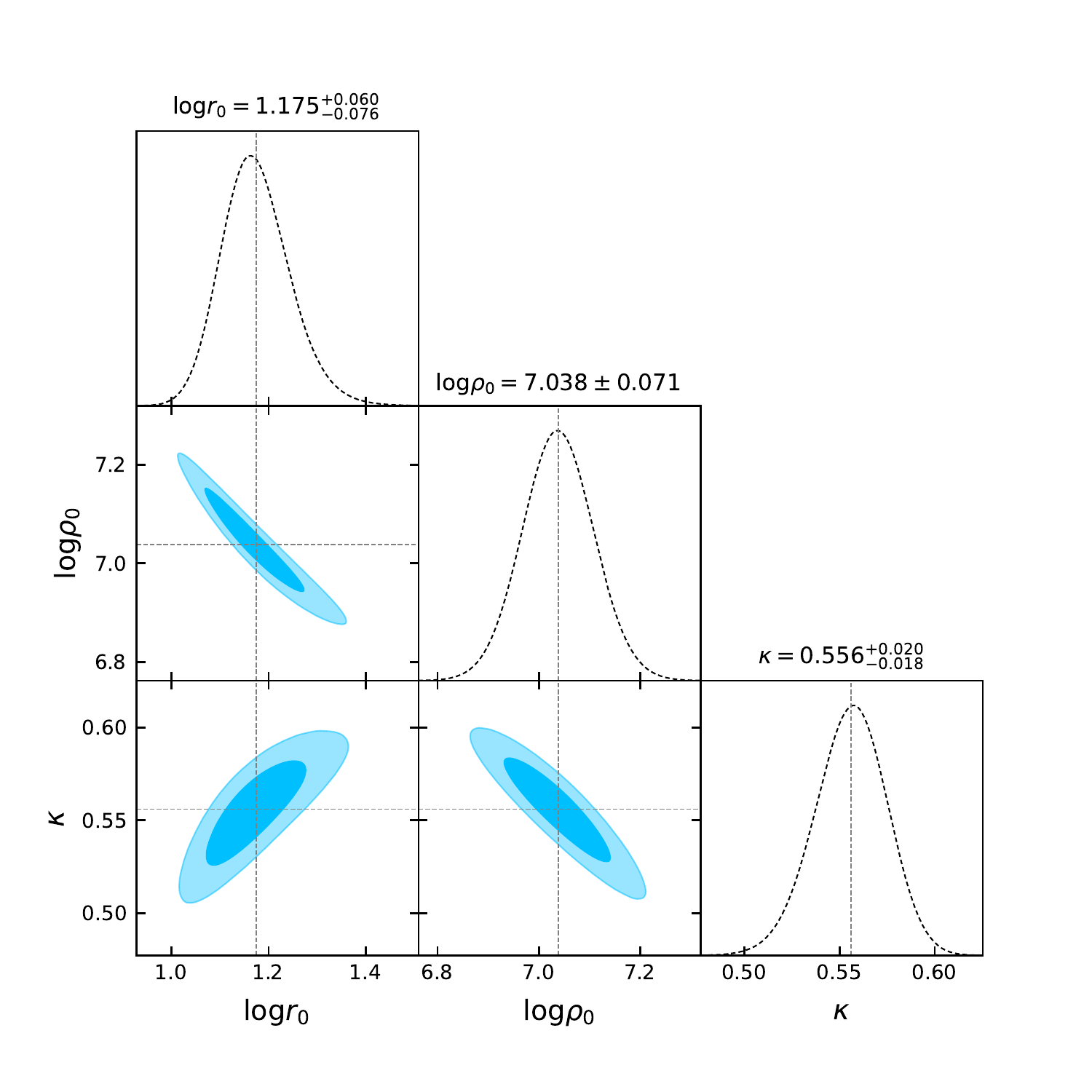}
    \includegraphics[width=.495\textwidth]{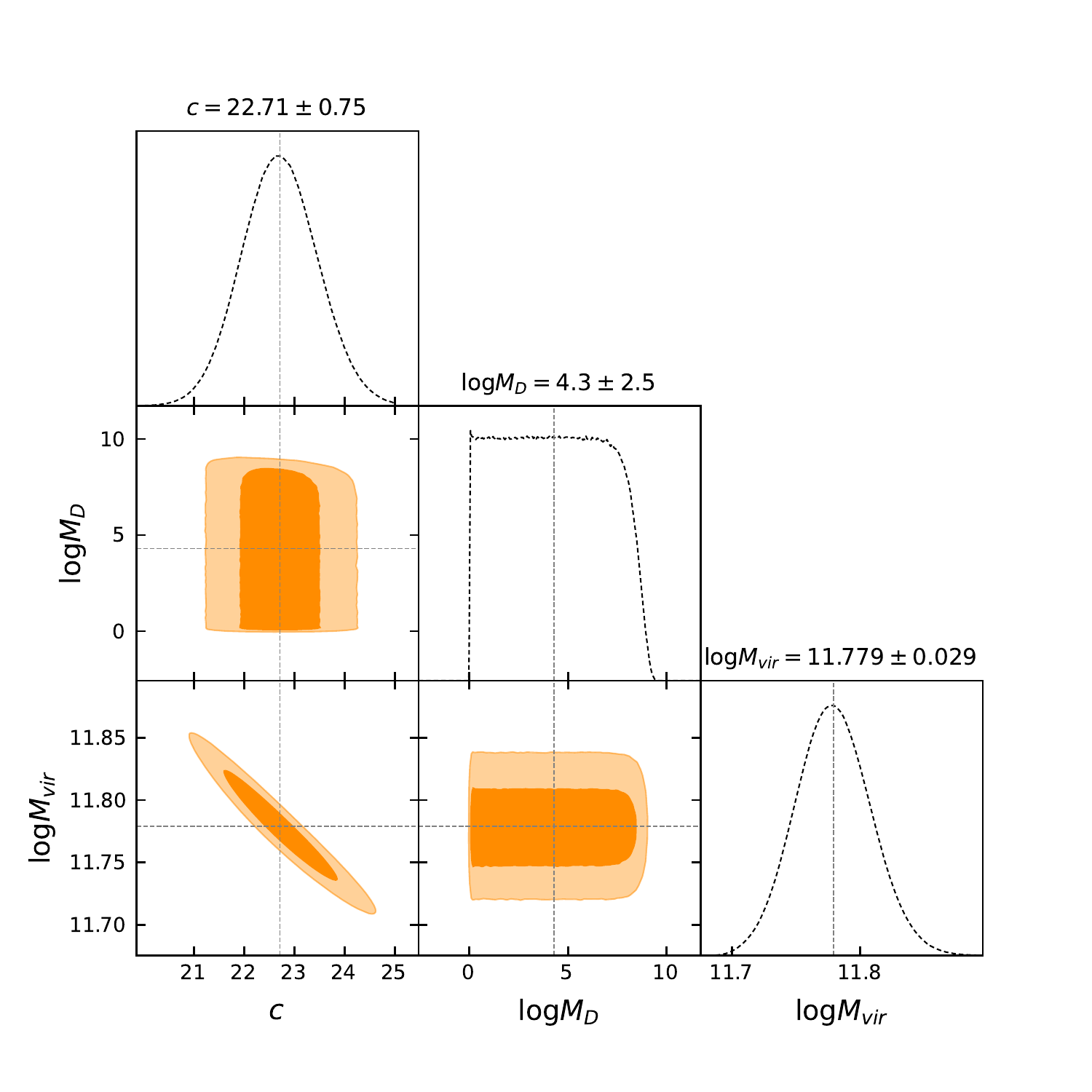}
    \includegraphics[width=.495\textwidth]{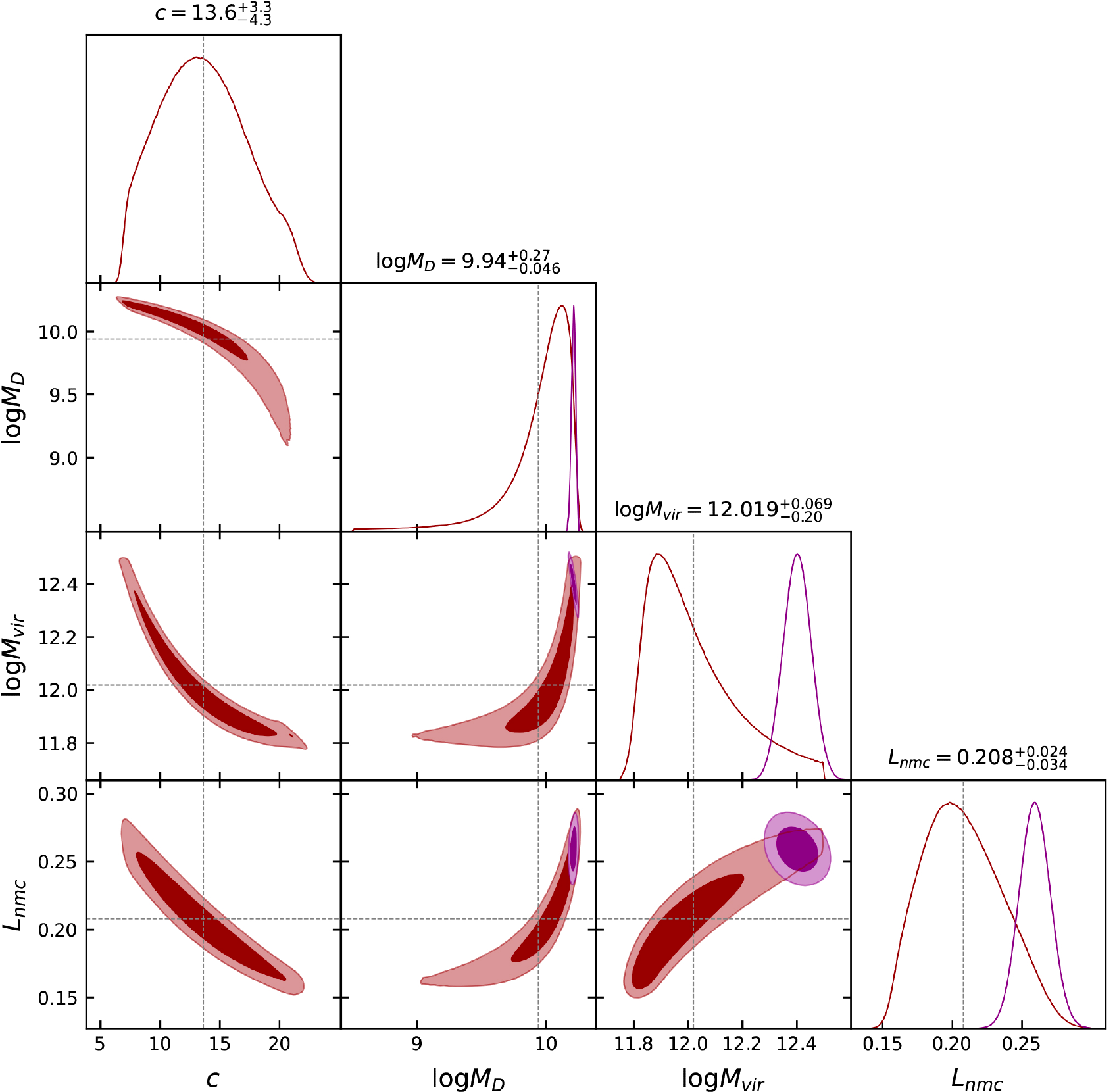}
    \end{center}
    \caption{Analysis of the stacked RC for bin 5 of the spiral galaxy sample by \cite{1996MNRAS.281...27P}. The top left panel illustrate the RC curve data (open symbols) and the best-fit model for the Burkert (cyan line), NFW (orange line) and NMC profile (red line). The outcome of the Bayesian MCMC parameter estimation are shown as corner plots for the Burkert profile (top right panel), for the NFW profile (bottom left) and for the NMC model (bottom right, with purple contours representing the posterior when the halo concentration is constrained by the relation of \citealt{Dutton:2014xda} given by Eq.~(\ref{c-mvir})). \label{fitbin5}}
\end{figure}

We fit the stacked RC data with the mass models described above, exploiting the \texttt{emcee} python package for Bayesian Monte Carlo Markov Chain parameter estimation (see \citealt{2013PASP..125..306F}). We present here representative outcomes concerning one velocity bin for each of the galaxy type: normal spirals, LSBs and dwarfs; the complete analysis for all the other velocity bins produces similar results and is reported in the Appendix and in Tables \ref{burkfit}-\ref{nfwfit}-\ref{nmcfit}. First, we consider Bin 5 from the \cite{1996MNRAS.281...27P} sample of spiral galaxies, whose outcome is reported in Fig.~\ref{fitbin5}. 

The results on the estimated virial mass are consistent for the three profiles. As to the disk mass, it is consistent between NMC and Burkert models, while for the NFW model only a rather loose upper limit can be derived. All in all, the NMC model curve performs appreciably better in terms of reduced $\chi_{\mathrm{red}}^2\approx 0.6$ with respect to the Burkert $\chi_{\mathrm{red}}^2 \approx 22.5$ and to the pure NFW model $\chi_{\mathrm{red}}^2 \approx 11$, as can be also appreciated graphically. The estimated value of the NMC length-scale is around $0.2$ kpc, roughly corresponding to a sixtieth of $r_s$. We also try to perform the fits of the NFW and NMC models by imposing the concentration parameter of the halo to satisfy the relation with the virial mass by \cite{Dutton:2014xda}. We find that both fits are not appreciably altered, but the posterior distribution of the fitted parameters in the NMC model are still consistent and somewhat narrowed.

\begin{figure}[ht]
    \begin{center}
    \includegraphics[width=.495\textwidth]{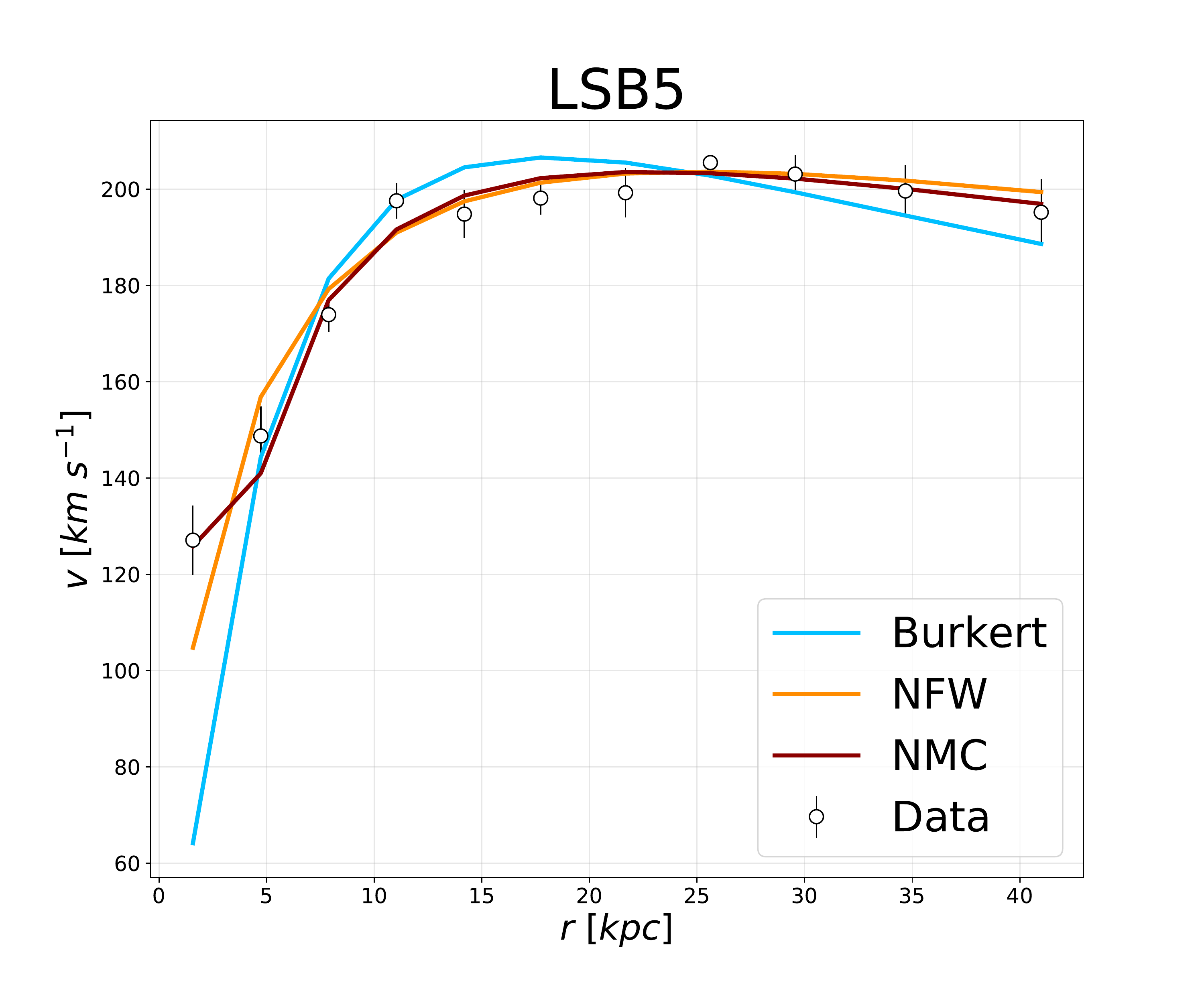}
    \includegraphics[width=.495\textwidth]{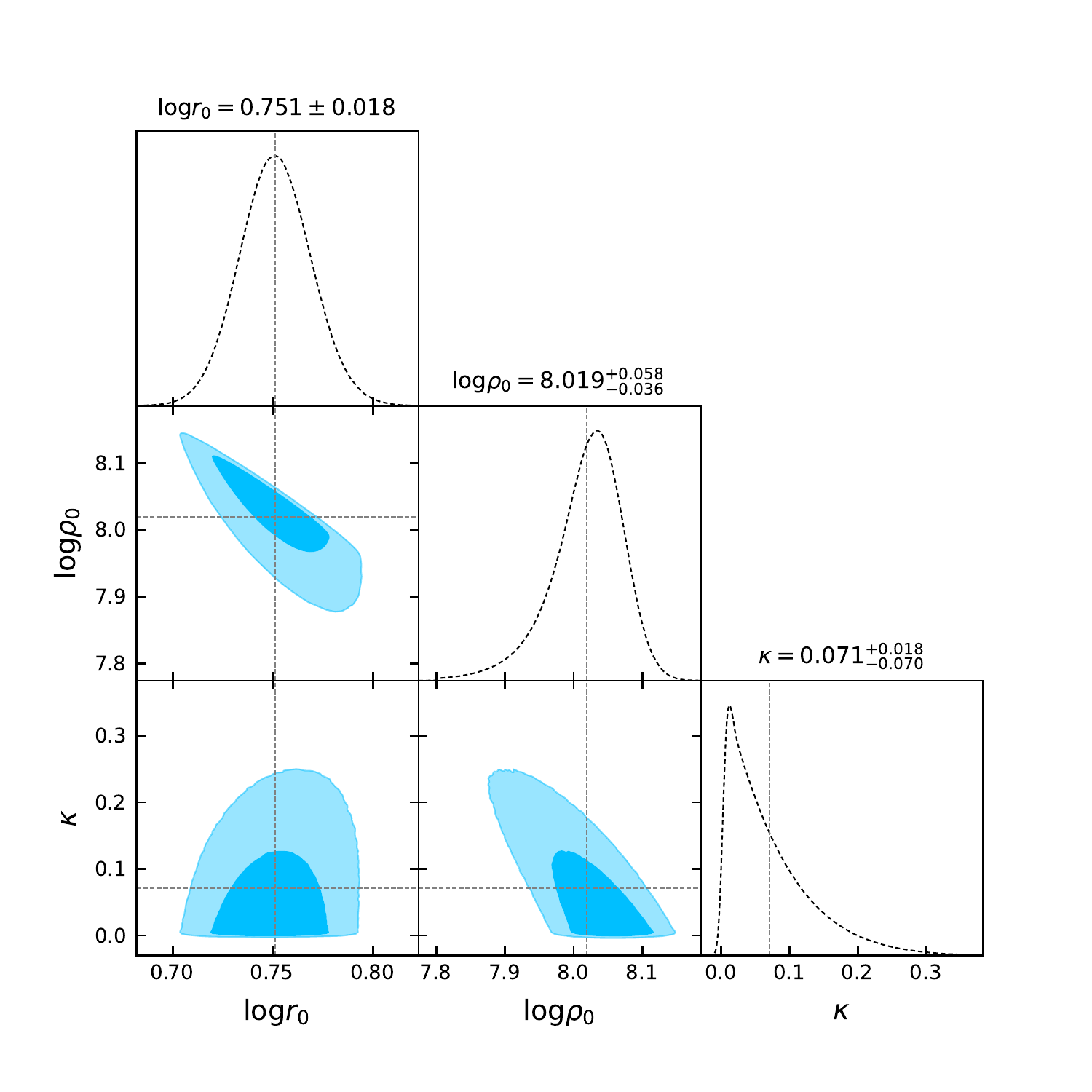}
    \includegraphics[width=.495\textwidth]{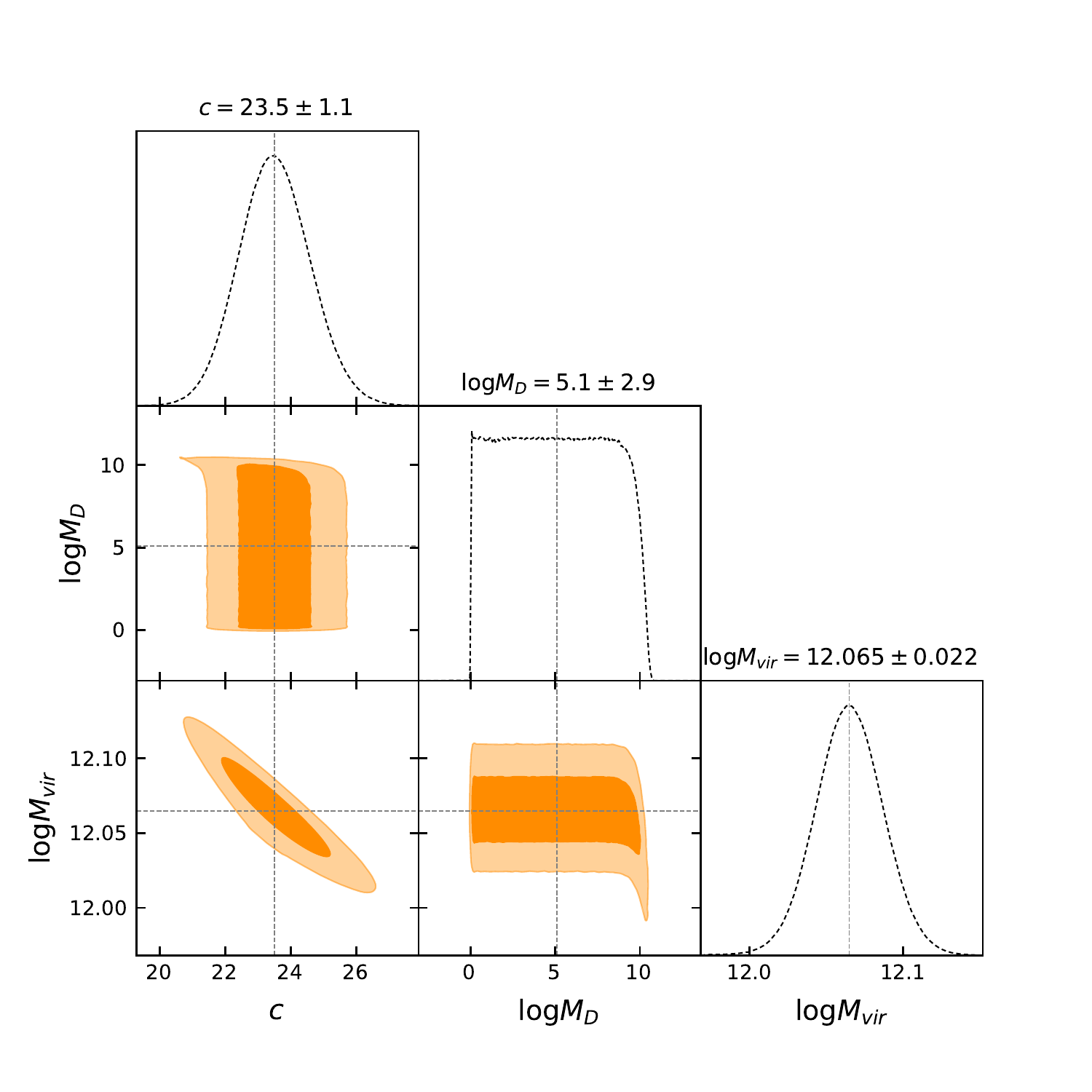}
    \includegraphics[width=.495\textwidth]{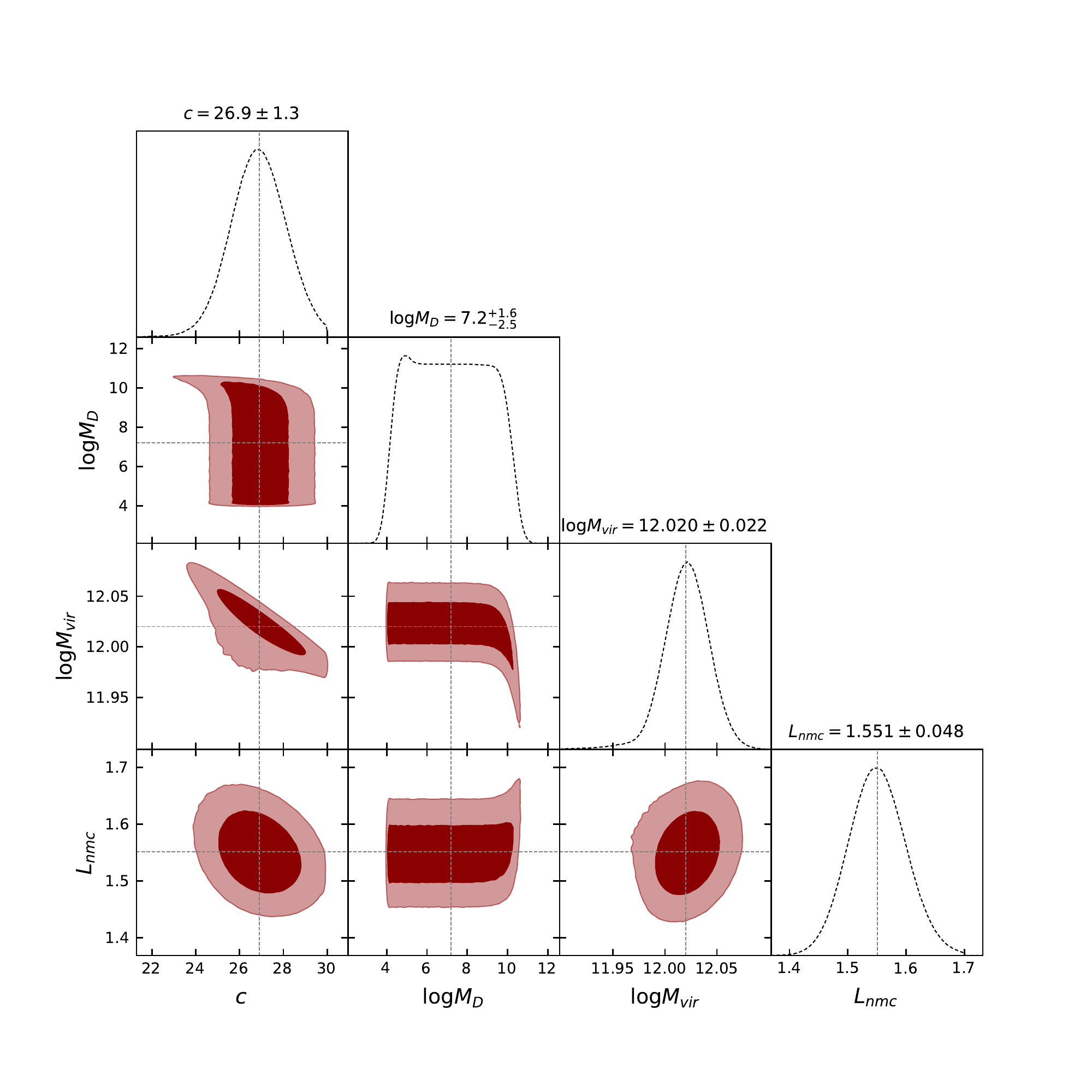}
    \end{center}
    \caption{The same of Fig.~\ref{fitbin5} for the bin 5 of LSB galaxies by \cite{Dehghani:2020cvl}.\label{fitLSB5}}
\end{figure}

Fig.~\ref{fitLSB5} refers to Bin 5 in the sample of LSB galaxies by \cite{Dehghani:2020cvl}. In this case, the Burkert model yields a reduced $\chi_{\mathrm{red}}^2 \approx 11$, the NFW fit yields $\chi_{\mathrm{red}}^2 \approx 3$ and the NMC model performs better yielding $\chi_{\mathrm{red}}^2 \approx 1.411$.
The disk mass in all the fits is poorly constrained, as expected since these LSB galaxies have an extremely extended disk mass distribution relative to the region probed by the RC.

\begin{figure}[ht]
    \begin{center}
    \includegraphics[width=.495\textwidth]{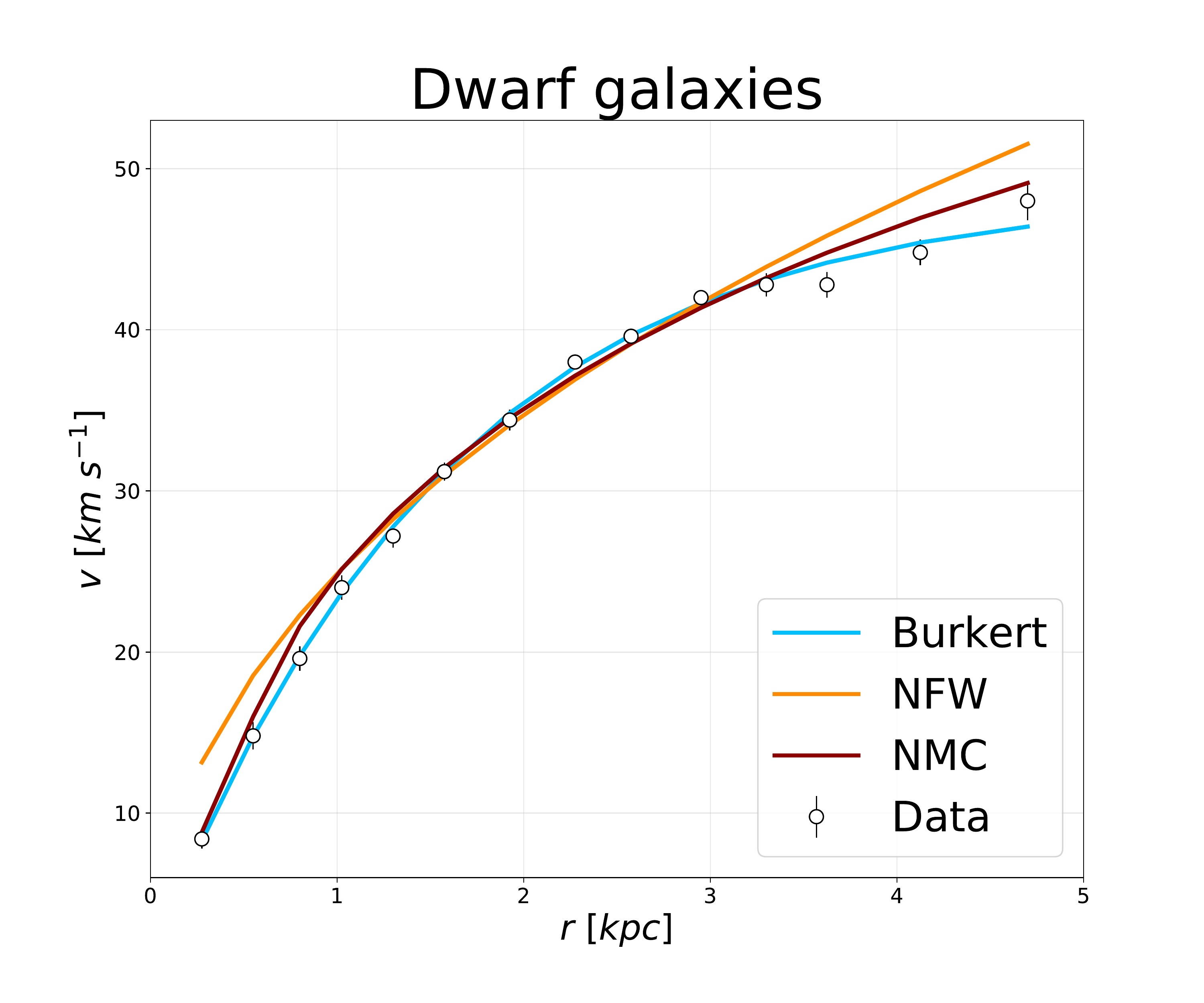}
    \includegraphics[width=.495\textwidth]{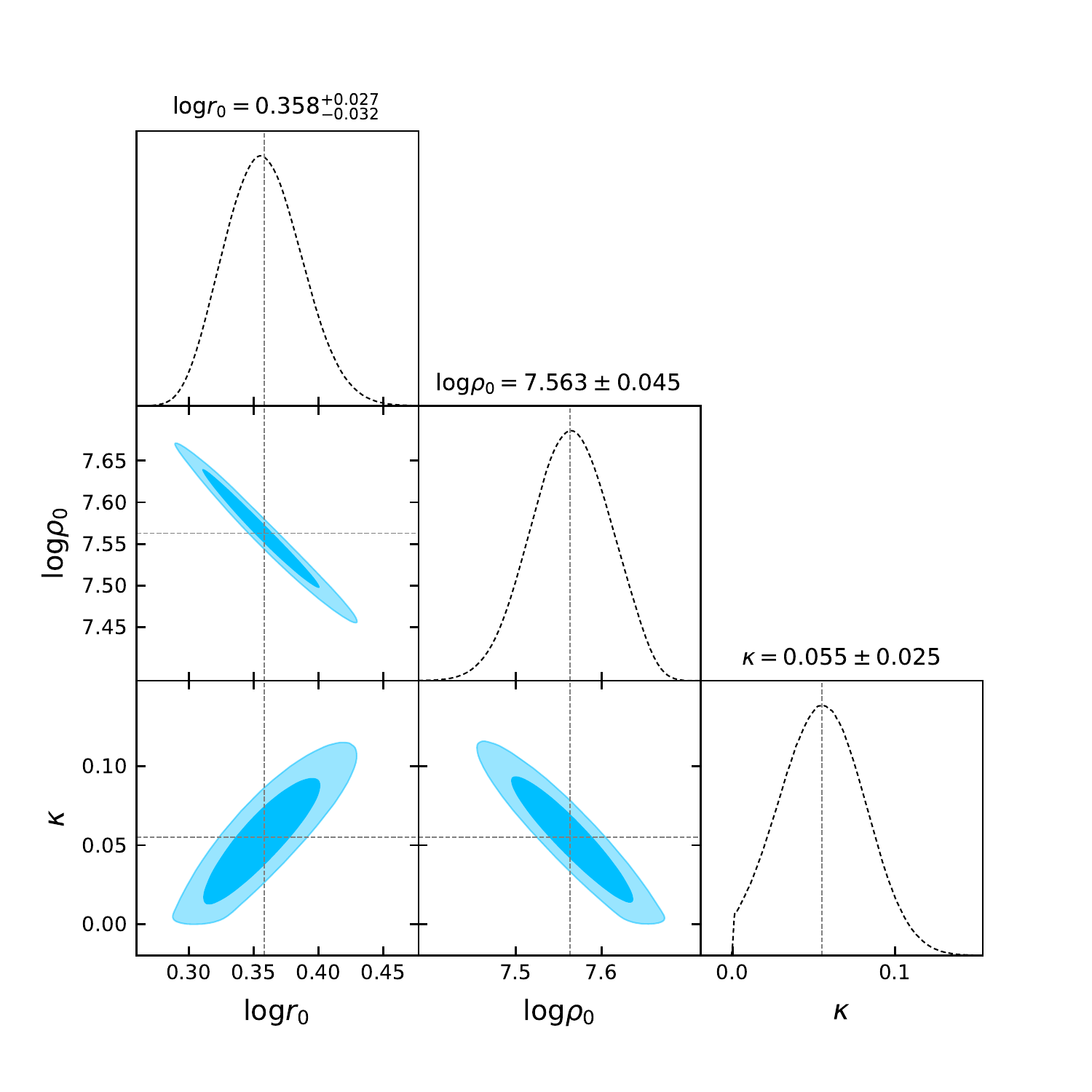}
    \includegraphics[width=.495\textwidth]{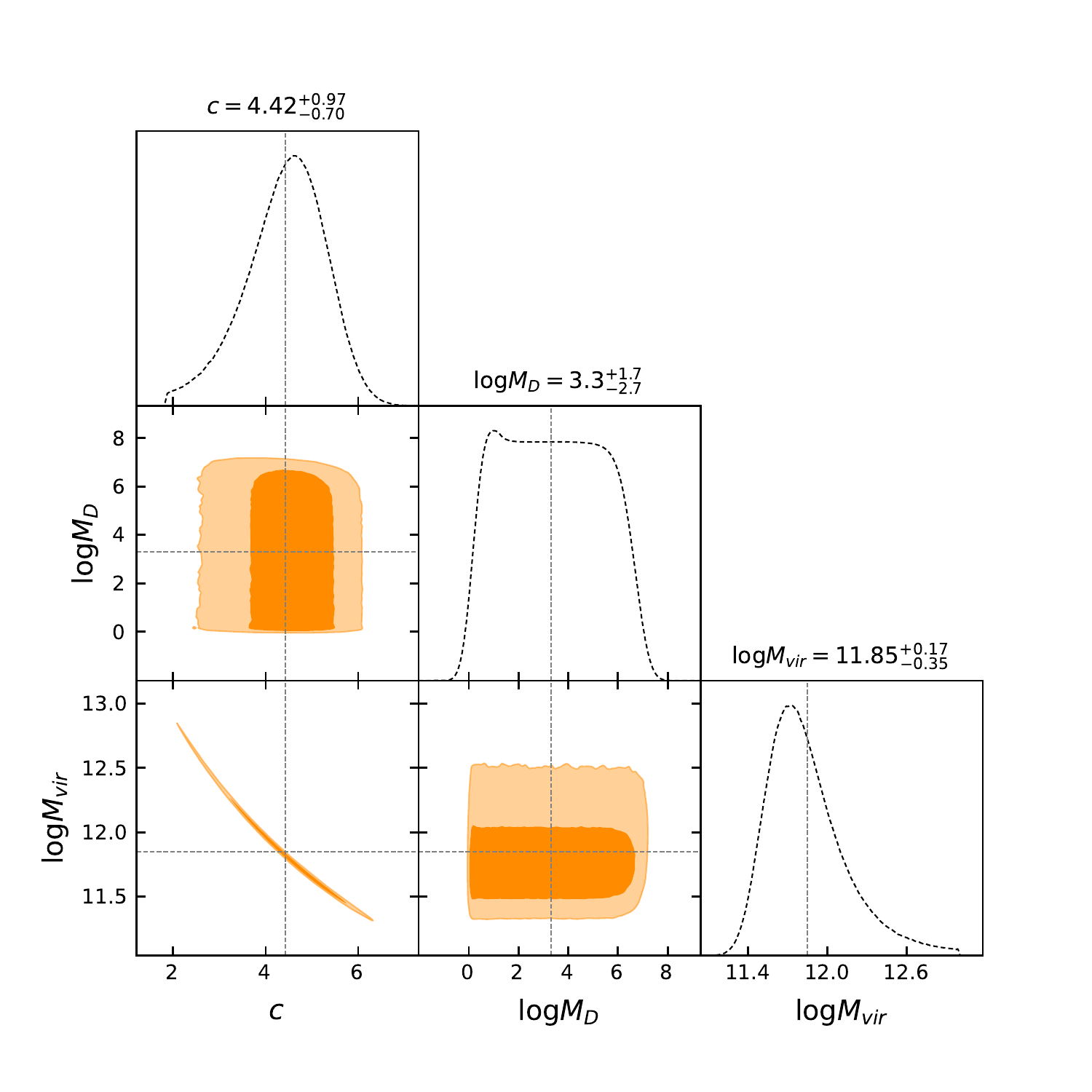}
    \includegraphics[width=.495\textwidth]{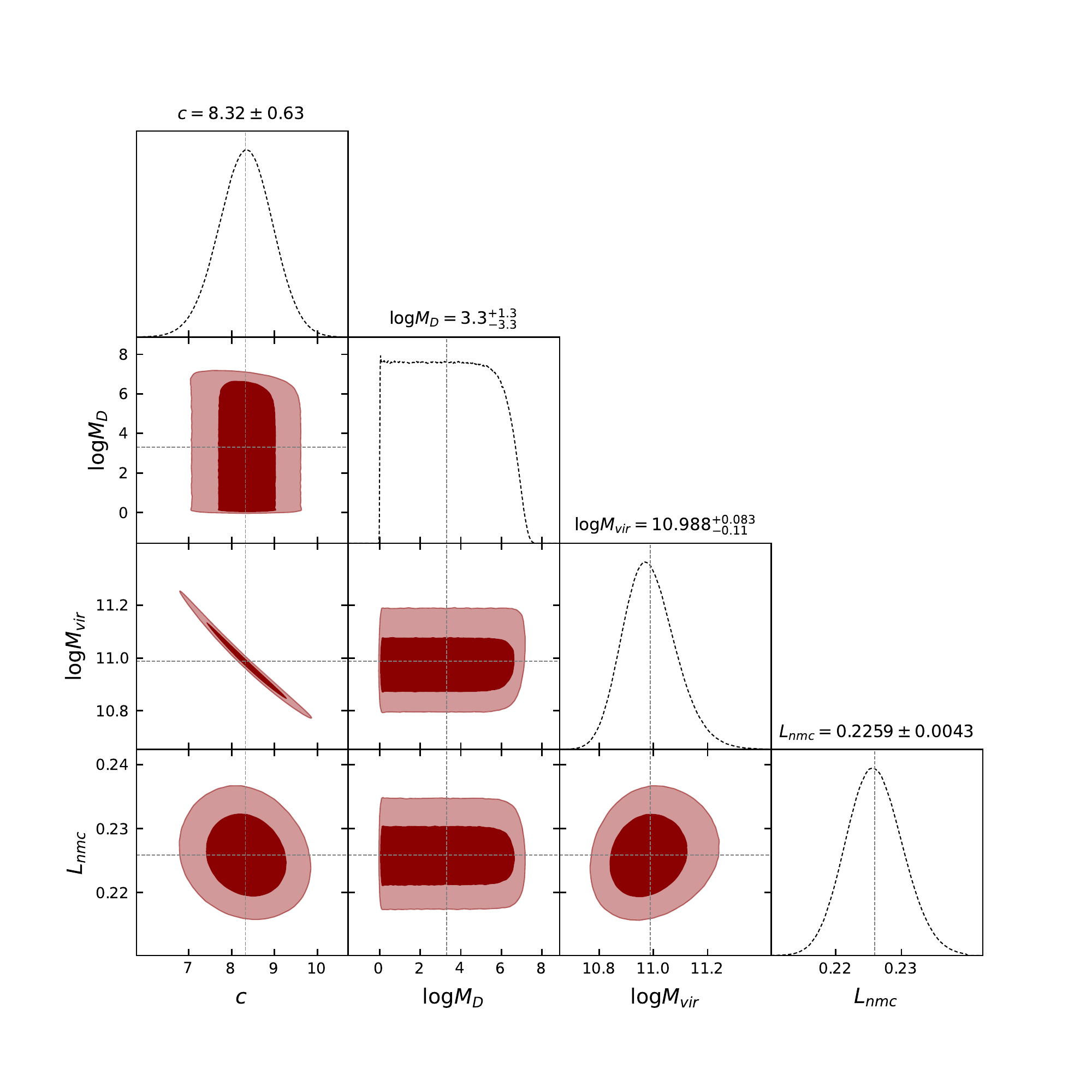}
    \end{center}
    \caption{The same of Fig.~\ref{fitbin5} for the dwarf galaxies by \cite{2017MNRAS.465.4703K}.\label{Dw}}
\end{figure}

Finally, in Fig.~\ref{Dw} the dwarf galaxy bin is analysed. Since it was originally designed on purpose, it is not surprising that in this case the Burkert profile yields the best description of the RC with a reduced $\chi_{\mathrm{red}}^2 \approx 0.8$. However, the NMC model performs decently with $\chi_{\mathrm{red}}^2 \approx 4$, and substantially better than the NFW profile for which $\chi_{\mathrm{red}}^2 \approx 14$. Note that such galaxies are strongly DM dominated in the region probed by the RC, hence the disk mass is vanishingly small and/or unconstrained by all models.

\begin{figure}[!htb]
    \begin{center}
        \includegraphics[width=1.\textwidth]{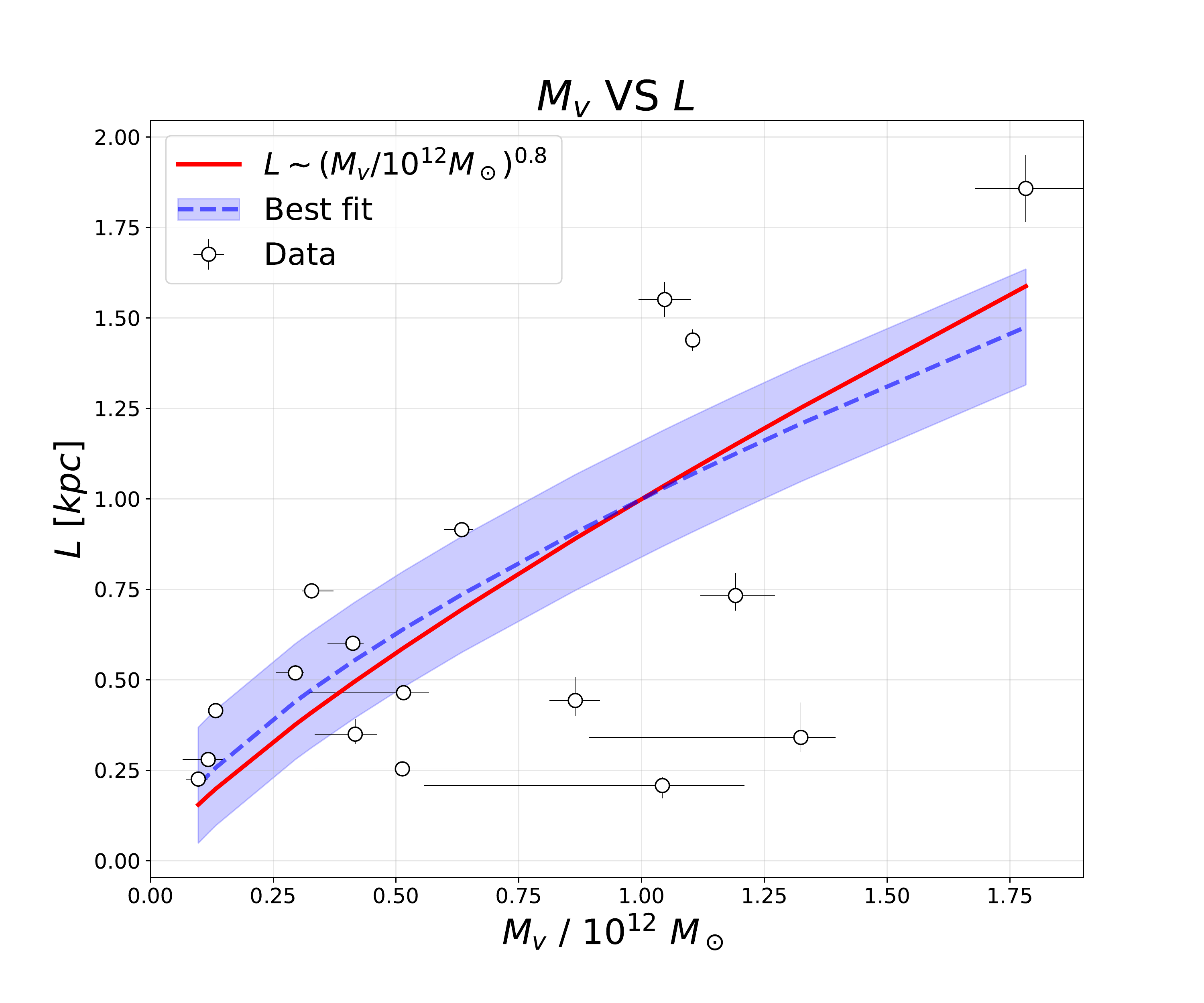}
        \caption{Scaling between $L$ and $M_{\rm v}$ found in the RC fit analysis of Sec.~(\ref{3|fitRCs}). The blue dashed line represents the best fit of data to a simple power function, resulting in a slope $m_{\mathrm{best}} = (0.67 \pm 0.16)$ (the shaded area represents a one-sigma confidence interval). The red solid line instead represents the generalization to baryonic-rich objects of the scaling $L \propto M_{\rm v}^{0.8}$ found in \citealt{Gandolfi:2021jai} in the DM-dominated dwarf galaxies regime.}
        \label{mv-lnmc}
    \end{center}
\end{figure}

An overall interesting result is that the NMC model predicts higher values of the length-scale $L$ in DM halos of higher virial masses --- see Fig.~(\ref{mv-lnmc}). This trending is well reproduced by the scaling $L(M_v) \propto M_v^{0.8}$, a result consistent with the findings of \cite{Gandolfi:2021jai} for DM-dominated dwarf galaxies.

As can be seen by looking at the overall results listed in the Appendix and recapped in Tables \ref{burkfit}-\ref{nfwfit}-\ref{nmcfit}, the NMC model yields RC fits that are always superior to the pure NFW one and in several instances comparable or even better than the Burkert model. Furthermore, we performed an F-test to compare the NFW and the NMC models, see Table \ref{nmcfit}. Overall, the test suggests that the addition of the parameter $L$ effectively improves the fits for the majority of the bins. Two caveats are in order here. First, the Burkert profile is phenomenological, and has been designed specifically to fit the RC of dwarf galaxies. Contrariwise, our NMC model is derived theoretically from first principles (though in a specific scenario), so the fact that its performances on RC fitting for different kind of galaxies improves substantially over the pure NFW shape is in itself encouraging. Second, we will show in the next Section that the NMC model will perform better than the Burkert profile in reproducing the RAR. 

\section{Testing the NMC with the RAR}\label{4|RAR}

The radial acceleration relation (or RAR) was originally proposed in \cite{McGaugh:2016leg} by exploiting the individual high-quality RCs of the SPARC sample (see \citealt{2016AJ....152..157L}). As argued in \cite{Lelli:2017vgz}, the RAR subsumes and generalizes a plethora of well-known dynamical laws of galaxies, such as the Baryonic Tully-Fisher relation (\citealt{1999ASPC..182..528M}; \citealt{Wheeler_2019}), the dichotomy between high and low surface brightness galaxies (\citealt{1997MNRAS.290..533D}; \citealt{1997ApJ...484..145T}), and others (\citealt{2013MNRAS.433L..30L}; \citealt{2016ApJ...827L..19L}; \citealt{1986RSPTA.320..447V}; \citealt{2004IAUS..220..233S}; \citealt{1976ApJ...204..668F}; \citealt{2016MNRAS.460.1382S}).

In \cite{Lelli:2017vgz}, an overall representation of the RAR was introduced in terms of the function
\begin{equation}
\label{fit2}
    g_{\mathrm{obs}}=\frac{g_{\mathrm{bar}}}{1-e^{-\sqrt{g_{\mathrm{bar}} / g_{\dagger}}}}+\hat{g} e^{-\sqrt{g_{\mathrm{bar}} g_{\dagger} / \hat{g}^{2}}},
\end{equation}
with $g_{\dagger}$ and $\hat{g}$ being fitting parameters. Eq.~(\ref{fit2}) with $\hat{g}=0$ accurately represents the RAR for spiral and irregulars, while the additive term depending on $\hat{g}$ describes the flattening of the RAR in the typical acceleration regime proper of dwarf spheroidal (dSphs) galaxies. All in all, the values $g_{\dagger} = (1.1 \pm 0.1) \times 10^{-10} \mathrm{~m} \mathrm{~s}^{-2}$ and $\hat{g} =(9.2 \pm 0.2) \times 10^{-12} \mathrm{~m} \mathrm{~s}^{-2}$ are derived from the analysis of the overall SPARC sample.
In the recent literature, it was argued that the parameter $g_{\dagger}$ could represent an acceleration scale governing the average internal dynamics of galaxies. Since the existence of such a scale in the standard cosmological model is far from trivial, this phenomenon has been interpreted as a possible sign of modified gravity (e.g., \citealt{Hossenfelder:2018vfs}; \citealt{Green:2019cqm}; \citealt{2019AN....340...95O}; \citealt{Islam:2019iua}; \citealt{Petersen:2020vks}).
More specifically, in the MOND framework the empirical constant $g_{\dagger}$ would be interpreted as the fundamental acceleration scale $a_0$. Indeed, the value of $g_{\dagger}$ derived both in \cite{McGaugh:2016leg} and \cite{Lelli:2017vgz} is compatible to the expected value of the MONDian characteristic acceleration scale $a_0 \sim 1.2 \times 10^{-10} \mathrm{~m} \mathrm{~s}^{-2}$. Notice, however, that such an interpretation is still highly debated, with some works supporting it (see, e.g., \citealt{2018A&A...615A...3L}; \citealt{2019MNRAS.487.2148G}) and some others ruling it out (e.g., \citealt{Marra:2020sts}; \citealt{Rodrigues:2020gbg}). 

Our aim here is to determine whether our NMC model can adequately reproduce the RAR, and whether it can do so with values on the NMC length-scale that are consistent with those derived from the previous analysis of stacked RC data. Toward this purpose, first notice that the RAR is a local scaling law that combines data at different radii in galaxies with different masses, which feature different contribution of stellar disc and bulge, gas and DM. Therefore we approach the problem via a semi-empirical method: we first build up mock RCs of galaxies with different properties, and then we sample them to derive the total and baryonic accelerations and construct the RAR, as detailed below.

\subsection{Mock RC modelling}\label{4.1|synthRCs}

Our procedure to build up mock RCs consists of the following steps.

\begin{itemize}

\item \underline{DM mass} 

We start by randomly drawing a very large number of total DM halo masses $M_{\mathrm{v}}$ within the range $8 < \log(M_{\mathrm{v}}/M_{\odot}) < 13.3$ according to the local halo mass  function (a uniform sampling does not impact appreciably the final outcomes). 

    \item \underline{Stellar mass} 
    
    We then derive the stellar mass associated to each galaxy by using the relation found by \cite{2013ApJ...770...57B} through abundance matching technique:
    \begin{equation}
        \log M_\star=\log\left(\varepsilon M_{1}\right)+f\left[\log\left(\frac{M_{\rm v}}{M_{1}}\right)\right]-f(0),
    \end{equation}
    \begin{align*}
        f(x)=-\log\left(10^{\alpha x}+1\right)+\delta \frac{\log[1+\exp (x)]^{\gamma}}{1+\exp \left(10^{-x}\right)},
    \end{align*}
    with $\log M_{1} = 11.514$ being a characteristic halo mass, and parameters $\log \varepsilon = -1.777, \alpha=-1.412, \delta=3.508, \gamma=0.316$. We allow for a log-normal scatter of $0.25$ dex.

    \item \underline{Gas mass} 
    
    We determine the gas mass by exploiting the relation found with the stellar mass by \cite{Papastergis:2012wh} and \cite{Peeples:2013rka}: 
    \begin{equation}
    \label{hydrogen}
        \log \left(\frac{M_{\text{HI}}}{M_\star}\right)=-0.43 \log \left(\frac{M_\star}{M_{\odot}}\right)+3.75,
    \end{equation}
    allowing a lognormal scatter of $0.15$ dex. Note that we are implicitly assuming that in local galaxies the majority of the intestellar medium consists in atomic hydrogen $\text{H}$I and that both the ionized and the molecular components are minor (see \citealt{Papastergis:2012wh}, \citealt{2011MNRAS.415...32S}). The total gas mass is $M_{\rm gas}\approx 1.33\, M_{\rm HI}$ to account for the contribution of He. 
    
    \item \underline{Stellar and gas radial distributions} 
    
    We assume that the gaseous and the stellar components are both distributed in a razor-thin exponential disk. We determine the stellar disk half-mass radius from the stellar mass via the relation by \cite{2003MNRAS.343..978S}
        \begin{equation}
        \log\left(\frac{R_\mathrm{e}}{\mathrm{kpc}}\right)=\frac{1}{2.47}\left(\log\left(\frac{\mathrm{M}_\star}{M_{\odot}}\right)-7.79\right). 
    \end{equation}
    applying for $M_\star < 10^{9} M_\odot$, and that by \cite{2015MNRAS.447.2603L}
    \begin{equation}
        R_{\mathrm{e}}=0.13\left(\frac{M_\star}{M_{\odot}}\right)^{0.14}\left(1.0+\frac{M_\star}{14.03 \cdot 10^{10} \mathrm{M}_{\odot}}\right)^{0.77}\, \mathrm{kpc}
    \end{equation}
    for $M_\star \geq 10^{9}\, M_\odot$. The stellar disk scale-length is given by $R_{\rm d}\approx R_e/1.678$. We allow for a lognormal scatter of both these relations around $0.1$ dex. The gas distribution scale-length is taken as $R_{\mathrm{gas}}=2\, R_{\mathrm{d}}$.
    
    \item \underline{Bulge mass} 
    
    We determine the bulge mass using the relation with the stellar mass by \cite{2009MNRAS.393.1531G} and \cite{2012ApJ...745...66M}:
    \begin{equation}
        \frac{M_{\text {B}}}{M_\star}=\frac{\log \left(M^\star/M_{\odot}\right)-9.5}{4.2},
    \end{equation}
    with a lognormal scatter of $0.1$ dex. The implied bulge-to-total mass ratio is $\sim 30 \%$ for Milky-Way like galaxies. We further assume that the bulge is present only if $M_\star \geq 3 \times 10^{9} M_{\odot}$.
    
    \item \underline{Bulge radial distribution} 
    
    We assume that the bulge mass is radially distributed according to an Hernquist profile  (see \citealt{1990ApJ...356..359H}) 
    \begin{equation}
        \rho(r)=\frac{M_{\text {B}} R_{1 / 4}}{2 \pi r\left(R+R_{1 / 4}\right)^{3}},
    \end{equation}
    where $R_{1 / 4}$ is the radius at which the enclosed bulge mass is a quarter of its total value. The half-mass radius $R_{1 / 2}=(1+\sqrt{2}) R_{1 / 4}$ is gauged on the basis of the scaling relation with the bulge mass by \cite{2009MNRAS.393.1531G}
    \begin{equation}
        \log\left(\frac{R_{1 / 2}}{\mathrm{kpc}}\right)=0.30 \log\left(\frac{M_{\text {B}}}{{M_{\odot}}}\right)-3.124,
    \end{equation}
    with a lognormal scatter of $0.1$ dex.

\item \underline{DM radial distribution} 

We radially distribute the DM mass according to various profiles, to test their performance on the RAR. For the NFW and the NMC models we use Eq.~(\ref{NFW}); the concentration parameter $c$ is determined according to the relation with the halo mass from \cite{Dutton:2014xda}
    \begin{equation}
    \label{c-mvir}
        \log c = 0.905-0.101 \log\left(M_{\rm v} /10^{12} h^{-1} M_{\odot}\right),
    \end{equation}
    with a lognormal scatter of 0.11 dex. 
    
    For the Burkert model we use Eq.~(\ref{burkert}) by setting the core radius $r_0$ from two conditions: (i) the mass within the virial radius must match $M_{\rm v}$; the core radius and core density must satisfy the universal core-column density relation $\rho_0\times r_0\approx 75\, M_\odot$ pc$^{-2}$, with a scatter of 0.2 dex (see e.g. \citealt{Salucci:2000ps}, \citealt{10.1111/j.1365-2966.2009.15004.x}, \citealt{burkert2015structure}, \citealt{2013ApJ...770...57B}, \citealt{Burkert:2020laq}).

    Finally, we consider the profile emerging from the hydrodynamical simulations by \cite{2014MNRAS.441.2986D} which take into account DM responses to baryonic effects (including stellar feedback); this is basically a generalized NFW profile 
    \begin{equation}
    \label{dc+14}
        \rho(r)=\frac{\rho_{s}}{\left(r/r_{\mathrm{s}}\right)^{\gamma}\left[1+\left(r/r_{\mathrm{s}}\right)^{\alpha}\right]^{(\beta-\gamma) / \alpha}}.
    \end{equation}
    with shape parameters linked to the  stellar-to-halo mass ratio $X\equiv M_\star/M_{\rm v}$ (see also \citealt{Stinson_2013}) as
    \begin{equation}
    \label{dc+14param}
        \begin{array}{l}
    \alpha=2.94-\log\left[\left(10^{X+2.33}\right)^{-1.08}+\left(10^{X+2.33}\right)^{2.29}\right], \\
    \beta=4.23+1.34 X+0.26 X^{2}, \\
    \gamma=-0.06+\log\left[\left(10^{X+2.56}\right)^{-0.68}+\left(10^{X+2.56}\right)\right]\; .
    \end{array}
    \end{equation}

\item\underline{Building up the mock RC}\label{3.3|synthesis}

For any galaxy of given virial mass $M_{\rm v}$, we have now specified all the mass components and the associated radial distribution $M_i(<R)$, so that the RC can be easily determined from $v_i^2(R)=G\, M_i(<R)/R$; the only exception is the NMC model for which the DM velocity has an additional term $v_{\rm DM}^2(r)=G\, M_{\rm DM}(<r)/r-\epsilon\, L^2\, r\, 4\pi\, G\, {\rm d\rho}/{\rm d}r$. The overall mock RC is then the sum of all the different contributions $v_{\rm tot}^2=\sum_i\, v_i^2$.

In Fig.~\ref{velcurv} we illustrate four representative mock RCs for galaxies with different halo masses $M_{\rm v}$, highlighting the diverse behavior when assuming the NFW, the Burkert, the Di Cintio or the NMC halo profiles. As for the baryonic components, in moving toward smaller halo masses, the inner contribution due to the bulge component becomes less prominent, while the gas contribution progressively increase to become even dominant over the stellar disk. As for the DM models, the halo shapes are rather different, with the Burkert profile yielding overall higher velocities in lower mass galaxies. In order to further test the realism of our mock RCs, we compared them to the stacked empirical RCs considered in Sec.~(\ref{3|fitRCs}). The outcome is shown in Fig.~(\ref{bincomp}) showing the compatibility between the mock curves and empirical, stacked ones.

\end{itemize}

\subsection{Building the RAR and results}

Once the mock RC for each mock galaxy has been characterized, we compute the gravitational acceleration following 
\begin{equation}
    |g_j(r)| = \frac{v^2_j(r)}{r},
\end{equation}
with the index $j=$ bar, tot specifying the baryonic contribution or the total value including DM. The RAR is then constructed by binning our mock galaxy sample in $g_{\rm bar}$ and extracting the average values and standard deviation of $g_{\rm tot}$. For fair comparison with the data, the sampled portion of the RC is restricted to twice the optical radius of each mock galaxy. It is clear that this procedure includes in a given bin of $g_{\rm bar}$ objects with different halo masses and at different radii; e.g., an object can display a low baryonic acceleration either because it has a small halo mass or because its RC is sampled at large radii.

\begin{figure}[!htb]
    \begin{center}
        \includegraphics[width=1.\textwidth]{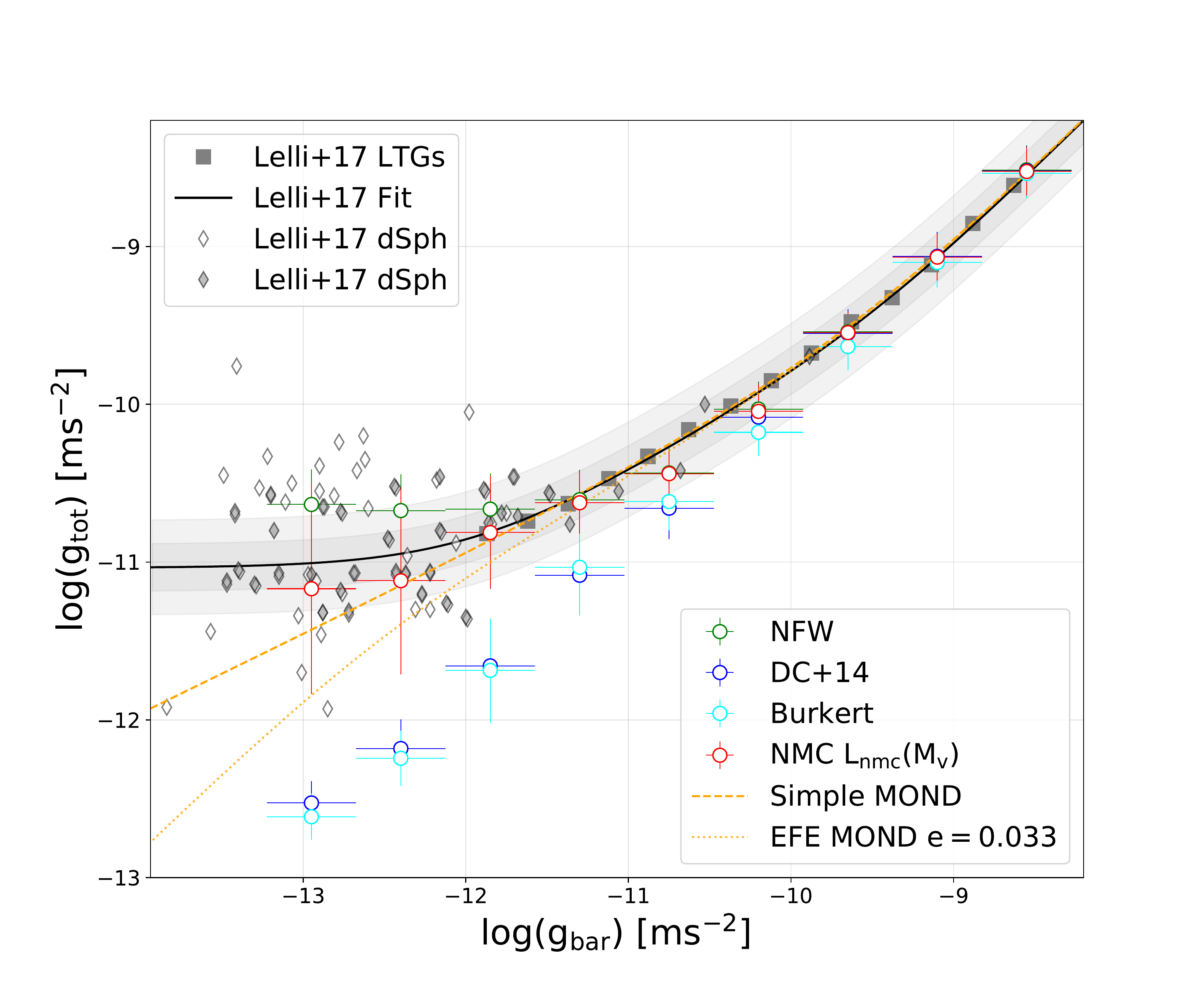}
        \caption{The Radial Acceleration Relation or RAR. The solid black line with shades illustrate the average results and its 2$\sigma$ and 3$\sigma$ variance from the analysis of the SPARC database by \cite{Lelli:2017vgz}; in particular, grey squares refer to the binned outcome for normal spiral galaxies and diamonds to measurements in individual dwarf spheroidal (filled symbols are more secure determinations). Such dwarf spheroidals have large error bars that are not displayed in this plot for visual clarity, thus the extension of the fit line through this cloud is much less certain than for the LTGs. The colored circles illustrate the prediction from our empirical modeling of RCs when adopting different halo profiles:  NFW (green), Di Cintio (blue), Burkert (cyan) and NMC with a mass dependent scaling for the coupling length-scale $L$ (red; see text for details). For reference, the MOND expectations without (dashed orange) and with (dotted orange) the external field effect (to the value $e=+0.033$ estimated in \citealt{Chae:2020omu}) is displayed.}
        \label{rar_models}
    \end{center}
\end{figure}

In Fig.~\ref{rar_models} we illustrate our results on the RAR, for the DM models listed above (color-coded). For comparison, we report as a black line with shaded area the determination by \cite{Lelli:2017vgz} represented by Eq.~(\ref{fit2}); binned data for spirals and irregulars are represented by grey squares and individual data for dSph galaxies are highlighted with diamonds (filled symbols are for more secure determinations). 
There is a substantial agreement of the RAR for all the DM models at high baryonic accelerations. This is because such a regime is mainly dominated by the contribution at small radii in high mass galaxies. There the total gravitational acceleration is anyway dominated by baryons, implying $g_{\rm tot}\approx g_{\rm bar}$ irrespective of the specific DM profile. 
However, a marked difference among the RAR for different DM models sets in toward lower baryonic accelerations. Such a regime is dominated by the behavior at small/intermediate radii in intermediate and low mass galaxies. There the total baryon acceleration $g_{\rm bar}$ is dominated by the stellar disk, while the total gravitational acceleration is contributed by both the disk and the halo $g_{\rm tot}\approx g_{\rm DM}+g_{\rm bar}$; thus depending on the DM model most of the contribution to $g_{\rm tot}$ may come either from the disk enforcing a behavior of the RAR similar to the high acceleration regime, or from the DM enforcing an upward deviation of the RAR. 

All in all, both the RAR associated to the Di Cintio and the Burkert models tend to appreciably deviate downward, to the point of becoming inconsistent with the measured RAR (especially in dSph) for $g_{\rm bar}\lesssim 10^{-11}$ m s$^{-2}$.
Contrariwise, the RAR of the NFW model displays the opposite behavior, with the corresponding curve flattening and progressively saturating to values slightly above the observed RAR, though still consistent with the upper outliers; nevertheless, one must keep in mind that NFW model suffers from the poor performances in fitting the individual RCs of many dSphs (e.g., \citealt{2010AdAst2010E...5D}), and the stacked dwarf galaxy RCs analysed in this paper. 
Finally, the NMC model can reproduce the average measured RAR when extrapolating down to smaller masses, the dependence 
$L(M_{\rm v})\propto M_{\rm v}^{0.8}$ found from the RC analysis of Sec.~(\ref{3|fitRCs}) and to be consistent in the dwarf irregular regime with Eq.~(9) of \cite{Gandolfi:2021jai}. We find that the RAR thus produced follows a profile intermediate between the NFW and the cored ones. Remarkably, the NMC one is the only model considered here that can simultaneously reproduce the RAR and decently fits the stacked RC of spirals, LSBs and dwarf galaxies.

For reference, we also illustrate the prediction on the RAR for the MOND framework. This may be derived from the relation $\mu(x)\,g_{\text{tot}} = g_{\text{bar}}$ where the simple interpolating function $\mu(x)=x/(1+x)$ with $x\equiv g_{\text{tot}}/a_0$ and $a_{0} \sim 1.2 \times 10^{-10} \mathrm{~m} \mathrm{~s}^{-2}$ is generally adopted (e.g., \citealt{2005MNRAS.363..603F}, \citealt{2006ApJ...638L...9Z}). The MOND outcome is quite close to the measured RAR at high acceleration, while it lacks the progressive flattening at low $g_{\rm bar}$. 
However, some authors pointed out that this simple parameterization of MOND is not accurate because of the so called external field effect (EFE; see \citealt{1983ApJ...270..384M}; \citealt{2010ApJ...718..380S}; \citealt{Candlish:2018pvu}) associated to the violation of the strong equivalence principle in the relativistic MOND theory and implying that a galaxy dynamics  depends also on the gravitational pull $g_{\rm ext}$ from external fields (e.g., external galaxies or large-scale surroundings). One can account for the EFE by modifying the interpolating function to read $\mu(x)=(x/1+x+e)\, [1+(2+e)\,e/x(1+e)]$, with $e=g_{\rm ext}/a_0$ being the strength of the effect with respect to the MOND acceleration scale (see \citealt{2021arXiv211002263T}).   
This parameter was estimated to be around $e\approx 0.033$ by \cite{Chae:2020omu} and \cite{Chae:2021dzt} from the analysis of individual galaxy RCs (\citealt{Desmond:2017ctk}).
The RAR from MOND theory including the EFE deviate downward at low accelerations, and can possibly account for some of the bottom outliers. However, to reproduce the observed RAR for the bulk of the galaxies would require to have negative values of $e$, which are not supported by observational estimates and known to be theoretically unfeasible in the MOND framework (see \citealt{Chae:2020omu}).
 
\begin{figure}[!htb]
    \begin{center}
        \includegraphics[width=1.\textwidth]{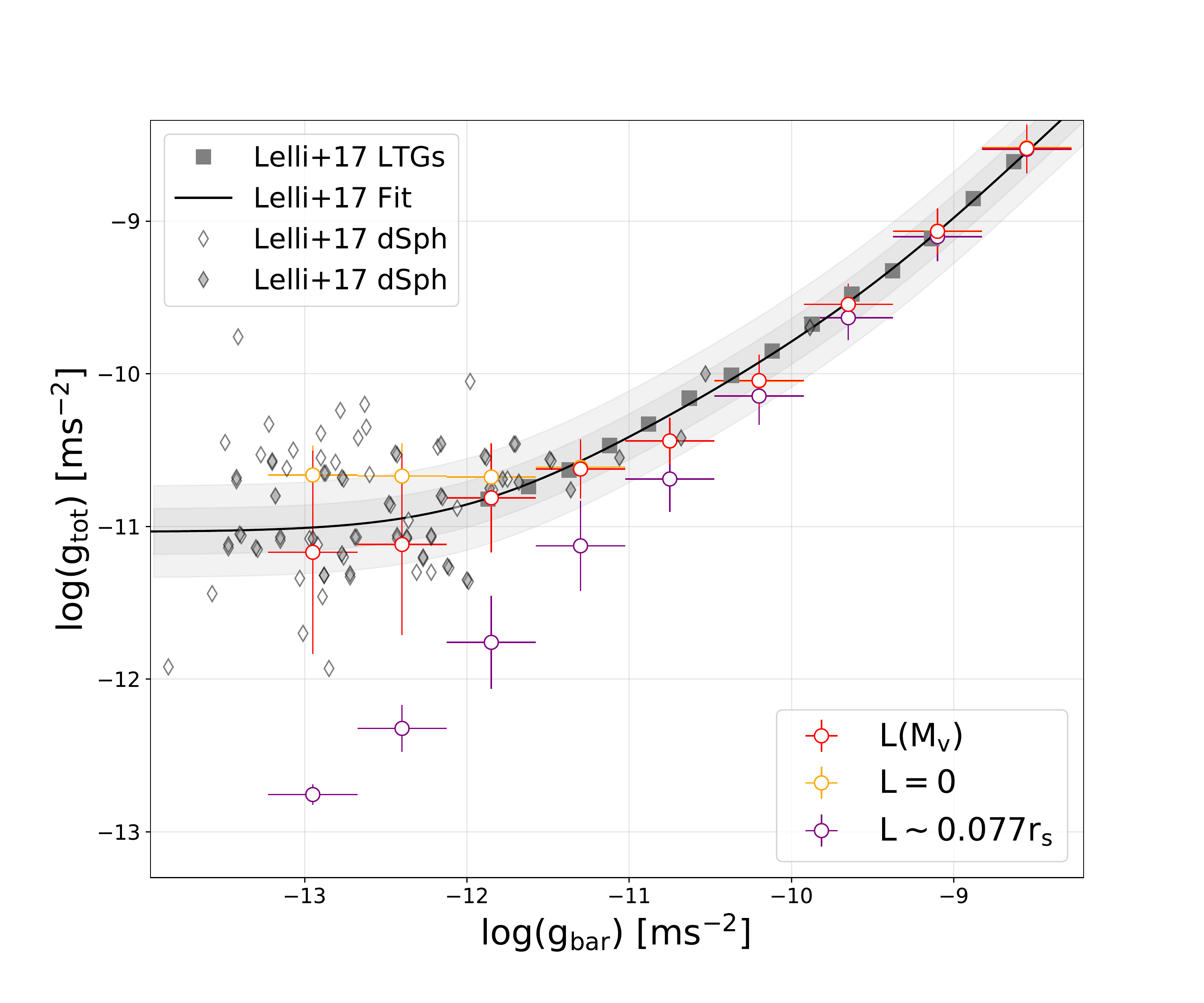}
        \caption{The RAR for the NMC model, with different coupling length-scales $L/r_s$. as expected, setting the coupling length to zero (orange) amounts in recovering the RAR reproduced by the NFW model. We also display the RAR for $L/r_s \sim 0.077$, i.e. the average value obtained from the RC analysis of large spiral galaxies (purple). Intermediate values for $L/r_s$ describe RARs that will lie between these two extremes. For reference the RAR obtained assuming a mass-dependent scaling for the coupling length-scale $L$ as in previous Figure is also reported (red).
        }
        \label{rar_nmc}
    \end{center}
\end{figure}

In Fig. \ref{rar_nmc} we plot the RAR expected from the NMC model for different values of the NMC length-scale $L/r_s$. Plainly, for vanishing $L/r_s$ the NFW outcome is recovered. For $L/r_s$ progressively increasing, the NMC model spans the dispersion of the outliers in the RAR at low baryonic accelerations.  

\section{Summary}\label{5|conclusion}

We have looked for empirical evidence of a non-minimal coupling (NMC) between dark matter (DM) and gravity in the dynamics of local spiral galaxies. 
In particular, taking up the work by \cite{Gandolfi:2021jai} we have considered a theoretically motivated non-minimal coupling that may arise dynamically from some collective behavior of the coarse-grained DM field (e.g., Bose–Einstein condensation) with averaging/coherence length $L$. In the Newtonian limit, this non-minimal coupling amounts to modify the Poisson equation by a term $L^{2} \nabla^{2} \rho$ proportional to the Laplacian of the DM density itself.  

We have then worked out how such a term, when acting as a perturbation over the standard NFW profile from gravity-only cold DM simulations, can substantially  alter the dynamical properties of galaxies, in terms of their total radial acceleration within the disk and rotation velocity. We have then tested such a model against dynamical data of local spiral galaxies. Our main results are as follows:

\begin{itemize}

\item We have found that the NMC model can fit the stacked RCs of local spiral galaxies with different average velocities at the optical radius, including dwarfs and low-surface brightness systems, at a level of precision superior to the NFW profile and comparable to (in some instances even better than) the phenomenological Burkert profile. 

\item We have shown that at the same time our NMC model, when extrapolating down to smaller masses the mass-dependent scaling of the coupling length-scale $L$ found from the RC analysis, can adequately reproduce in shape and scatter the radial acceleration relation (RAR) down to the dwarf spheroidal galaxy range, a task which constitutes a serious challenge for alternative DM profiles even inclusive of baryonic effects.

\end{itemize}

A possible future extension of the present work may include to trace the physical origin of the NMC in DM halos, especially in terms of the mechanism determining the NMC length-scale $L$ in different galaxies and originating the mass dependence required to fit the RAR at very low baryonic accelerations. In this vein, full $N-$body simulations incorporating the NMC hypothesis could be exploited to study time-dependent conditions and the overall formation of cosmic structures in such a framework. 
\\
\\
We warmly thank T. Ronconi and P. Salucci for the helpful insights and stimulating discussions. AL is supported by the EU H2020-MSCA-ITN-2019 Project 860744 BiD4BESt: Big Data applications for black hole Evolution STudies, and by the PRIN MIUR 2017 prot. 20173ML3WW: Opening the ALMA window on the cosmic evolution of gas, stars and supermassive black holes.

\clearpage

\bibliography{bibliography}{}
\bibliographystyle{aasjournal}

\clearpage

\clearpage

\clearpage

\clearpage

\clearpage
\begin{figure}[!htb]
    \begin{center}
    \includegraphics[width=.495\textwidth]{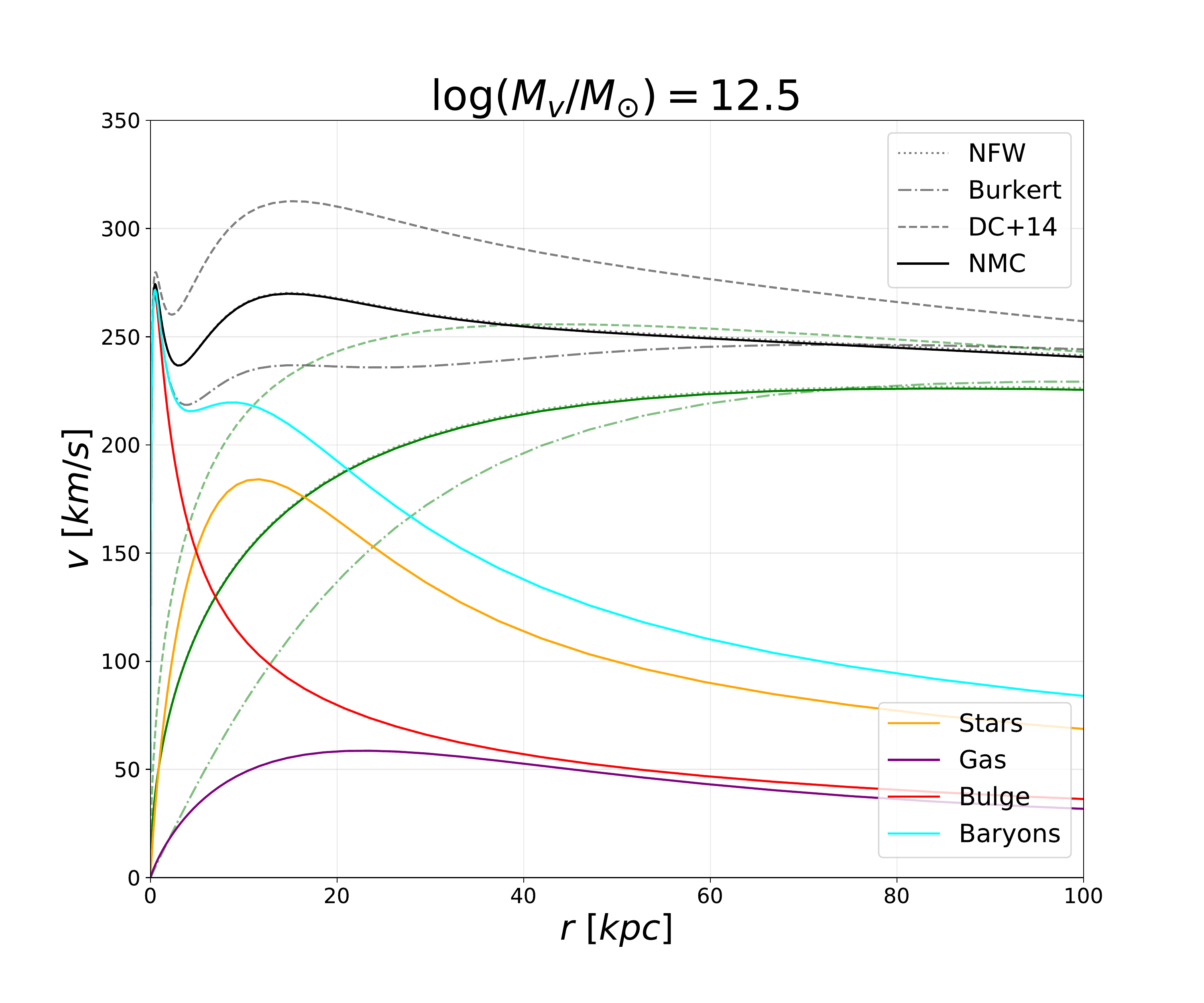}
    \includegraphics[width=.495\textwidth]{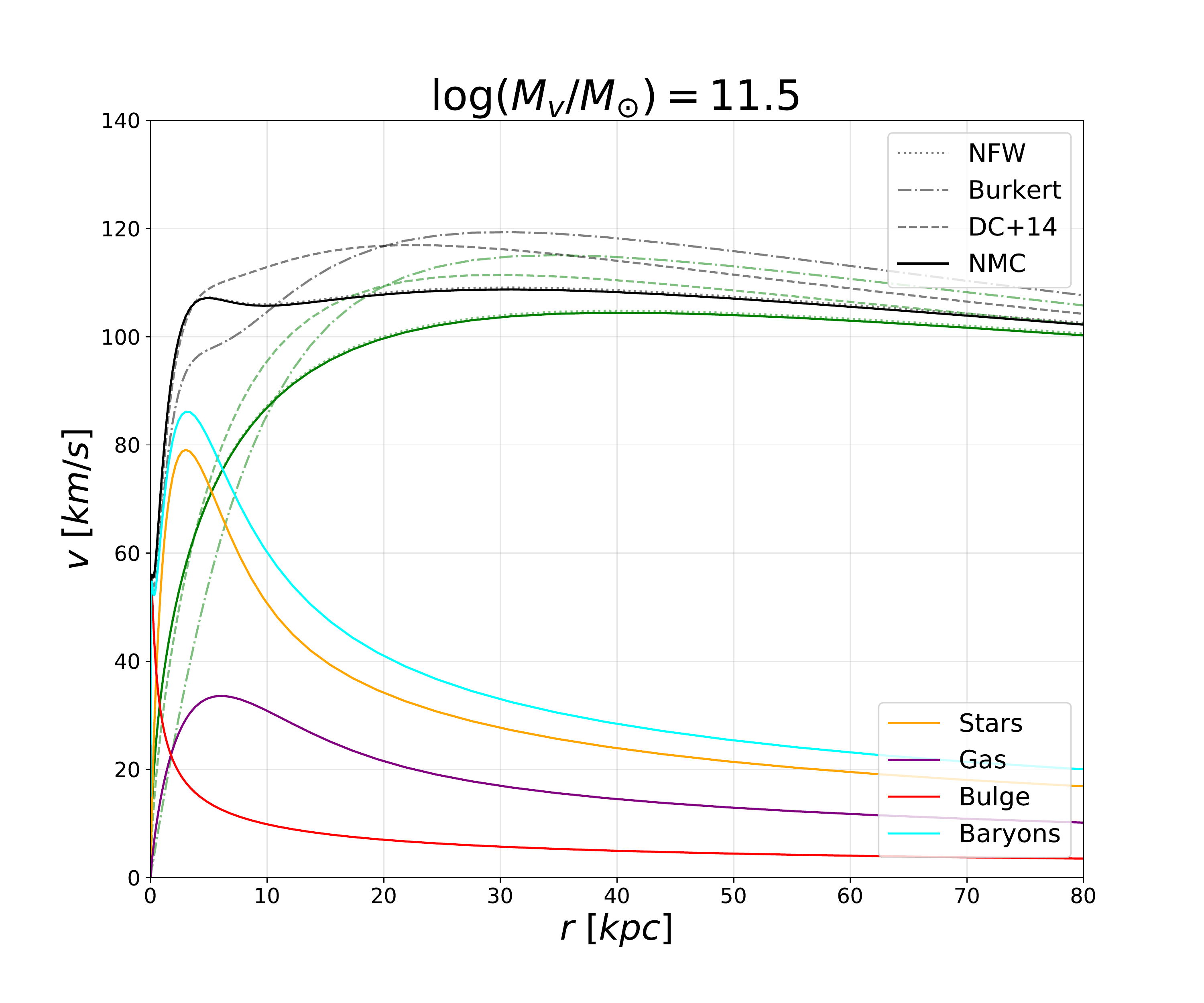}
    \includegraphics[width=.495\textwidth]{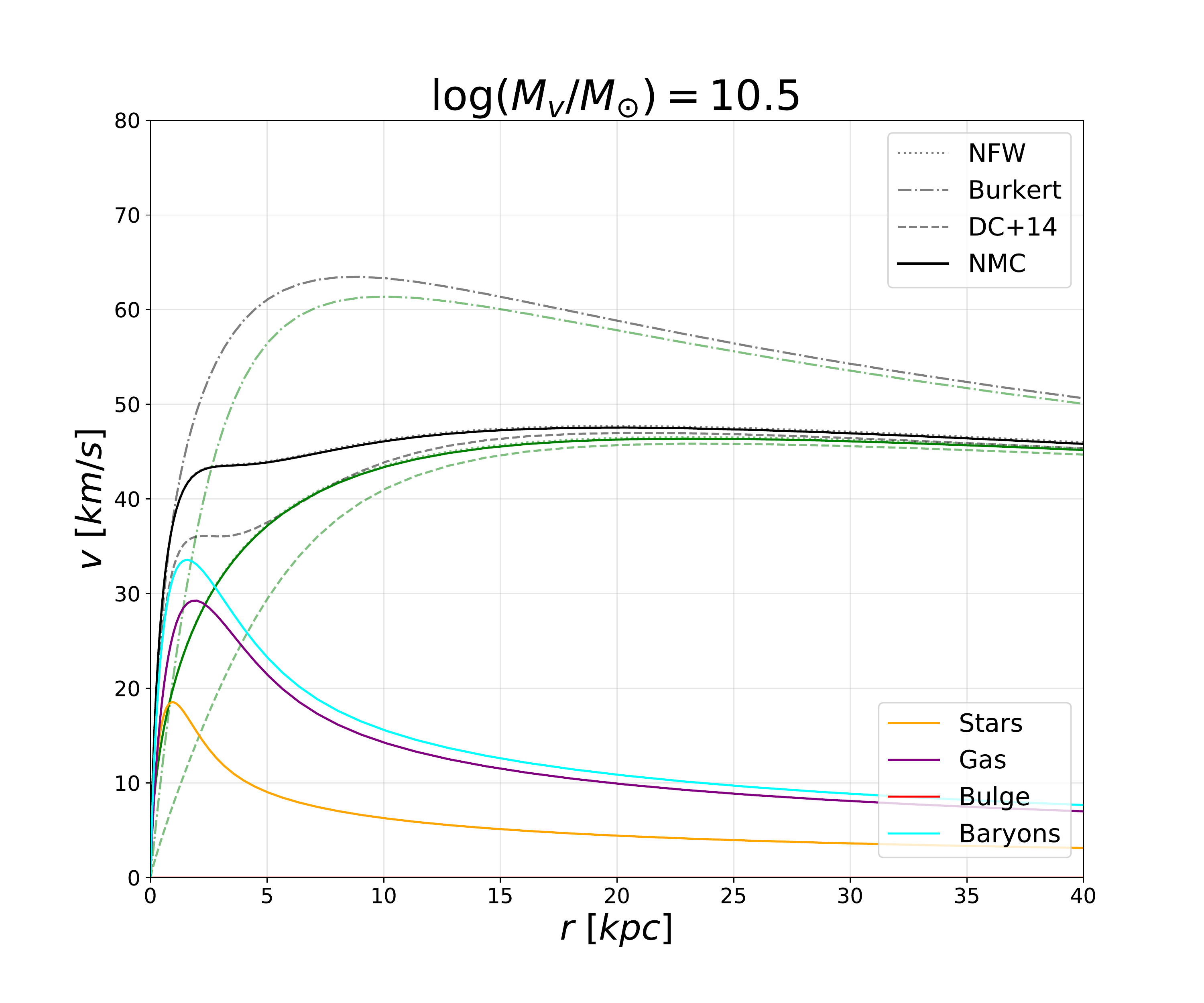}
    \includegraphics[width=.495\textwidth]{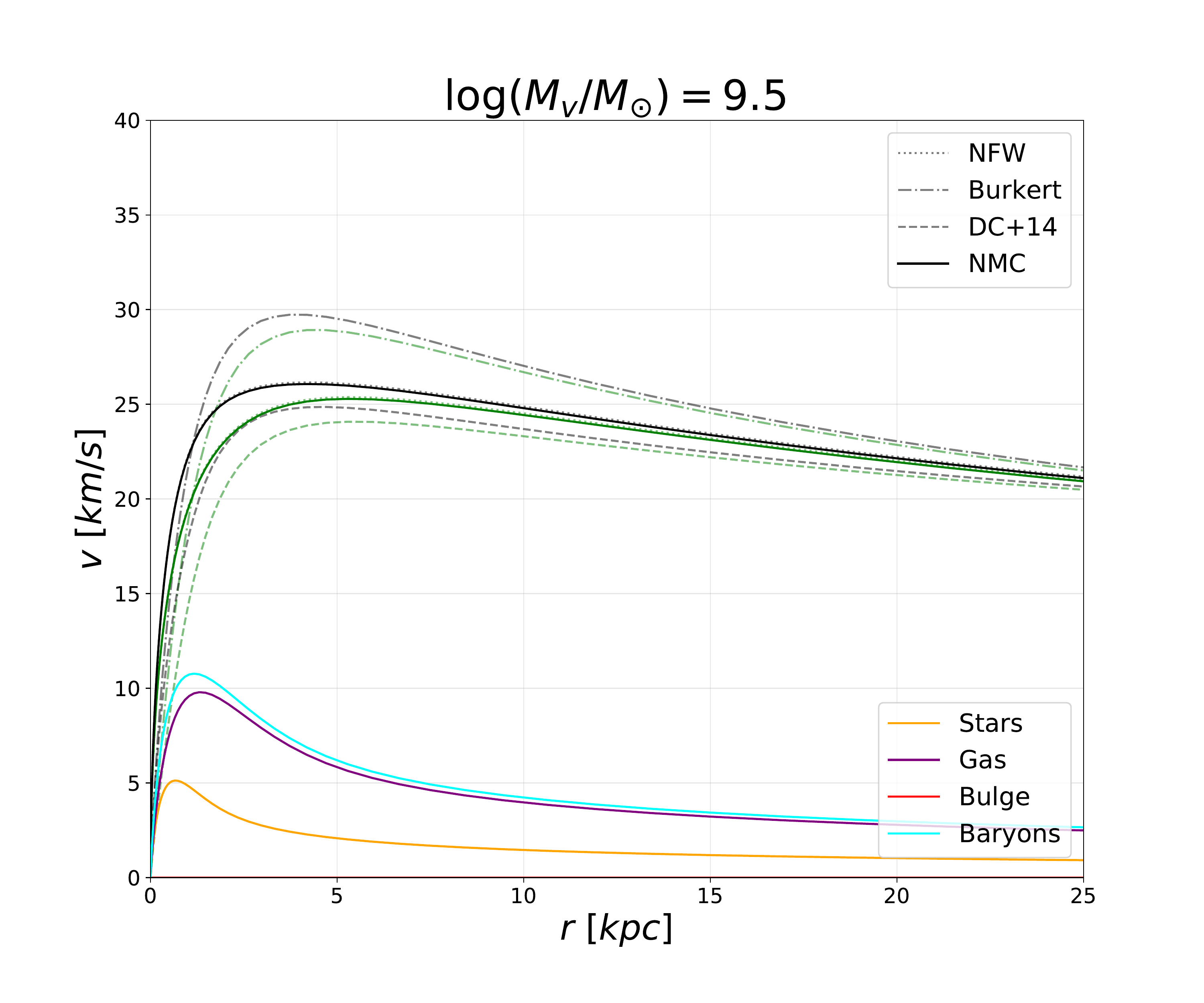}
    \end{center}
    \caption{Four representative mock RCs for different DM halo masses. In each panel, the contributions from stellar disk (orange), gas disk (purple), bulge (red), overall baryons=bulge$+$stars$+$gas (cyan), DM halo (green), and total (black) are shown. For the green and black colors, dotted lines refer to the NFW profile, dot-dashed lines to the Burkert profile, dashed lines to the DC+14 profile and solid lines refer to the NMC profile.\label{velcurv}}
\end{figure}

\clearpage
\begin{figure}[!htb]
    \begin{center}
    \includegraphics[width=.495\textwidth]{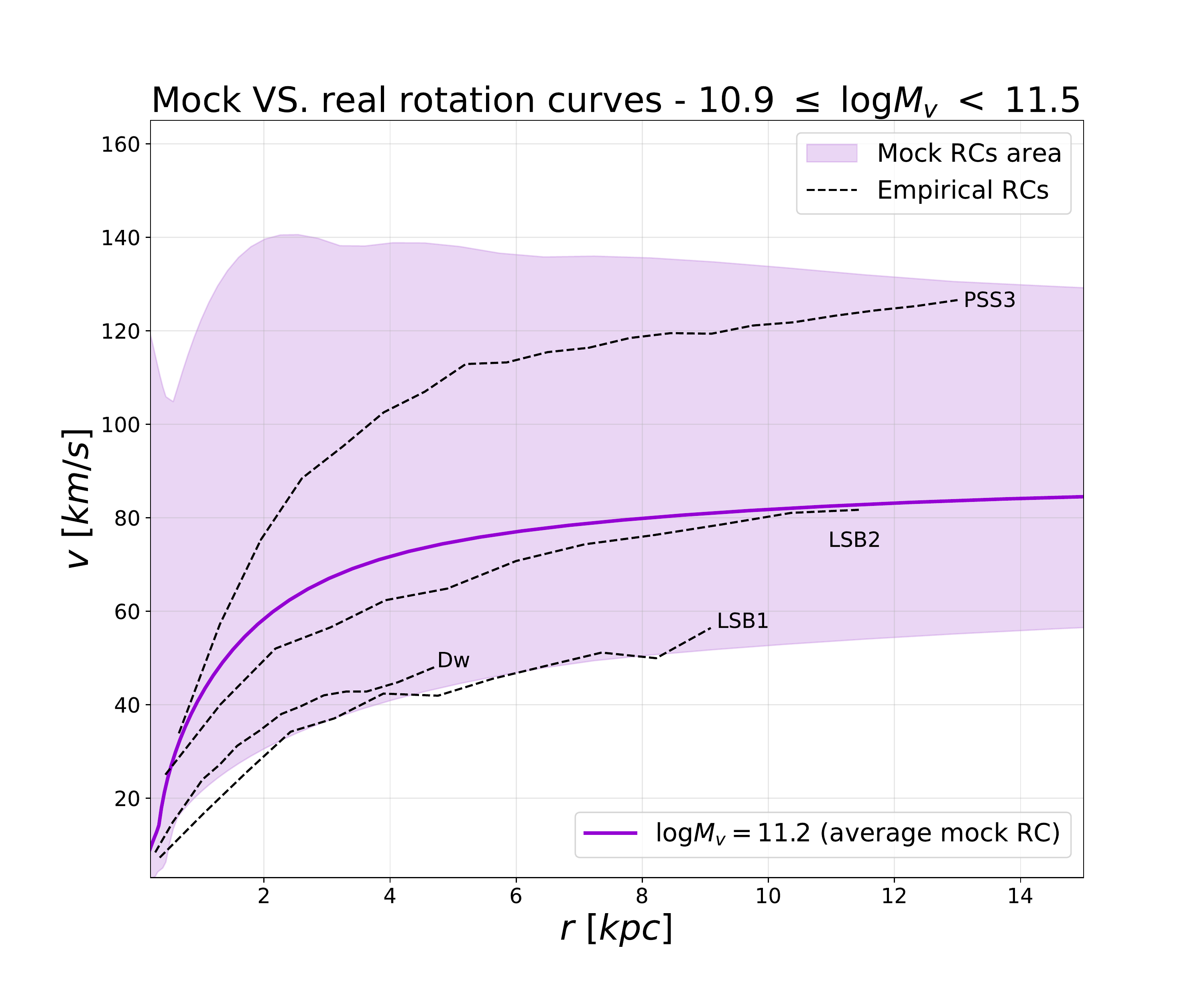}
    \includegraphics[width=.495\textwidth]{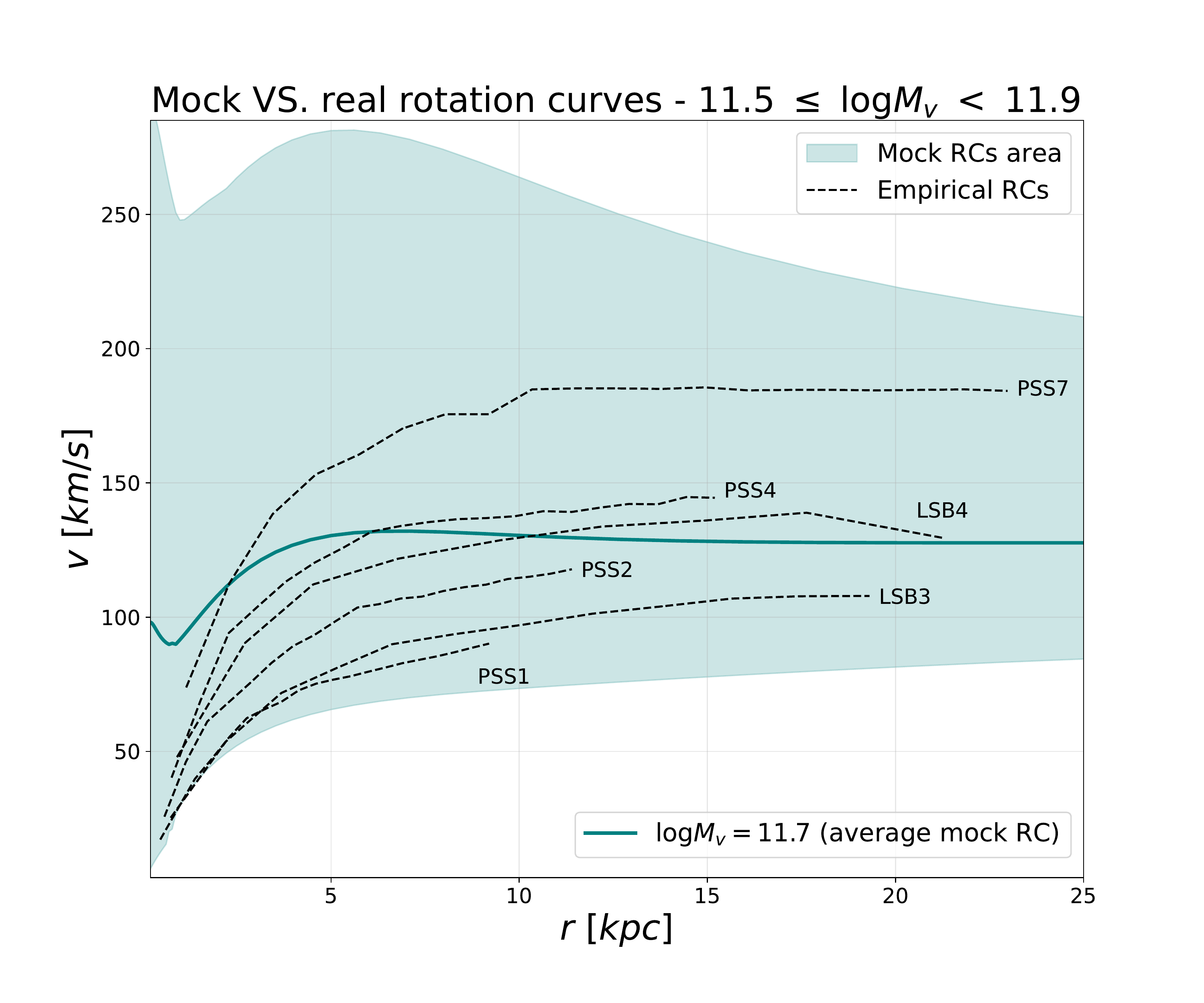}
    \includegraphics[width=.495\textwidth]{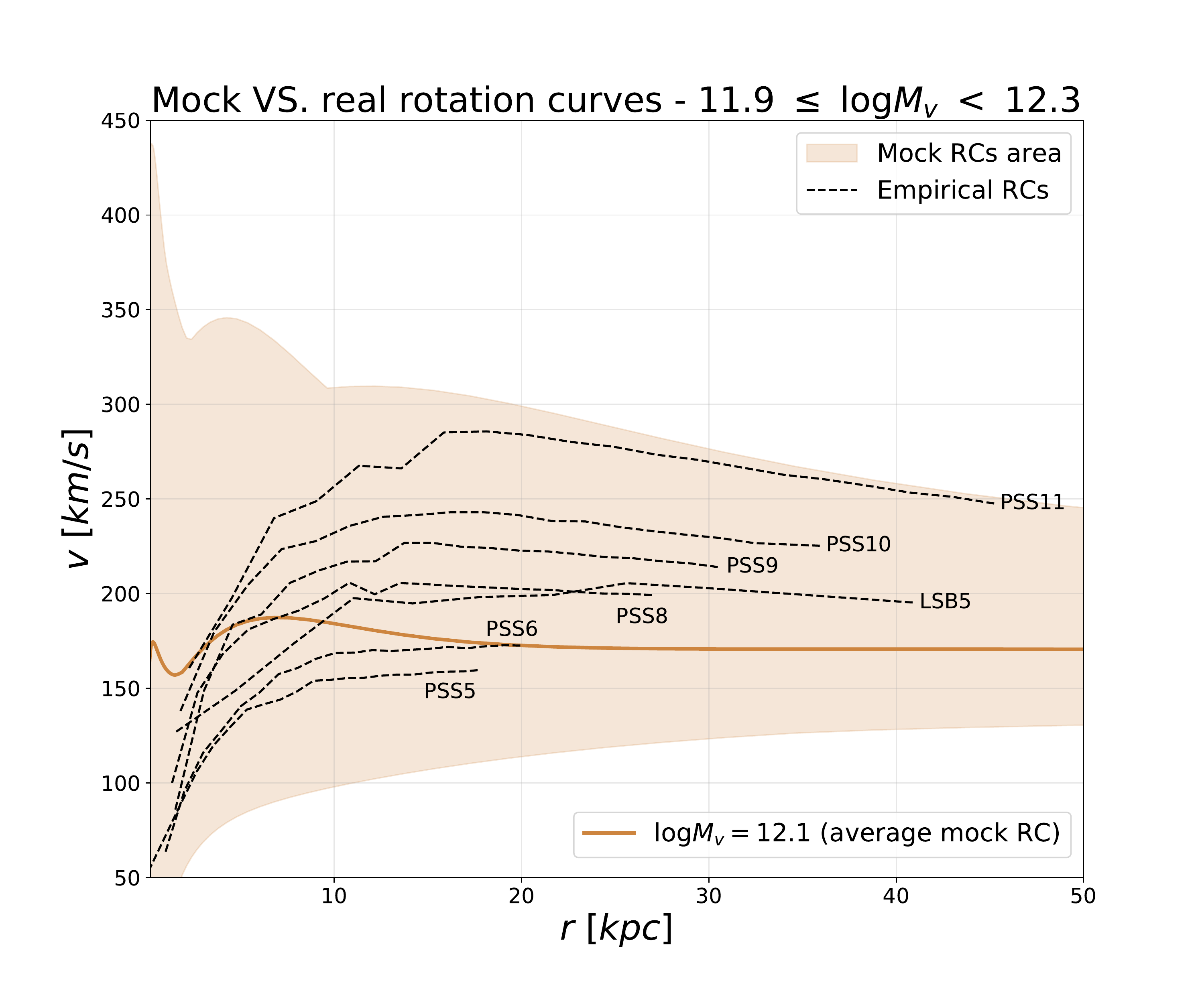}
    \end{center}
    \caption{Comparison between our mock RCs with the empirical stacked RCs utilized in Sec.~(\ref{3|fitRCs}). The stacked RCs (dashed black lines) are divided in three virial mass bins, and for each bin we have generated $10^3$ mock curves that are varying in the range outlined by the shaded area. For each bin, the average mock RC is plotted as a solid line. \label{bincomp}}
\end{figure}

\clearpage

\clearpage

\clearpage
\begin{deluxetable*}{ccccc}
\tablenum{1}
\tablecaption{Samples considered for the analysis of stacked RC in Sect. \ref{3|fitRCs}: PSS stands for the sample of normal spirals by \cite{1996MNRAS.281...27P}, LSB stands for the sample of low surface brightness spirals from \cite{Dehghani:2020cvl}, and Dw for the sample of dwarfs by \cite{2017MNRAS.465.4703K}. For each bin the optical radius $r_\mathrm{opt}$ and optical velocities $v_\mathrm{opt}$ are reported. \label{binlist}}
\tablewidth{0pt}
\tablehead{
\colhead{Sample/Bin} & \colhead{$r_\mathrm{opt}$ [$\mathrm{kpc}$]} & \colhead{$v_\mathrm{opt}$ [km s$^{-1}$]}
}
\startdata
\vspace{-0.35cm} &&&\\
\vspace{0.0cm} PSS 1 & $4.6$ & $75$ \\
\vspace{0.0cm} PSS 2 & $5.7$ & $104$ \\
\vspace{0.0cm} PSS 3 & $6.5$ & $116$ \\
\vspace{0.0cm} PSS 4 & $7.6$ & $135$ \\
\vspace{0.0cm} PSS 5 & $8.9$ & $154$ \\
\vspace{0.0cm} PSS 6 & $10.1$ & $169$ \\
\vspace{0.0cm} PSS 7 & $11.5$ & $185$ \\
\vspace{0.0cm} PSS 8 & $13.5$ & $205$ \\
\vspace{0.0cm} PSS 9 & $15.3$ & $225$ \\
\vspace{0.0cm} PSS 10 & $18.$ & $243$ \\
\vspace{0.0cm} PSS 11 & $22.7$ & $279$ \\
\vspace{0.0cm} LSB 1 & $5.5$ & $44$ \\
\vspace{0.0cm} LSB 2 & $6.9$ & $73$ \\
\vspace{0.0cm} LSB 3 & $11.8$ & $101$ \\
\vspace{0.0cm} LSB 4 & $14.5$ & $141$ \\
\vspace{0.0cm} LSB 5 & $25.3$ & $206$ \\
\vspace{0.0cm} Dw & $2.5$ & $40$ \\
\vspace{-0.35cm} &&&\\
\enddata
\end{deluxetable*}
\begin{deluxetable*}{ccccc}
\tablenum{2}
\tablecaption{Results of the MCMC parameter estimation from the fits to the stacked RCs of Sect.\ref{3|fitRCs} when using the Burkert profile. \label{burkfit}}
\tablewidth{0pt}
\tablehead{
\colhead{Sample/Bin} & \colhead{$\log r_0$ [kpc]} & \colhead{$\log \rho_0$ [M$_{\odot}$ kpc$^{-3}$]} & \colhead{$\kappa\equiv v_{\rm d}^2(r_{\rm opt})/v_{\rm tot}^2(r_{\rm opt})$} & \colhead{$\chi^2_{\mathrm{red}}$}
}
\startdata
\vspace{-0.35cm} &&&\\
\vspace{0.0cm} PSS 1 & $0.596\pm 0.049$ & $7.609\pm 0.080$ & $0.113^{+0.046}_{-0.039}$ & 0.210 \\
\vspace{0.0cm} PSS 2 & $0.790\pm 0.050$ & $7.486^{+0.062}_{-0.069}$ & $0.249^{+0.028}_{-0.024}$ & 0.436 \\
\vspace{0.0cm} PSS 3 & $0.696^{+0.066}_{-0.060}$ & $7.638^{+0.094}_{-0.13}$ & $0.314^{+0.057}_{-0.036}$ & 0.477 \\
\vspace{0.0cm} PSS 4 & $0.796^{+0.073}_{-0.062}$ & $7.556^{+0.093}_{-0.13}$ & $0.376^{+0.058}_{-0.033}$ & 0.589 \\
\vspace{0.0cm} PSS 5 & $1.175^{+0.060}_{-0.076}$ & $7.038\pm 0.071$ & $0.556^{+0.020}_{-0.018}$ & 22.466 \\
\vspace{0.0cm} PSS 6 & $1.179^{+0.047}_{-0.053}$ & $7.058\pm 0.050$ & $0.545\pm 0.011$ & 1.290 \\
\vspace{0.0cm} PSS 7 & $1.297^{+0.077}_{-0.11}$ & $6.892\pm 0.098$ & $0.632^{+0.023}_{-0.020}$ & 0.686 \\
\vspace{0.0cm} PSS 8 & $3.15^{+1.2}_{-0.64}$ & $6.301^{+0.032}_{-0.053}$ & $0.791\pm 0.011$ & 3.851 \\
\vspace{0.0cm} PSS 9 & $1.517^{+0.093}_{-0.17}$ & $6.65\pm 0.12$ & $0.722^{+0.022}_{-0.018}$ & 1.525 \\
\vspace{0.0cm} PSS 10 & $2.26^{+0.35}_{-0.54}$ & $6.167^{+0.046}_{-0.11}$ & $0.836^{+0.015}_{-0.011}$ & 2.279 \\
\vspace{0.0cm} PSS 11 & $1.963^{+0.095}_{-0.76}$ & $6.30^{+0.24}_{-0.43}$ & $0.823^{+0.055}_{-0.024}$ & 2.279 \\
\vspace{0.0cm} LSB 1 & $0.664^{+0.062}_{-0.099}$ & $7.03^{+0.16}_{-0.12}$ & $0.151^{+0.077}_{-0.088}$ & 0.971 \\
\vspace{0.0cm} LSB 2 & $1.259^{+0.076}_{-0.16}$ & $6.601\pm 0.074$ & $0.534\pm 0.027$ & 3.710 \\
\vspace{0.0cm} LSB 3 & $1.272^{+0.062}_{-0.079}$ & $6.536\pm 0.076$ & $0.518^{+0.032}_{-0.030}$ & 0.370 \\
\vspace{0.0cm} LSB 4 & $3.28^{+1.7}_{-0.65}$ & $5.911^{+0.047}_{-0.075}$ & $0.750\pm 0.018$ & 4.882 \\
\vspace{0.0cm} LSB 5 & $0.751\pm 0.018$ & $8.019^{+0.058}_{-0.036}$ & $0.071^{+0.018}_{-0.070}$ & 12.268 \\
\vspace{0.0cm} Dw & $0.358^{+0.027}_{-0.032}$ & $7.563\pm 0.045$ & $0.055\pm 0.025$ & 0.760 \\
\vspace{-0.35cm} &&&\\
\enddata
\end{deluxetable*}

\clearpage
\begin{deluxetable*}{cccccc}
\tablenum{3}
\tablecaption{Results of the MCMC parameter estimation from the fits to the stacked RCs of Sect.\ref{3|fitRCs} when using the NFW profile. \label{nfwfit}}
\tablewidth{0pt}
\tablehead{
\colhead{Sample/Bin} & \colhead{c} & \colhead{$\log M_{\rm d}$ [M$_\odot$]} & \colhead{$\log M_{\rm v}$ [M$_\odot$]} & \colhead{$\chi^2_{\mathrm{red}}$}
}
\startdata
\vspace{-0.35cm} &&&\\
\vspace{0.0cm} PSS 1 & $6.43^{+1.2}_{-0.78}$ & $6.8\pm 1.0$ & $12.17^{+0.13}_{-0.29}$ & 4.265 \\
\vspace{0.0cm} PSS 2 & $7.4^{+2.5}_{-1.5}         $ & $8.67^{+0.61}_{-0.082}     $ & $12.45^{+0.15}_{-0.43}     $ & 3.931\\
\vspace{0.0cm} PSS 3 & $5.4\pm 1.7$ & $9.728^{+0.062}_{-0.034}$ & $12.75^{+0.23}_{-0.57}$ & 5.730 \\
\vspace{0.0cm} PSS 4 & $6.4^{+2.2}_{-1.9}         $ & $9.980^{+0.063}_{-0.036}   $ & $12.66^{+0.16}_{-0.49}     $ & 4.542\\
\vspace{0.0cm} PSS 5 & $22.71\pm 0.75             $ & $4.3\pm 2.5                $ & $11.779\pm 0.029           $ & 10.913\\
\vspace{0.0cm} PSS 6 & $10.2^{+2.0}_{-1.7}$ & $10.208^{+0.098}_{-0.062}$ & $12.347^{+0.067}_{-0.18}$ & 1.166\\
\vspace{0.0cm} PSS 7 & $9.0^{+4.3}_{-2.7}         $ & $10.54^{+0.10}_{-0.053}    $ & $12.435^{+0.038}_{-0.41}   $ & 2.470\\
\vspace{0.0cm} PSS 8 & $26.2^{+1.9}_{-1.7}$ & $10.19^{+0.28}_{-0.091}$ & $11.939^{+0.039}_{-0.034}$ & 1.281\\
\vspace{0.0cm} PSS 9 & $15.0^{+5.4}_{-3.8}        $ & $10.79^{+0.17}_{-0.073}    $ & $12.186^{+0.032}_{-0.13}   $ & 1.392\\
\vspace{0.0cm} PSS 10 & $29.1^{+2.6}_{-2.2}$ & $10.47^{+0.33}_{-0.081}$ & $12.091^{+0.042}_{-0.036}$ & 1.109 \\
\vspace{0.0cm} PSS 11 & $18.6^{+7.1}_{-4.8}$ & $11.19^{+0.17}_{-0.10}$ & $12.233^{+0.046}_{-0.080}$ & 0.531 \\
\vspace{0.0cm} LSB 1 & $3.51^{+0.67}_{-1.3}$ & $7.94^{+0.61}_{-0.28}$ & $11.63^{+0.25}_{-0.37}$ & 4.335 \\
\vspace{0.0cm} LSB 2 & $11.27\pm 0.68             $ & $4.4\pm 2.5                $ & $11.229^{+0.043}_{-0.052}  $ & 0.456\\
\vspace{0.0cm} LSB 3 & $3.85^{+0.84}_{-1.8} $ & $9.901^{+0.096}_{-0.037}$ & $12.28^{+0.22}_{-0.42}$ & 6.382 \\
\vspace{0.0cm} LSB 4 & $12.7^{+2.0}_{-1.5}        $ & $10.31^{+0.11}_{-0.074}    $ & $11.514\pm 0.061           $ & 1.502 \\
\vspace{0.0cm} LSB 5 & $23.5\pm 1.1$ & $5.1\pm 2.9$ & $12.065\pm 0.022$ & 2.564 \\
\vspace{0.0cm} Dw & $4.42^{+0.97}_{-0.70}      $ & $3.3^{+1.7}_{-2.7}         $ & $11.85^{+0.17}_{-0.35}     $ & 14.519 \\
\vspace{-0.35cm} &&&\\
\enddata
\end{deluxetable*}
\vspace{-2em}

\clearpage
\begin{deluxetable*}{cccccccc}
\tablenum{4}
\tablecaption{Results of the MCMC parameter estimation from the fits to the stacked RCs of Sect.\ref{3|fitRCs} when using the NMC profile. In addition to the fit parameter estimates, we report the F-ratio between the NFW and NMC models calculated as in Eq.~(11.50) of \citealt{Bevington:1305448}, i.e. $F = (\chi_{\mathrm{NFW}}^2-\chi_{\mathrm{NMC}}^2) / \chi_{\mathrm{NMC,red}}^2$. Values of $F$ are reported alongside the associated p-values. Here, the null hypothesis is $L=0$.  \label{nmcfit}}
\tablewidth{0pt}
\tablehead{
\colhead{Sample/Bin} & \colhead{c} & \colhead{$\log M_{\rm d}$ [M$_\odot$]} & \colhead{$\log M_{\rm v}$ [M$_\odot$]} & \colhead{$L$ [kpc]} & \colhead{$\chi^2_{\mathrm{red}}$} & \colhead{F} & \colhead{p-value}
}
\startdata
\vspace{-0.35cm} &&&\\
\vspace{0.0cm} PSS 1 & $9.14^{+1.0}_{-0.84}$ & $6.2^{+1.0}_{-1.9}$ & $11.71^{+0.10}_{-0.15}$ & $0.254^{+0.016}_{-0.012}$ & 1.742 & 25.6 & $ 10^{-4} $\\
\vspace{0.0cm} PSS 2 & $13.7^{+2.4}_{-0.68}$ & $7.9^{+1.3}_{-1.6}$ & $11.712^{+0.043}_{-0.16}$ & $0.4645 \pm 0.0084$ & 0.803 & 67.2 & $ < 10^{-5} $\\
\vspace{0.0cm} PSS 3 & $22.1^{+2.0}_{-0.42}$ & $7.0 \pm 1.7$ & $11.470^{+0.026}_{-0.057}$ & $0.5192 \pm 0.0067$ & 0.511 & 174.6 & $ < 10^{-5} $\\
\vspace{0.0cm} PSS 4 & $23.7^{+2.3}_{-0.32}$ & $7.1 \pm 1.8$ & $11.615^{+0.023}_{-0.054}  $ & $0.6011\pm 0.0091 $ & 0.786 & 82.2 & $ < 10^{-5} $ \\
\vspace{0.0cm} PSS 5 & $13.6^{+3.3}_{-4.3}$ & $9.95^{+0.27}_{-0.047}$ & $12.018^{+0.069}_{-0.20}$ & $0.208^{+0.024}_{-0.035}$ & 0.615 & 285.7 & $ < 10^{-5} $\\
\vspace{0.0cm} PSS 6 & $14.2 \pm 2.9$ & $10.01^{+0.27}_{-0.077}$ & $12.122^{+0.053}_{-0.14}$ & $0.314^{+0.097}_{-0.040}$ & 1.098 & 2.0 & 0.2\\
\vspace{0.0cm} PSS 7 & $32.7^{+1.4}_{-1.2}$ &  $6.9^{+1.5}_{-2.3}$ & $11.802 \pm 0.025$ & $0.915 \pm 0.013$ & 1.088 & 22.6 & $ 2 \cdot 10^{-4} $\\
\vspace{0.0cm} PSS 8 & $32.5 \pm 1.4$ &  $6.0\pm 2.3$ & $11.937\pm 0.026$ & $0.443^{+0.065}_{-0.042}$ & 0.591 & 20.9 & $ 3 \cdot 10^{-4} $\\
\vspace{0.0cm} PSS 9 & $31.2^{+1.8}_{-1.1}$ & $6.4 \pm 2.5$ & $12.076 \pm 0.026$ & $0.733^{+0.063}_{-0.042}$ & 0.854 & 11.7 & $ 3.5 \cdot 10^{-3} $\\
\vspace{0.0cm} PSS 10 & $44.4 \pm 1.6$ & $6.9 \pm 1.7$ & $12.043 \pm 0.017$ & $1.439 \pm 0.030$ & 1.139 & 0.6 & 0.5\\
\vspace{0.0cm} PSS 11 & $42.4^{+2.3}_{-2.6}$ & $7.1\pm1.8$ & $12.251 \pm 0.025$ & $1.858 \pm 0.093$ & 0.952 & - & -\\
\vspace{0.0cm} LSB 1 & $6.05^{+1.0}_{-0.88}$ & $5.2\pm 1.9$ & $11.07^{+0.11}_{-0.19}$ & $0.280^{+0.010}_{-0.013}$ & 1.980 & 21.2 & $ 3 \cdot 10^{-4} $\\
\vspace{0.0cm} LSB 2 & $12.98^{+0.87}_{-0.65}$ & $5.4 \pm 2.0$ & $11.123^{+0.038}_{-0.051}$ & $0.415^{+0.011}_{-0.013}$ & 1.512 & - & -\\
\vspace{0.0cm} LSB 3 & $9.4^{+1.7}_{-2.1}$ & $9.29^{+0.45}_{-0.17}$ & $11.620^{+0.054}_{-0.085}$ & $0.350^{+0.042}_{-0.028}$ & 0.923 & 101.5 & $ < 10^{-5} $\\
\vspace{0.0cm} LSB 4 & $23.7^{+1.2}_{-1.0}$ & $6.1 \pm 2.3$ & $11.516^{+0.029}_{-0.026}$ & $0.746^{+0.012}_{-0.014}$ & 1.352 & 2.9 & $ 0.1 $\\
\vspace{0.0cm} LSB 5 & $26.9\pm 1.3$ & $7.2^{+1.6}_{-2.5}$ & $12.020\pm 0.022$ & $1.551 \pm 0.048$ & 1.411 & 14.9 & $ 10^{-3} $\\
\vspace{0.0cm} Dw & $8.32 \pm 0.63$ & $3.3^{+1.3}_{-3.3}$ & $10.988^{+0.083}_{-0.11}$ & $0.2259\pm 0.0043$ & 3.987 & 45.9 & $ < 10^{-5} $\\
\vspace{-0.35cm} &&&\\
\enddata
\end{deluxetable*}

\clearpage

\appendix
In this Appendix we present the fits to stacked RCs for the full galaxy samples considered in this work (see Table~\ref{binlist}).
\clearpage
\begin{figure}[ht]
    \begin{center}
    \includegraphics[width=.495\textwidth]{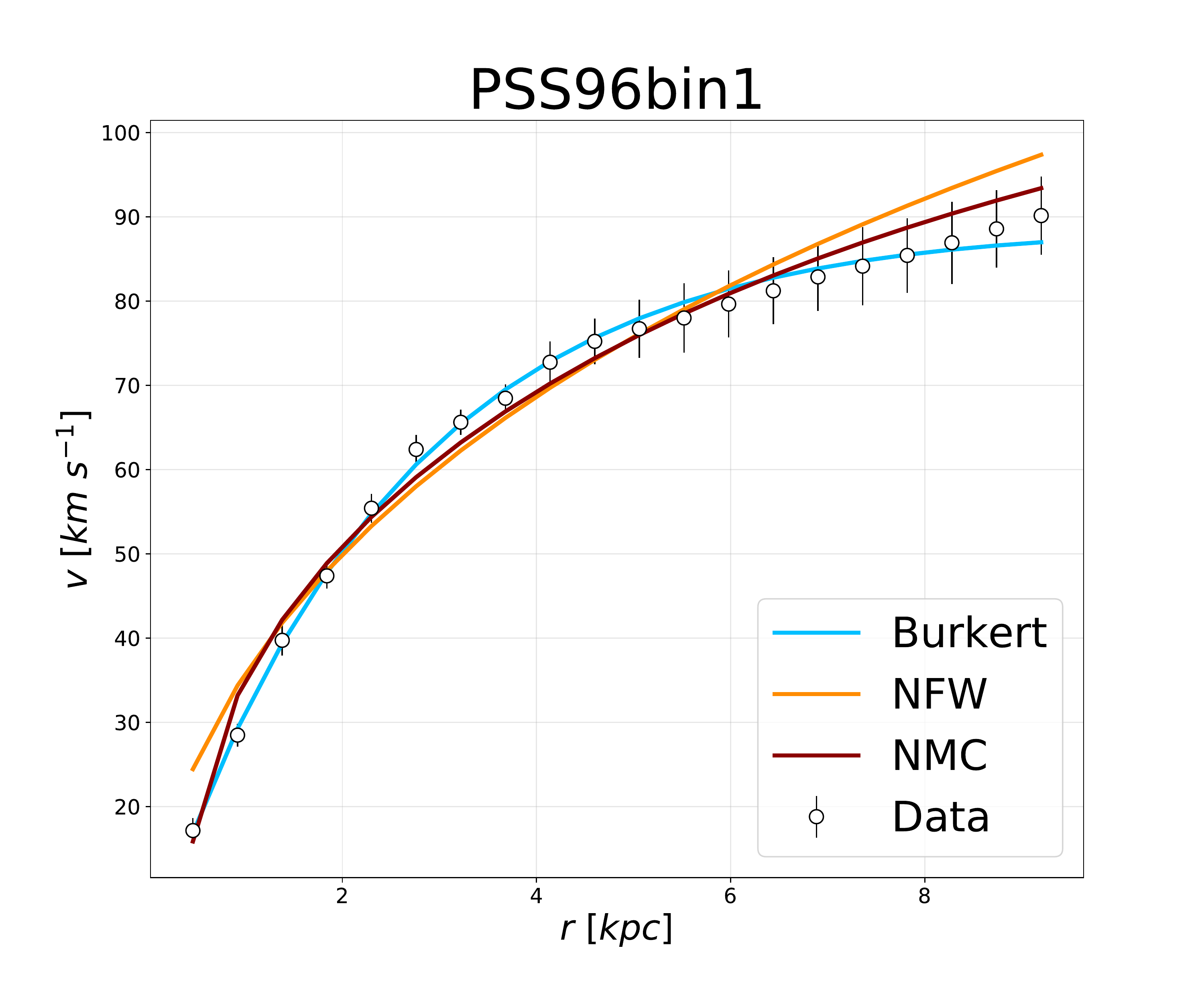}
    \includegraphics[width=.495\textwidth]{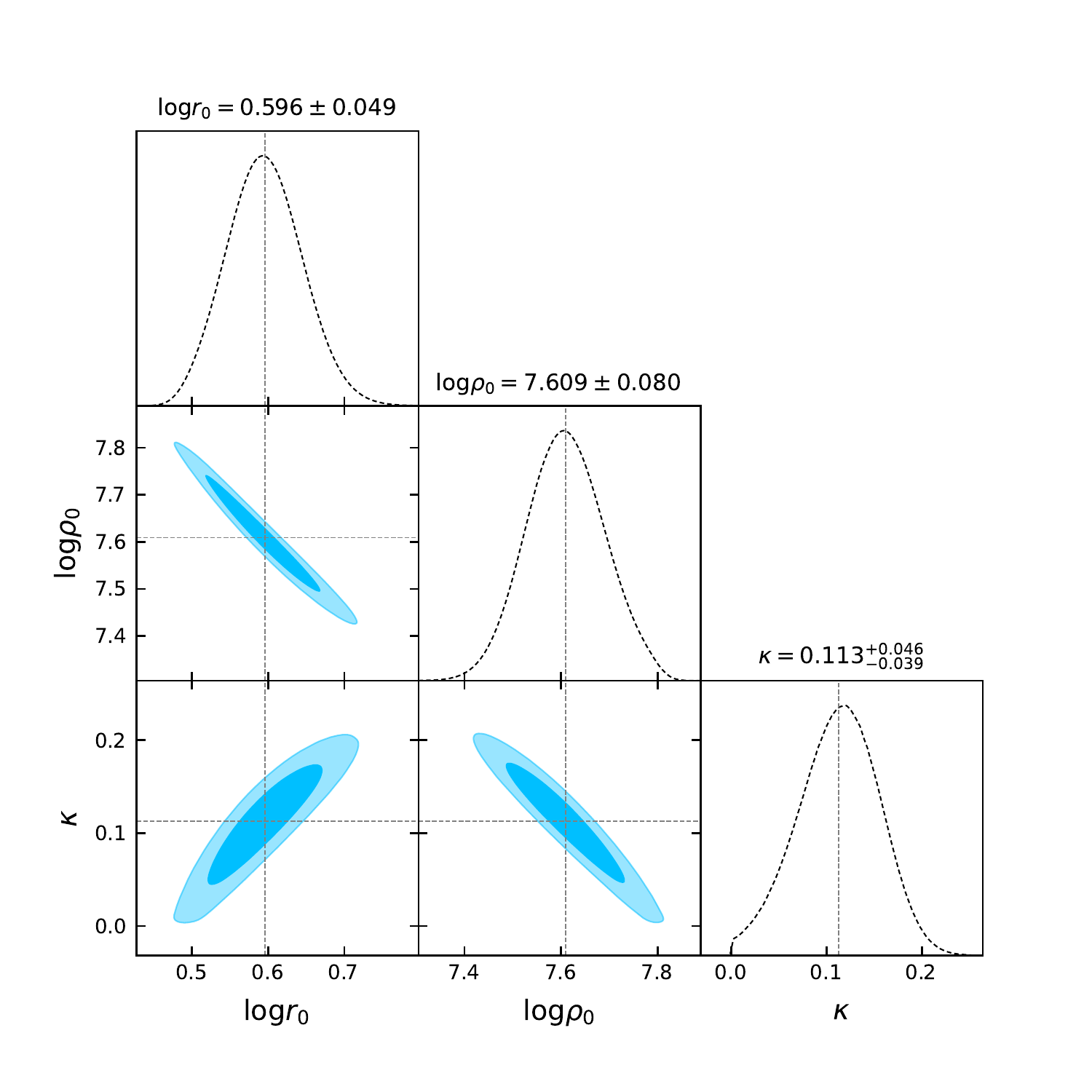}
    \includegraphics[width=.495\textwidth]{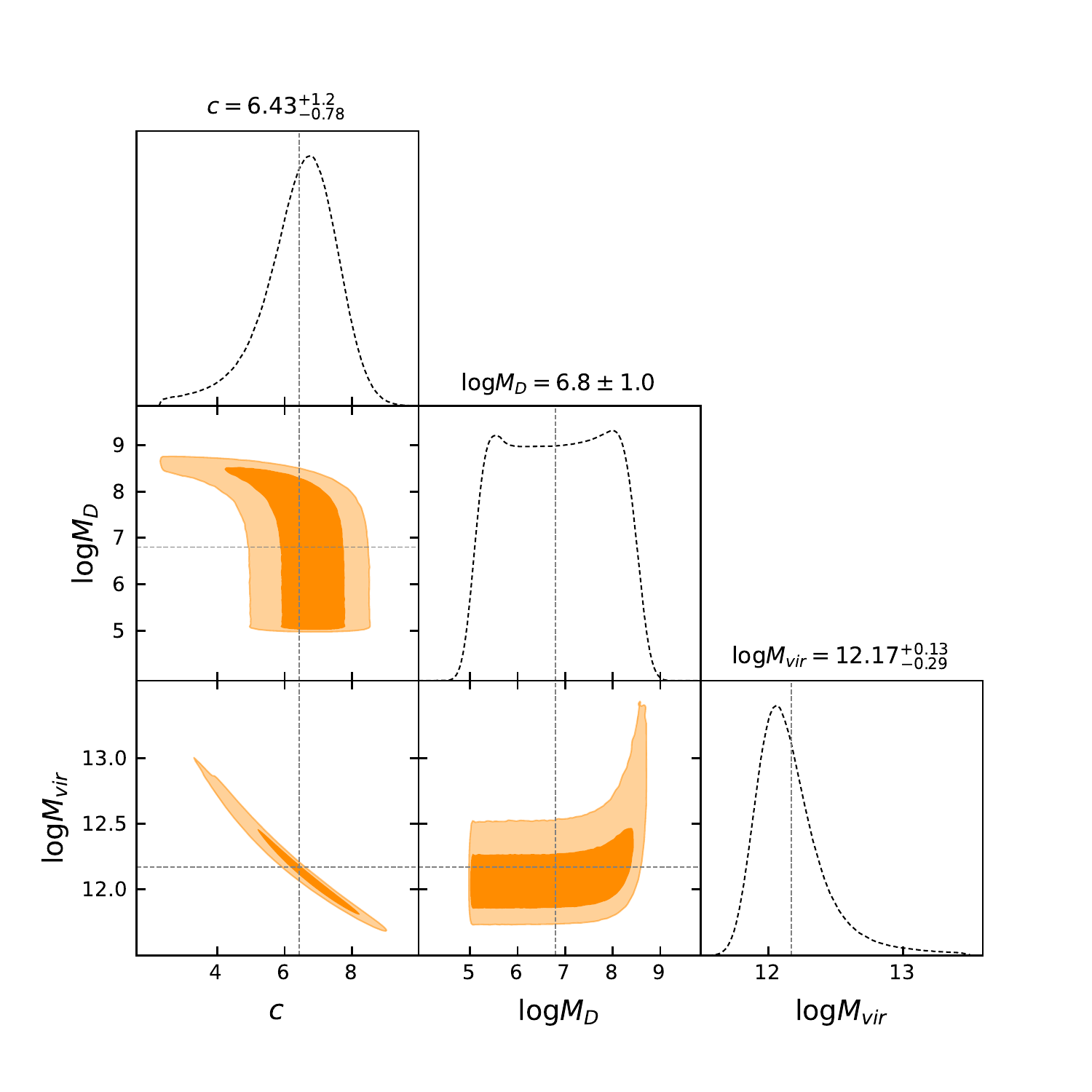}
    \includegraphics[width=.495\textwidth]{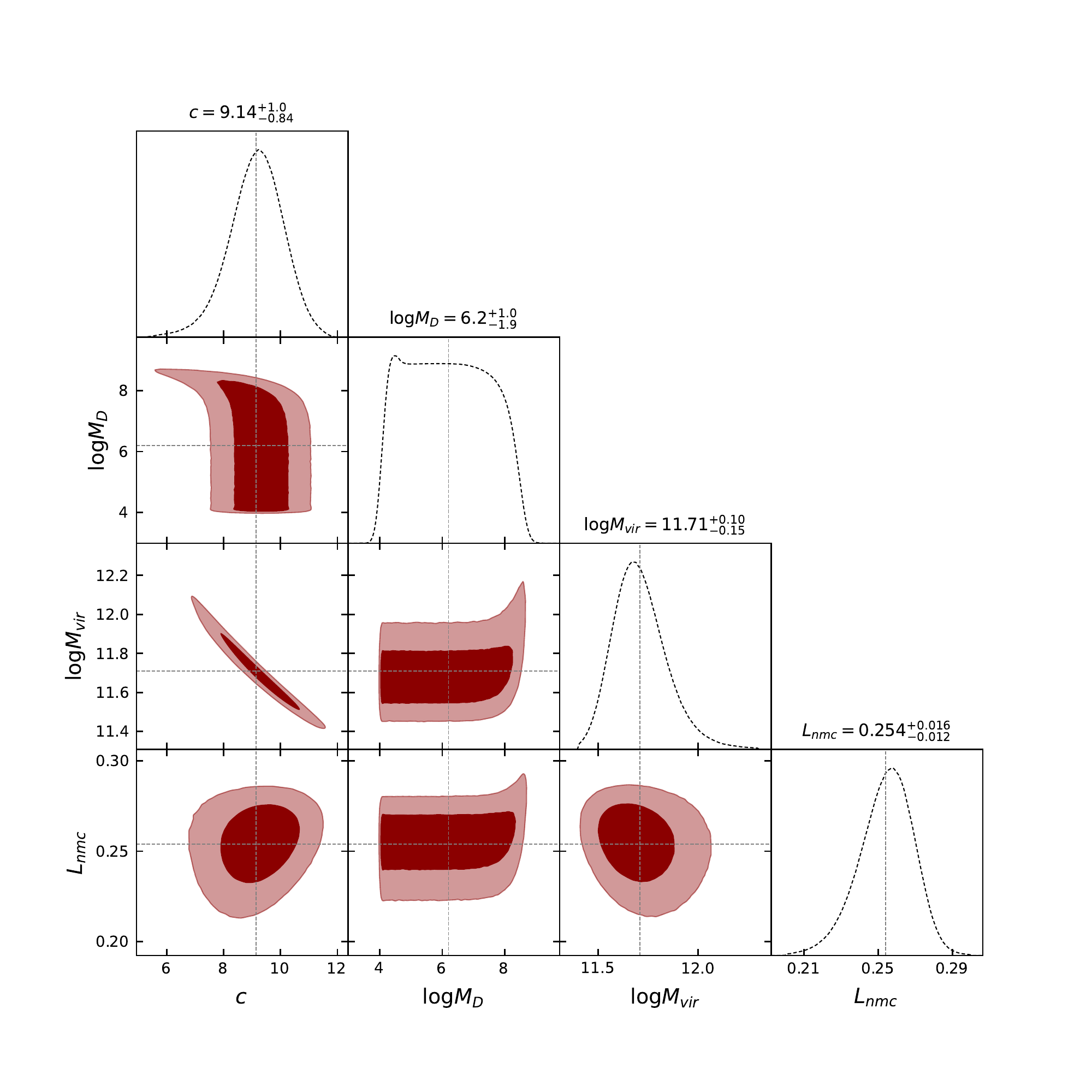}
    \end{center}
    \caption{Bin 1 of PSS96}
\end{figure}
\clearpage
\begin{figure}[ht]
    \begin{center}
    \includegraphics[width=.495\textwidth]{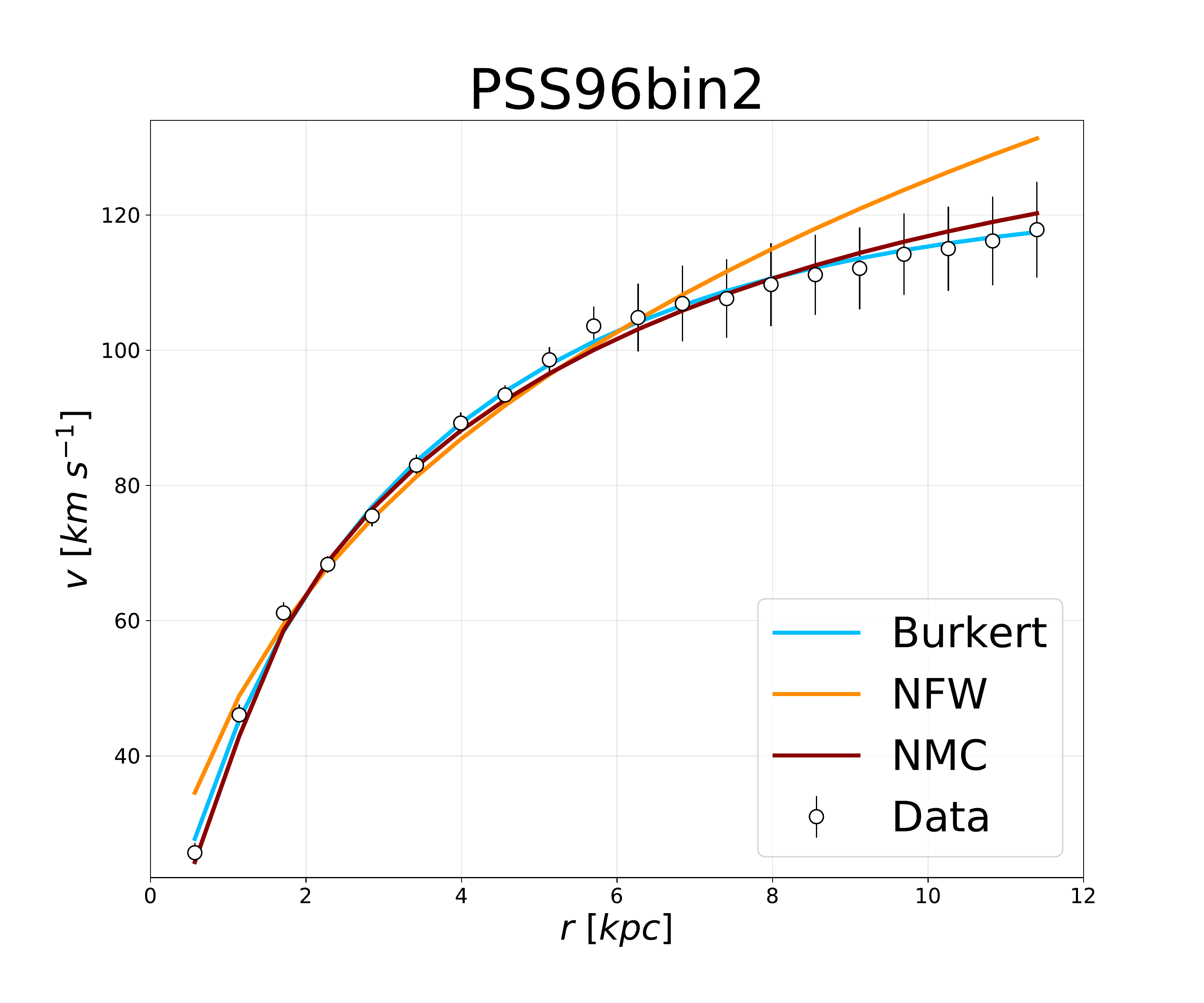}
    \includegraphics[width=.495\textwidth]{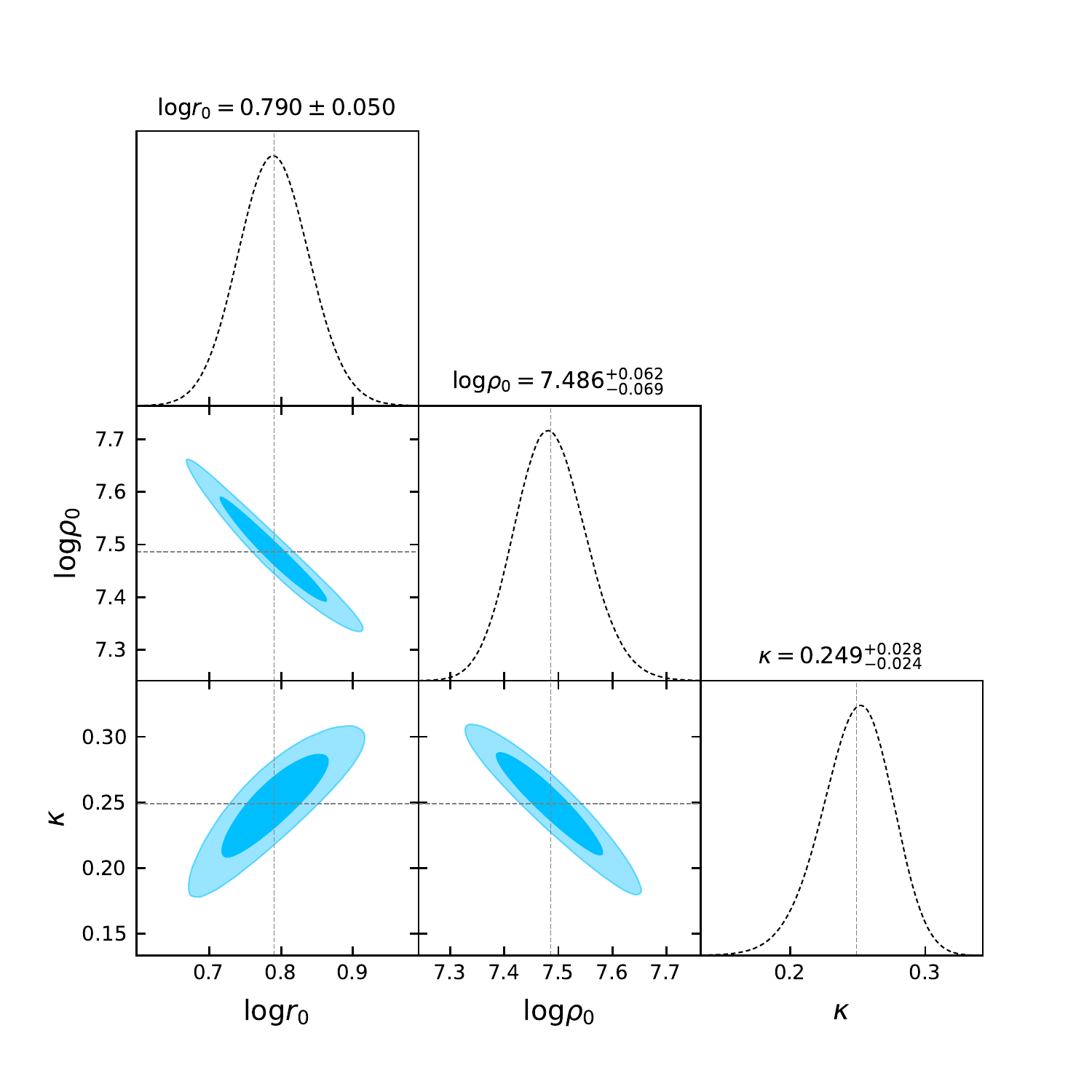}
    \includegraphics[width=.495\textwidth]{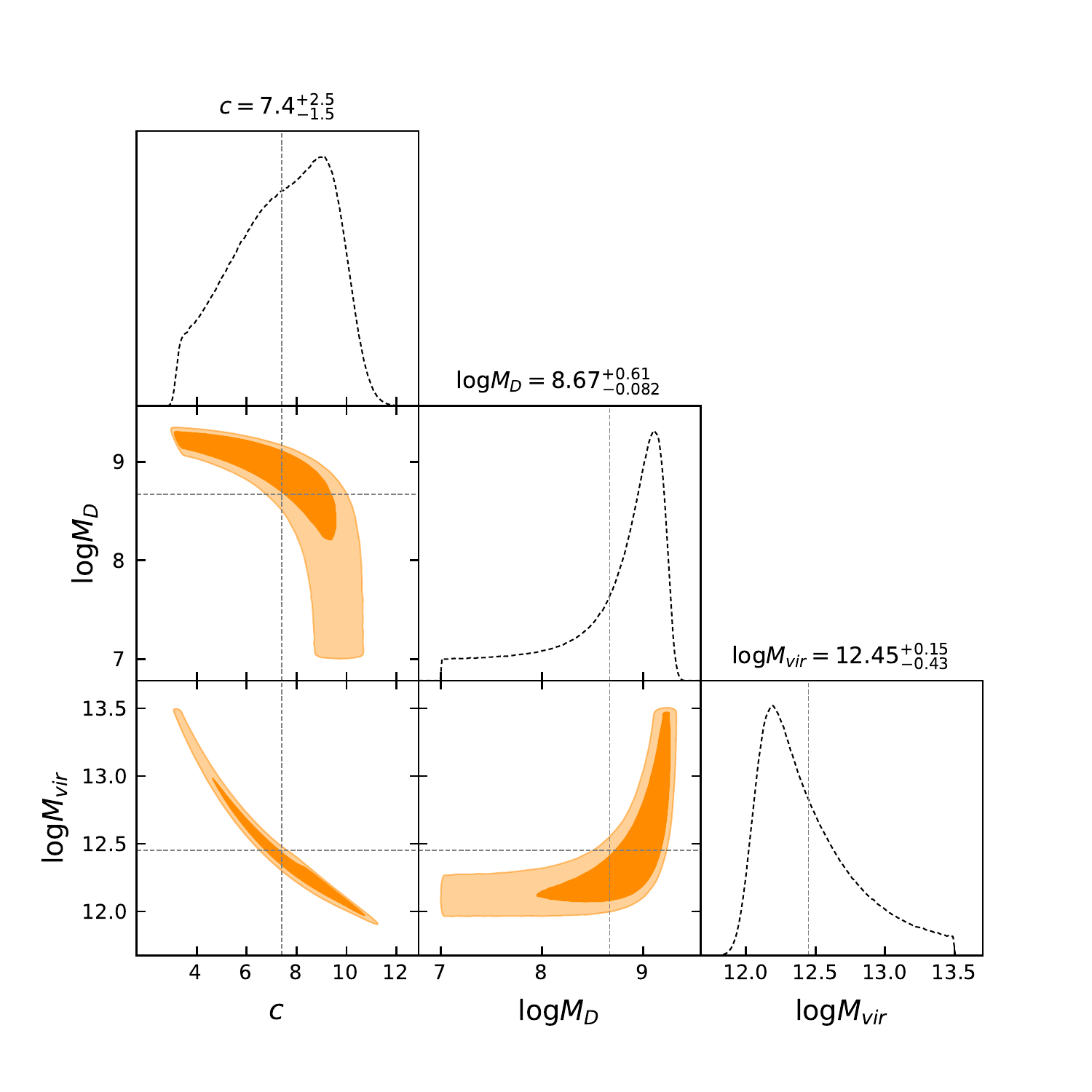}
    \includegraphics[width=.495\textwidth]{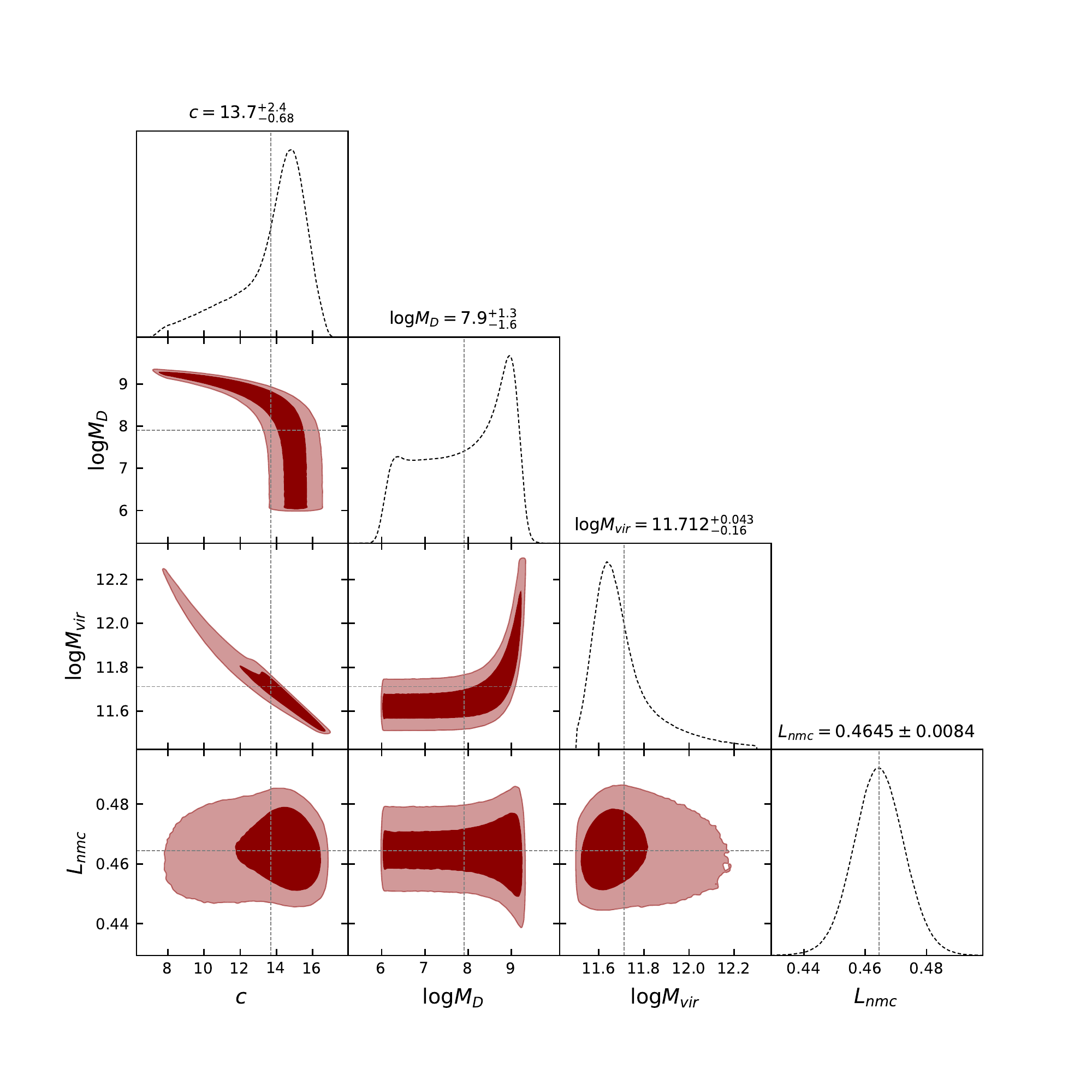}
    \end{center}
    \caption{Bin 2 of PSS96}
\end{figure}
\clearpage
\begin{figure}[ht]
    \begin{center}
    \includegraphics[width=.495\textwidth]{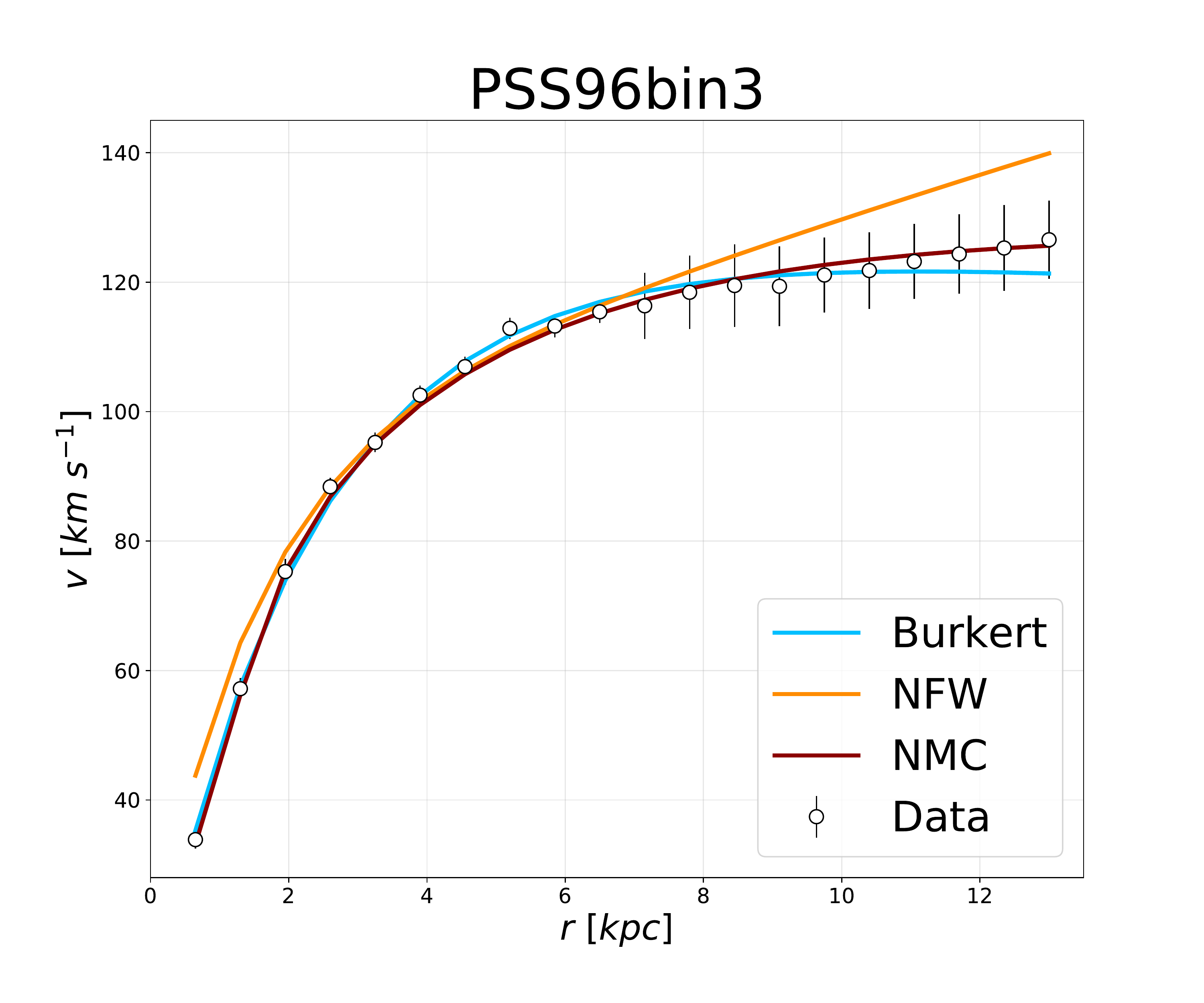}
    \includegraphics[width=.495\textwidth]{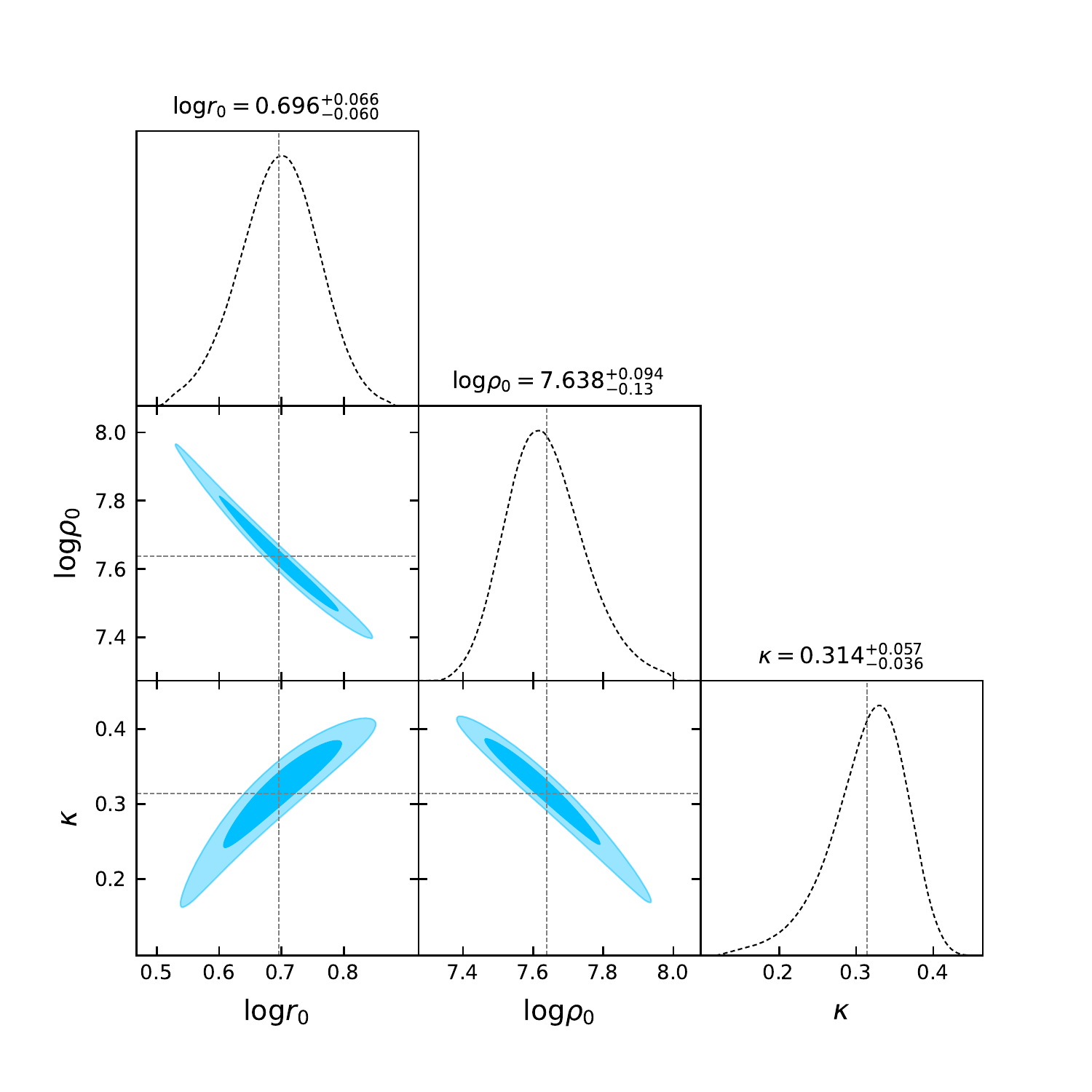}
    \includegraphics[width=.495\textwidth]{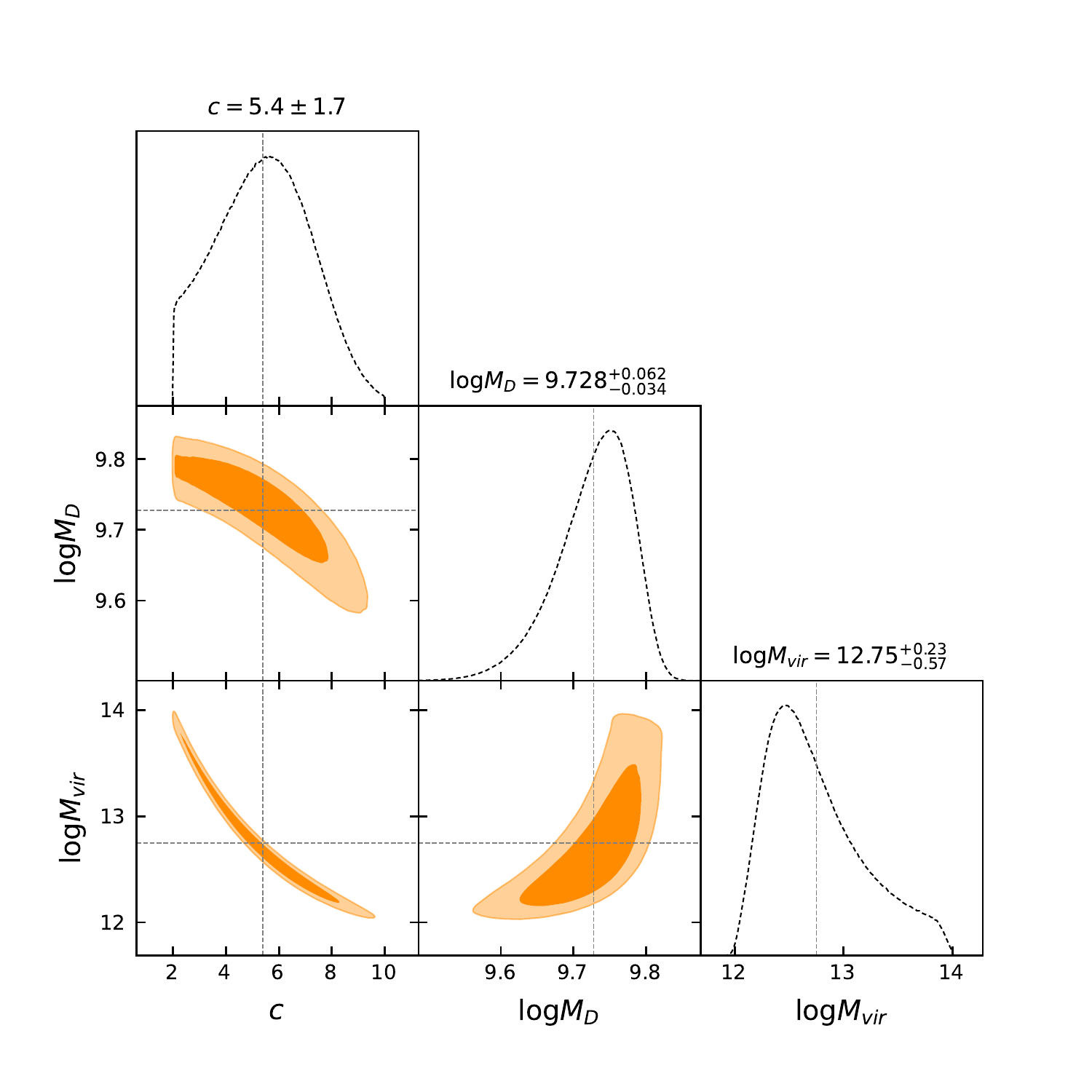}
    \includegraphics[width=.495\textwidth]{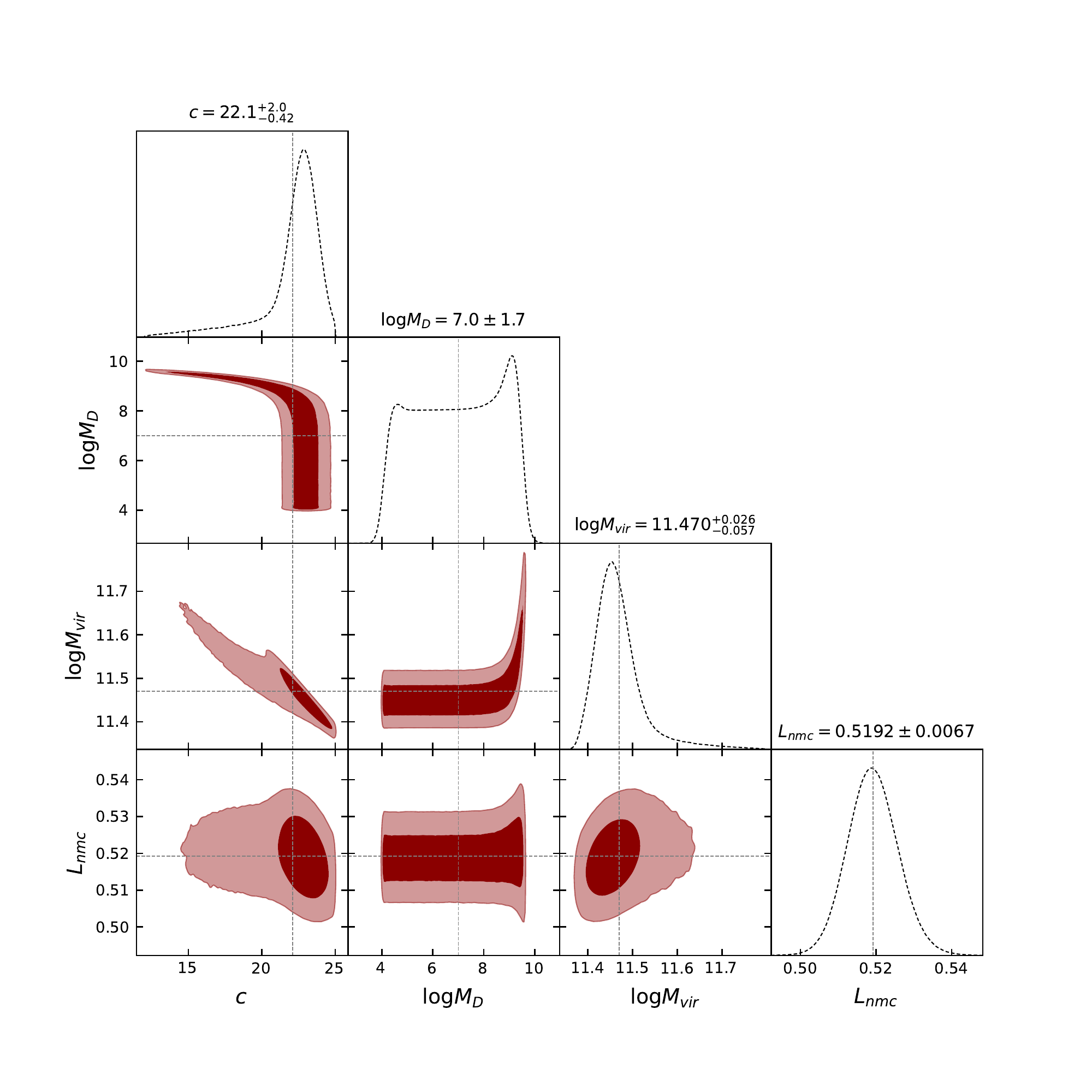}
    \end{center}
    \caption{Bin 3 of PSS96}
\end{figure}
\clearpage
\begin{figure}[ht]
    \begin{center}
    \includegraphics[width=.495\textwidth]{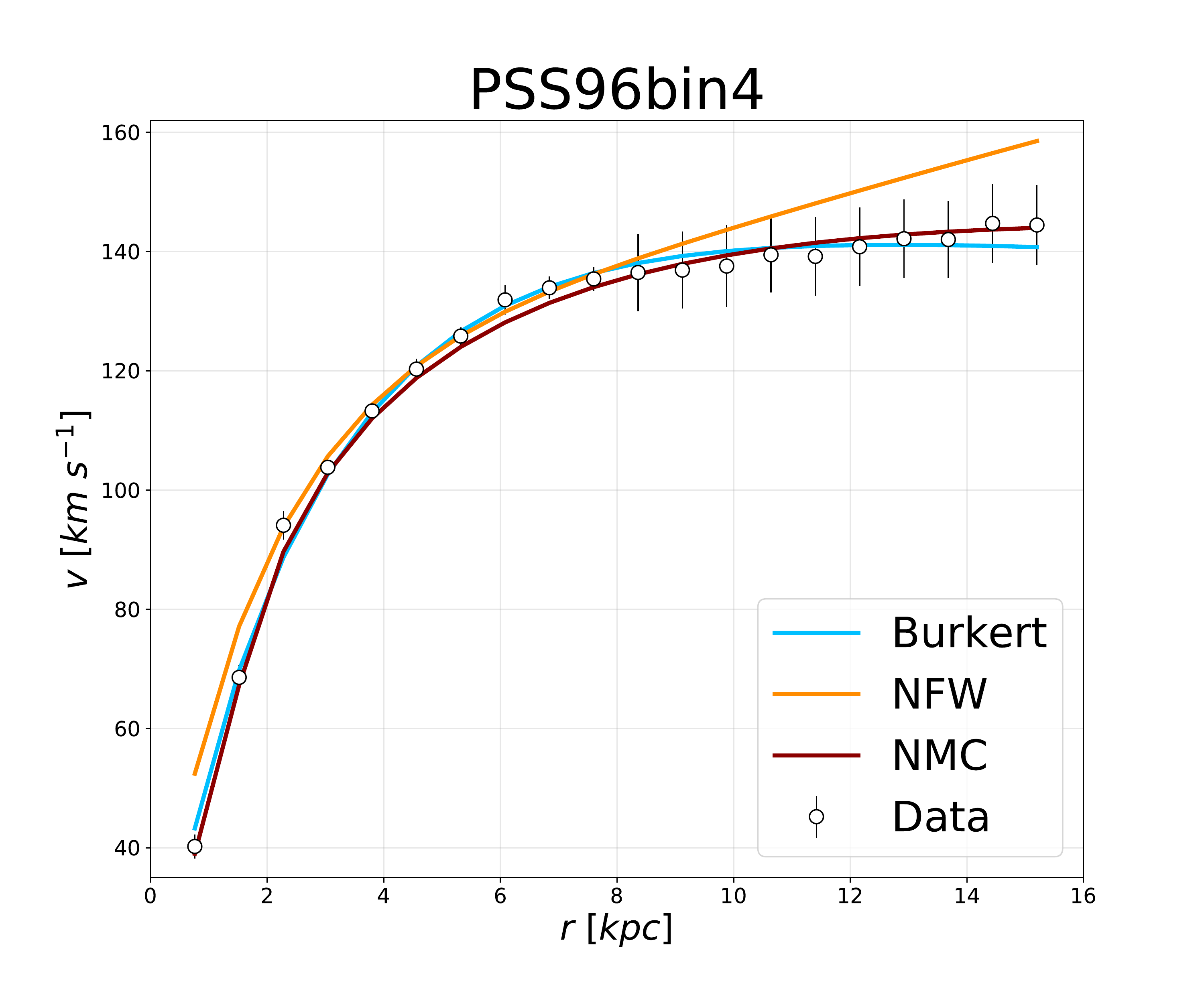}
    \includegraphics[width=.495\textwidth]{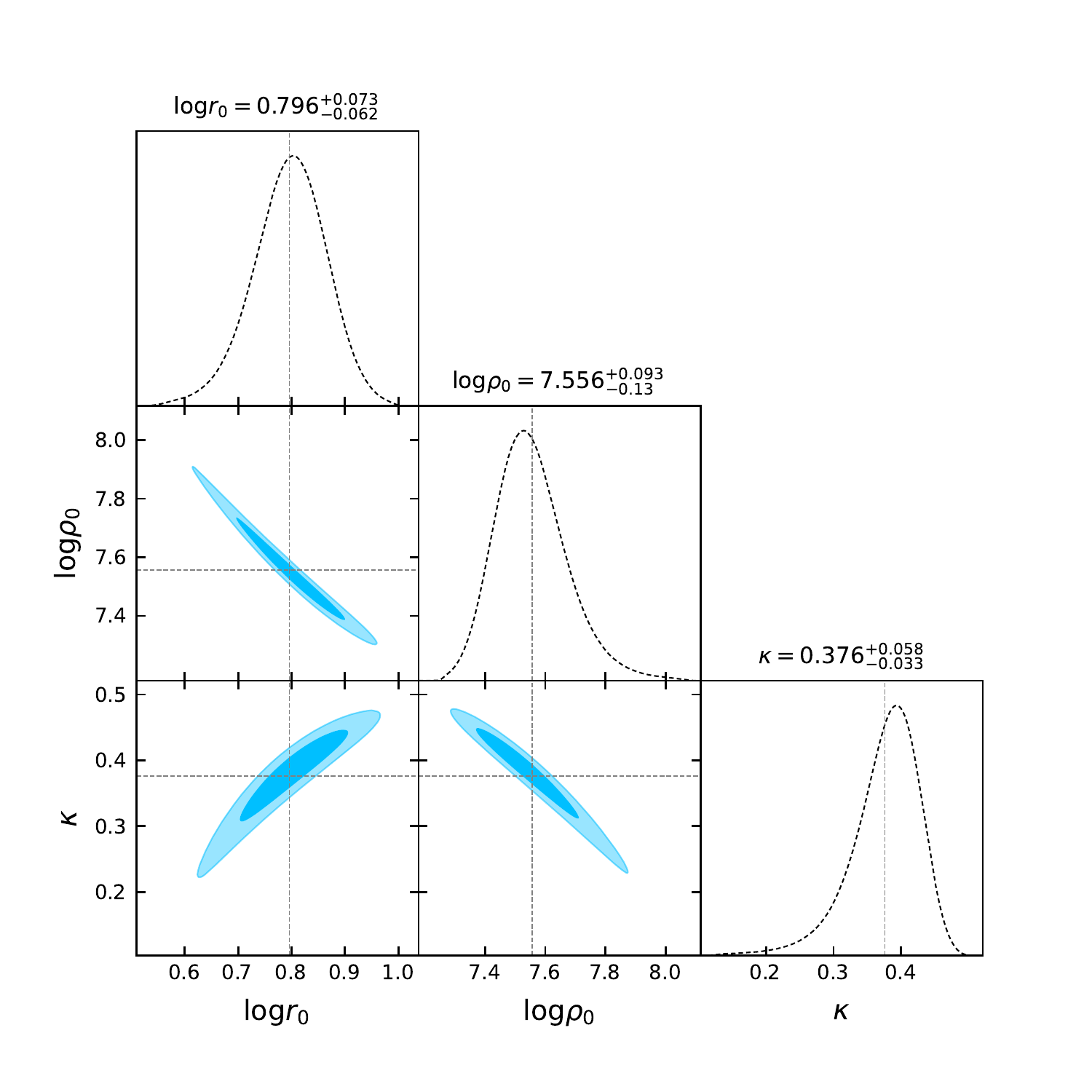}
    \includegraphics[width=.495\textwidth]{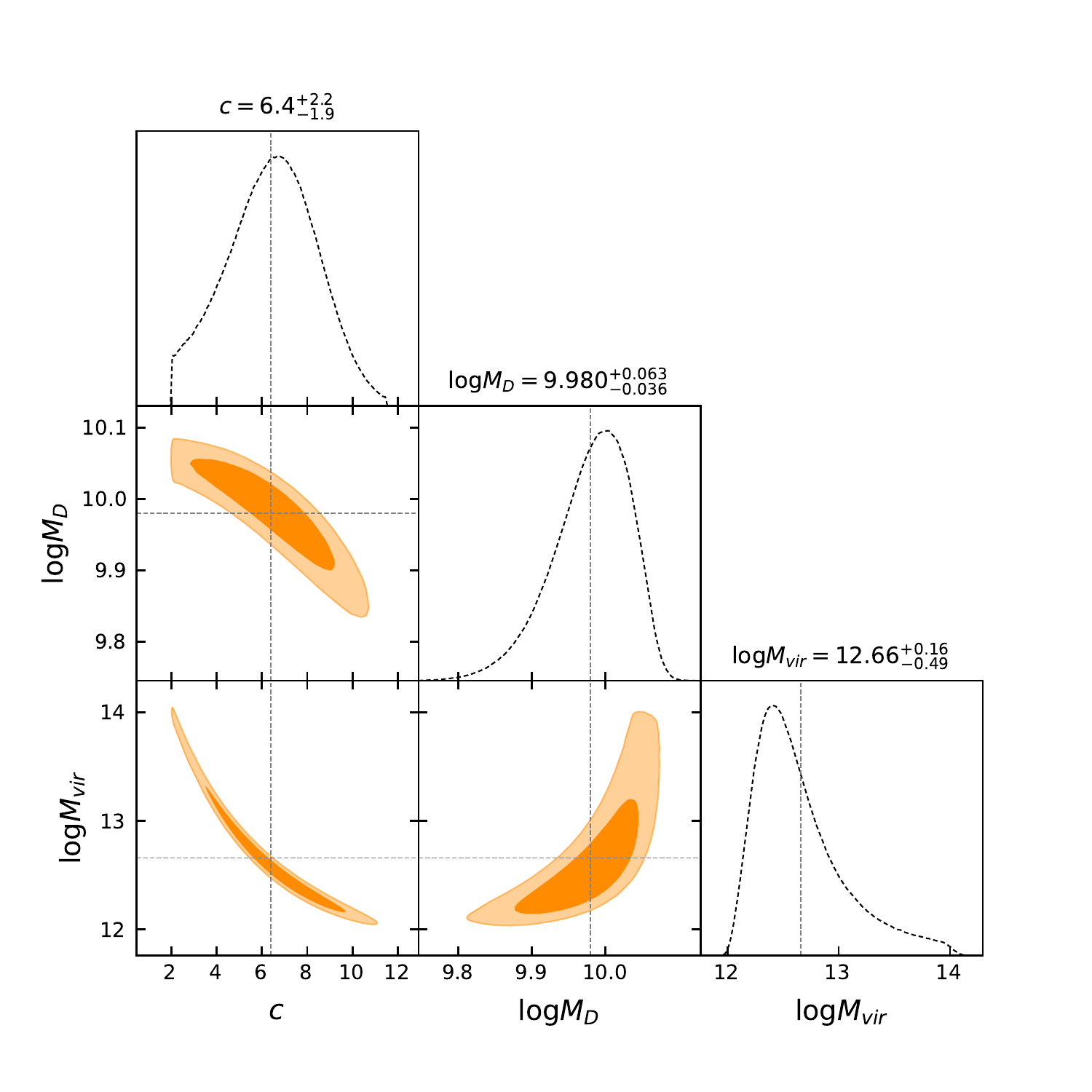}
    \includegraphics[width=.495\textwidth]{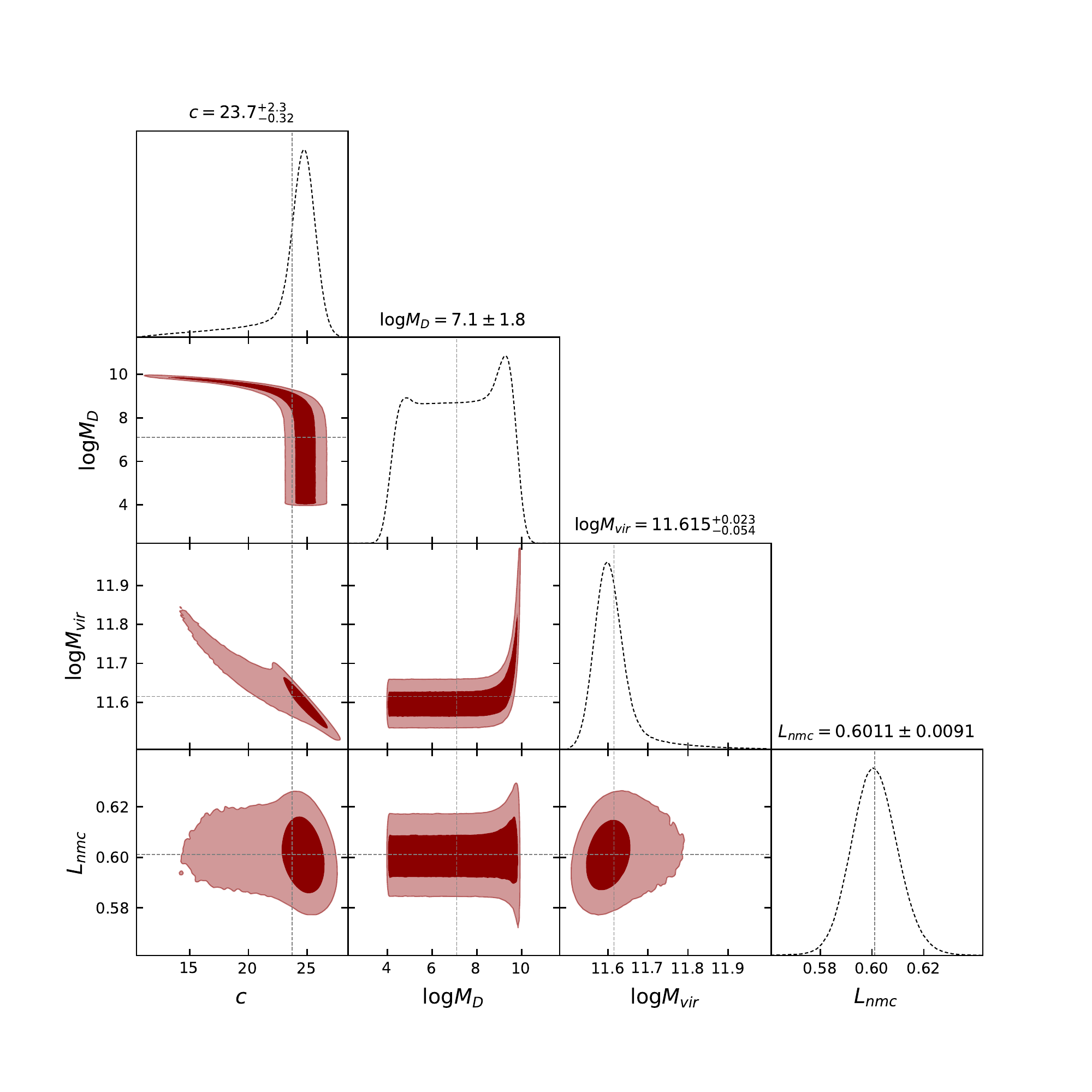}
    \end{center}
    \caption{Bin 4 of PSS96}
\end{figure}
\clearpage
\begin{figure}[ht]
    \begin{center}
    \includegraphics[width=.495\textwidth]{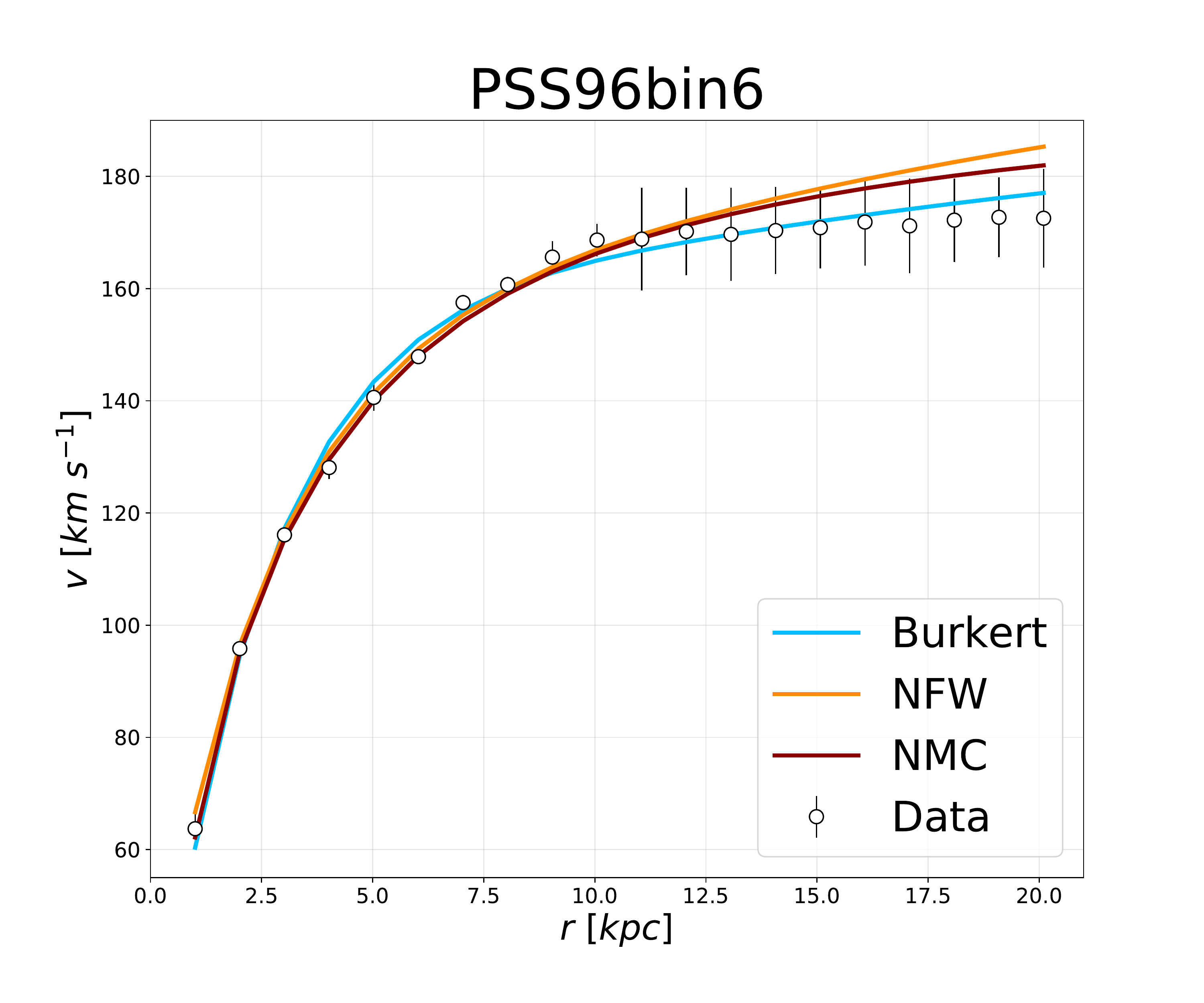}
    \includegraphics[width=.495\textwidth]{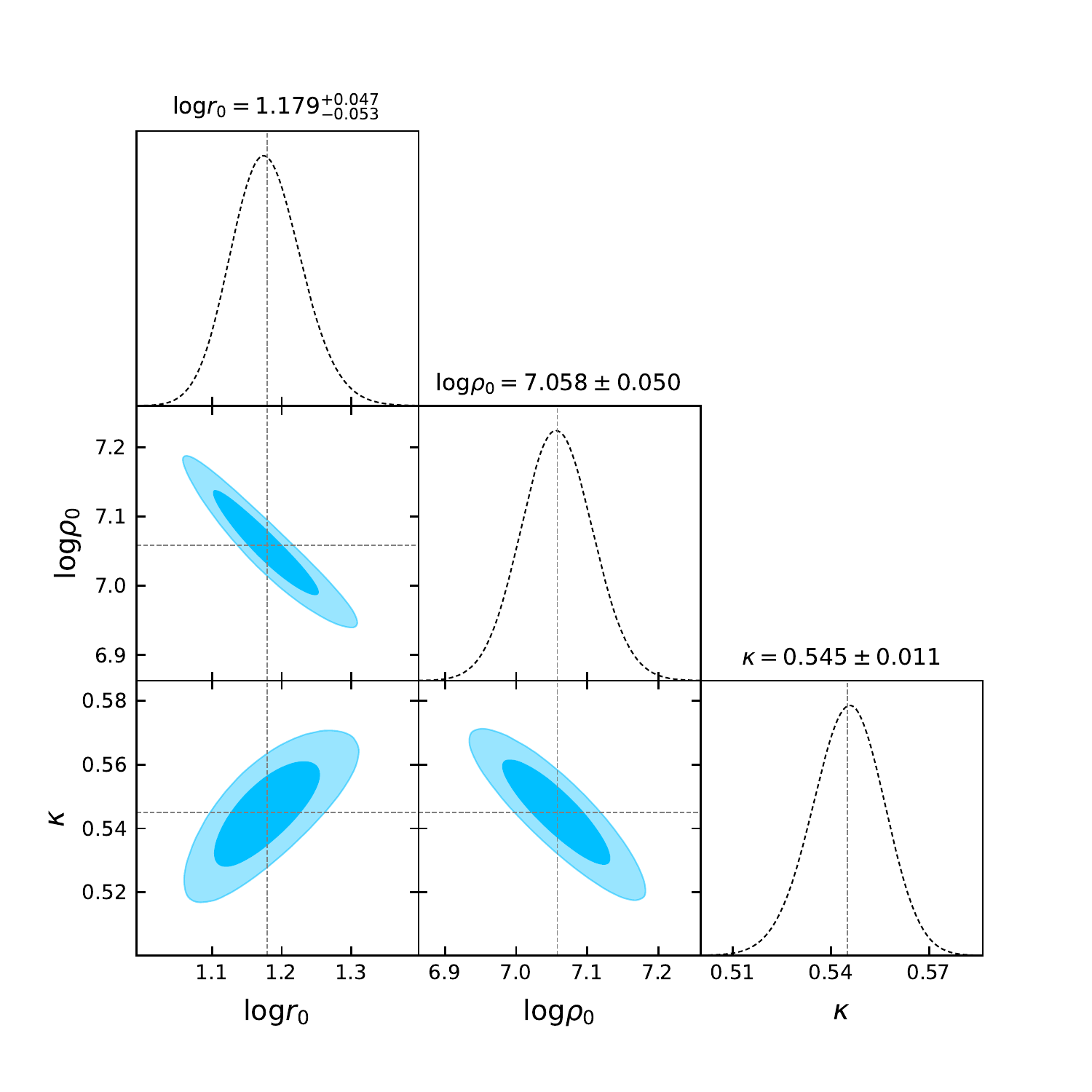}
    \includegraphics[width=.495\textwidth]{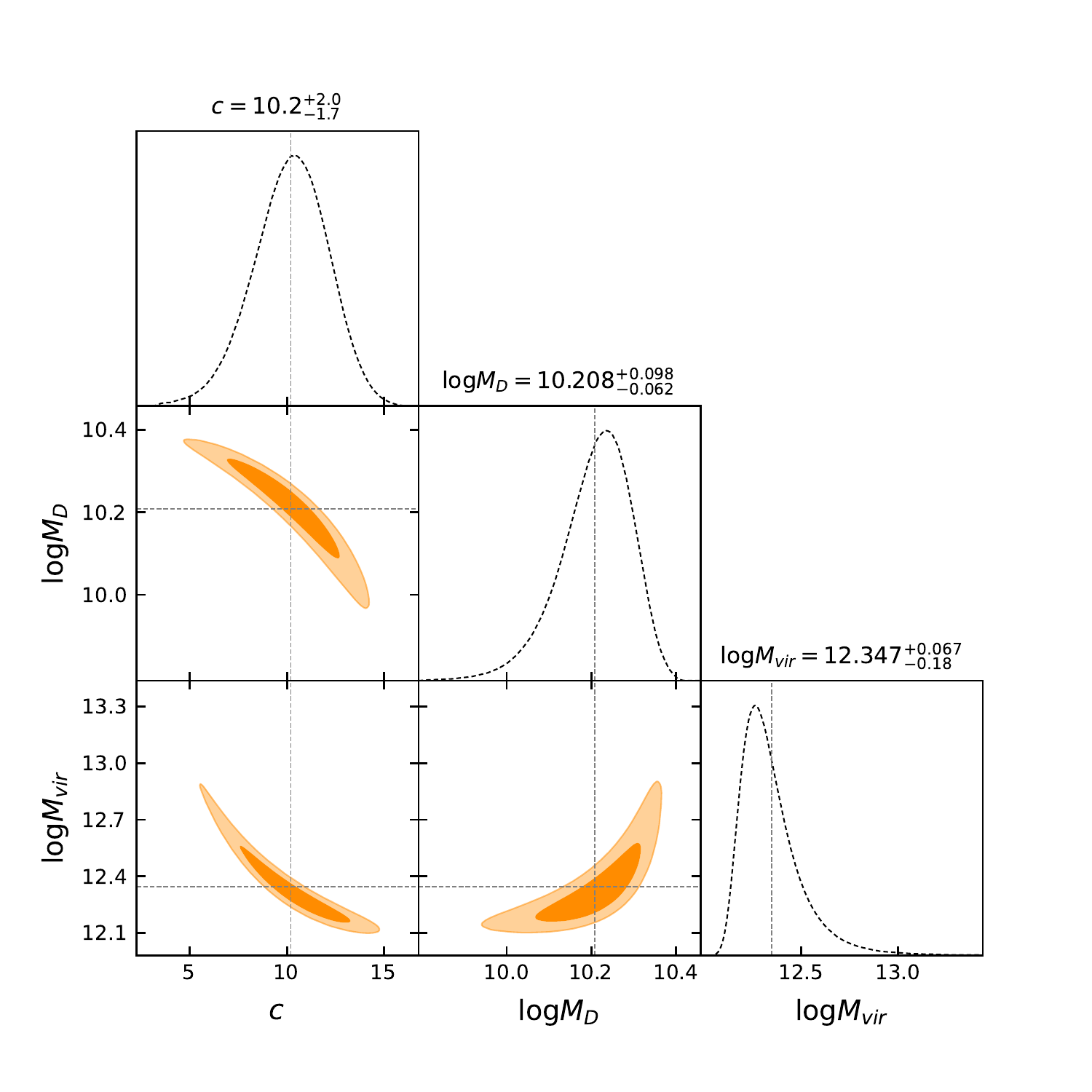}
    \includegraphics[width=.495\textwidth]{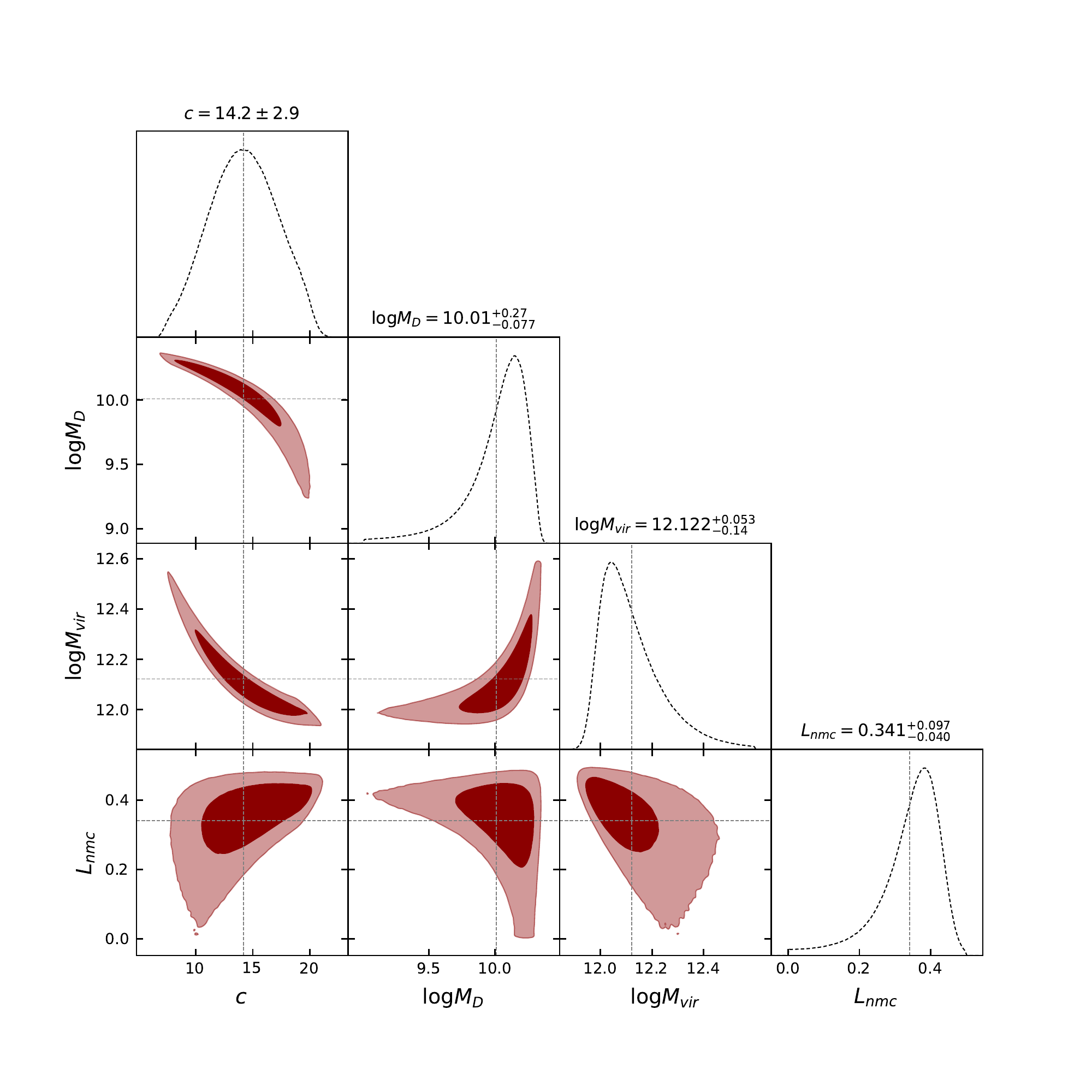}
    \end{center}
    \caption{Bin 6 of PSS96}
\end{figure}
\clearpage
\begin{figure}[ht]
    \begin{center}
    \includegraphics[width=.495\textwidth]{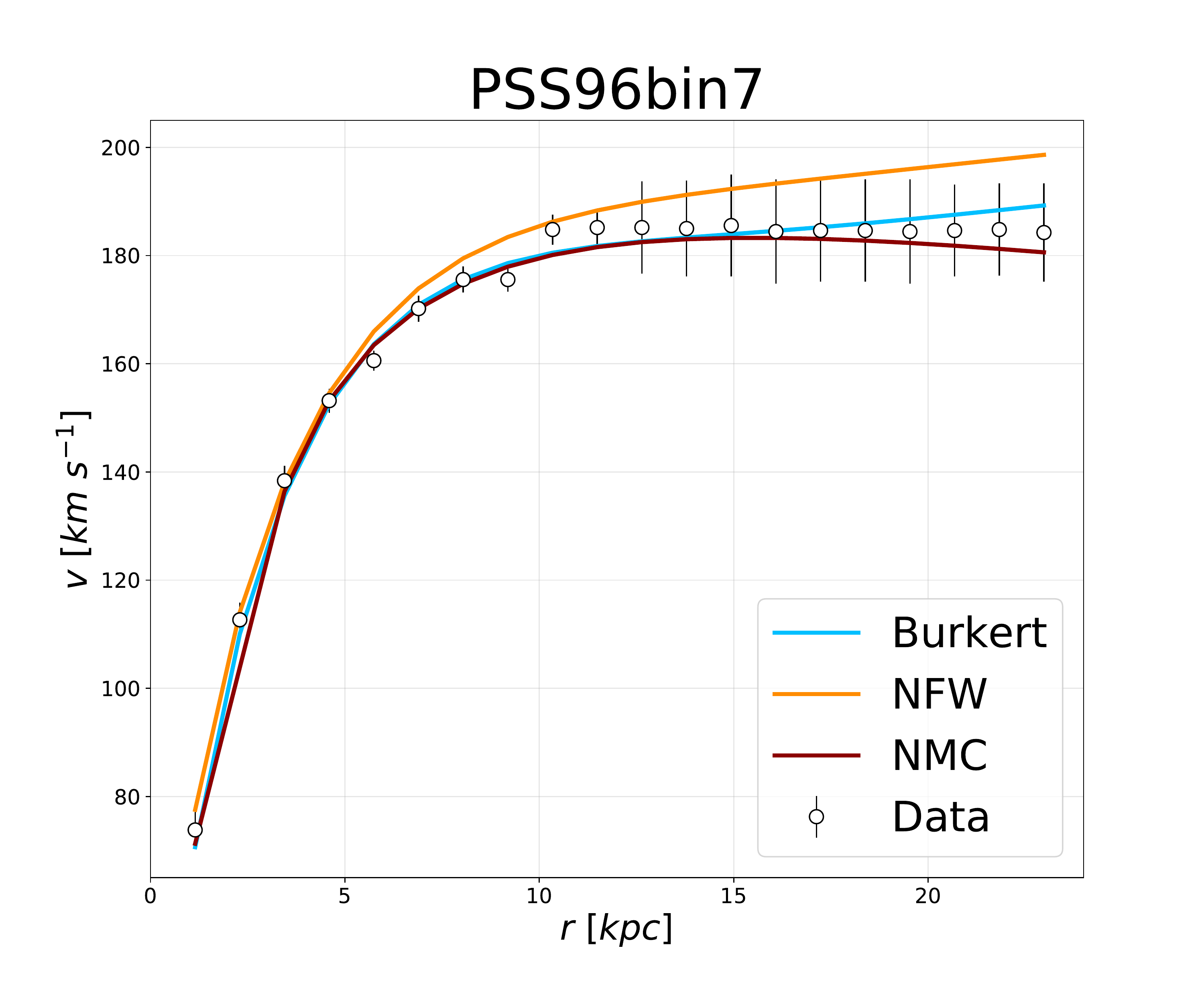}
    \includegraphics[width=.495\textwidth]{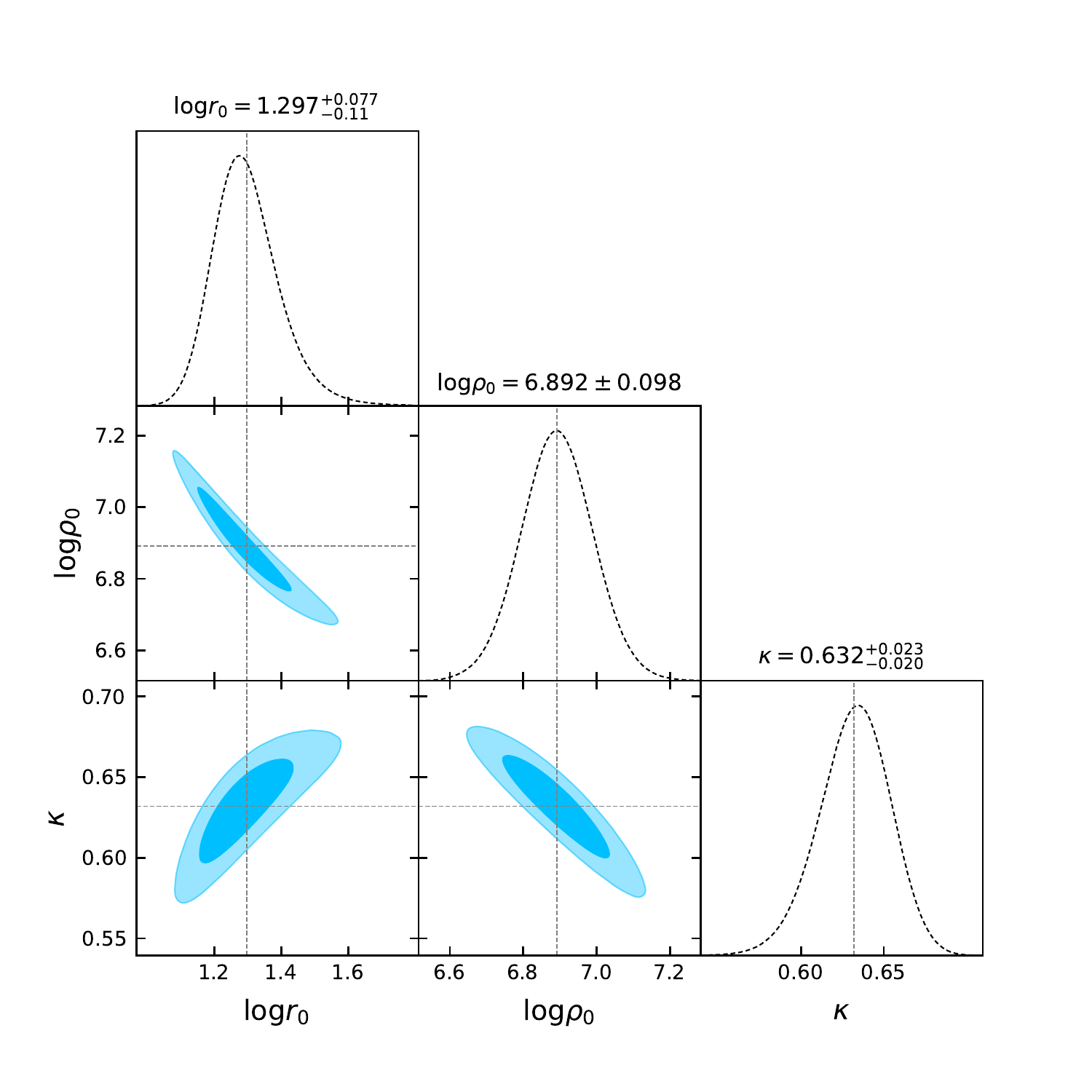}
    \includegraphics[width=.495\textwidth]{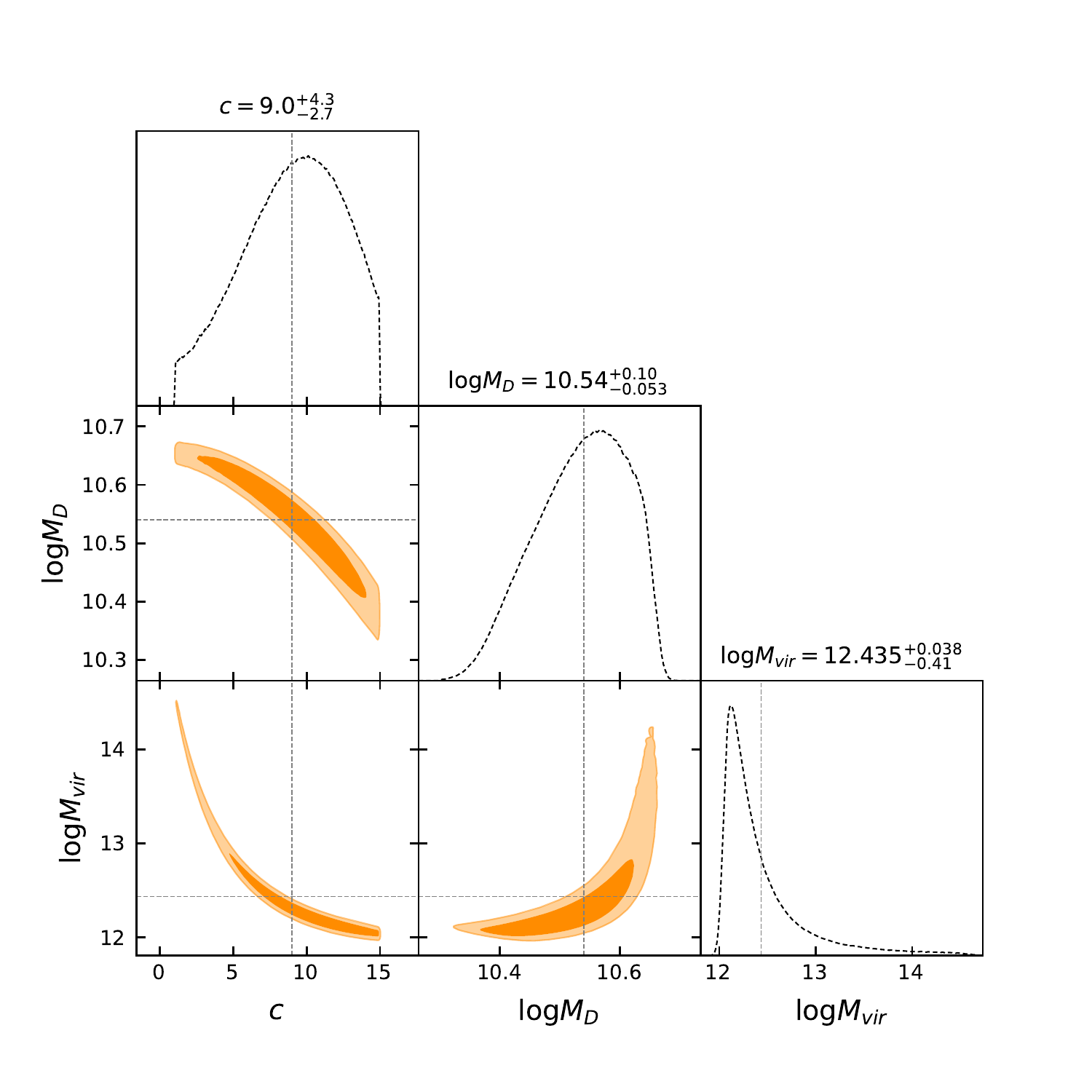}
    \includegraphics[width=.495\textwidth]{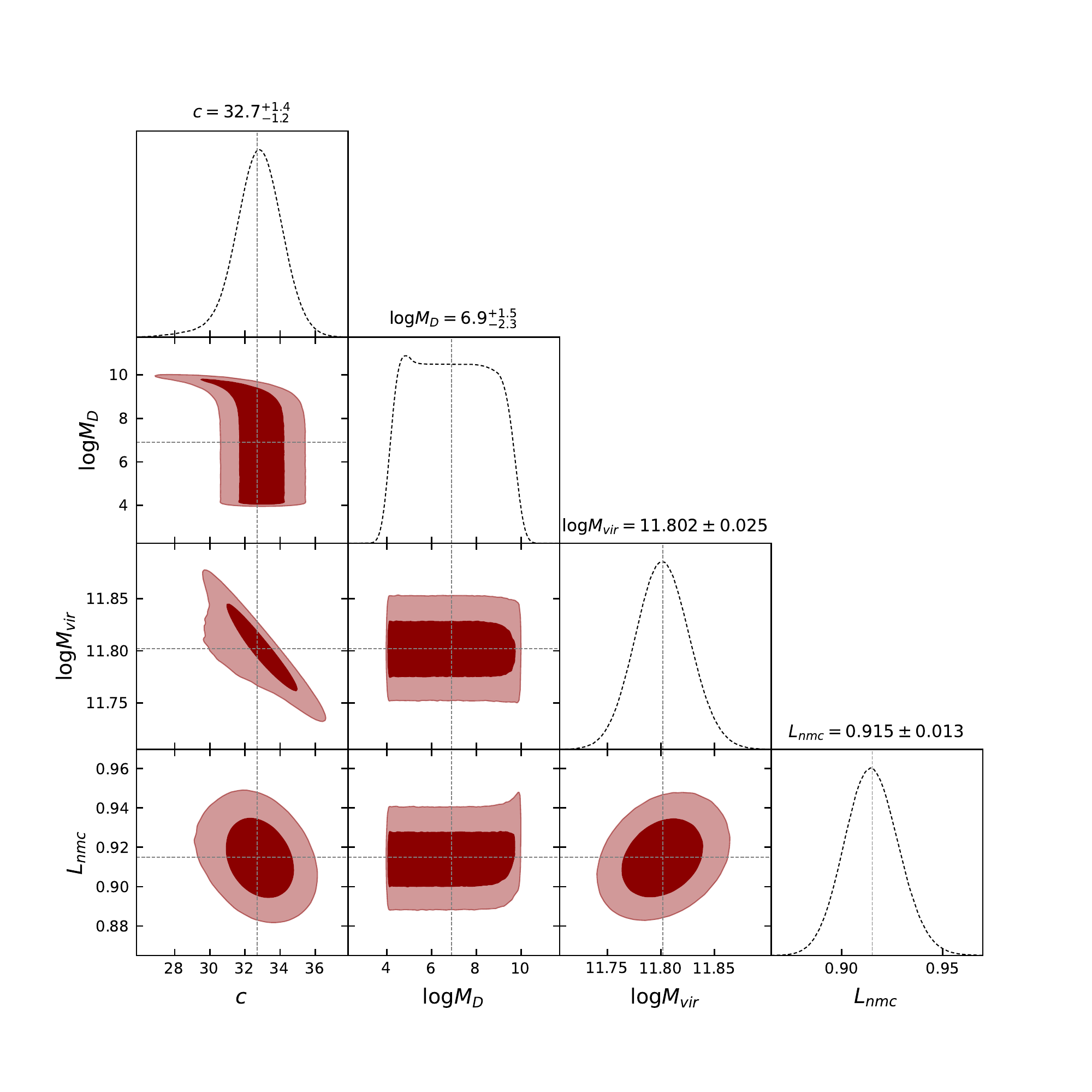}
    \end{center}
    \caption{Bin 7 of PSS96}
\end{figure}
\clearpage
\begin{figure}[ht]
    \begin{center}
    \includegraphics[width=.495\textwidth]{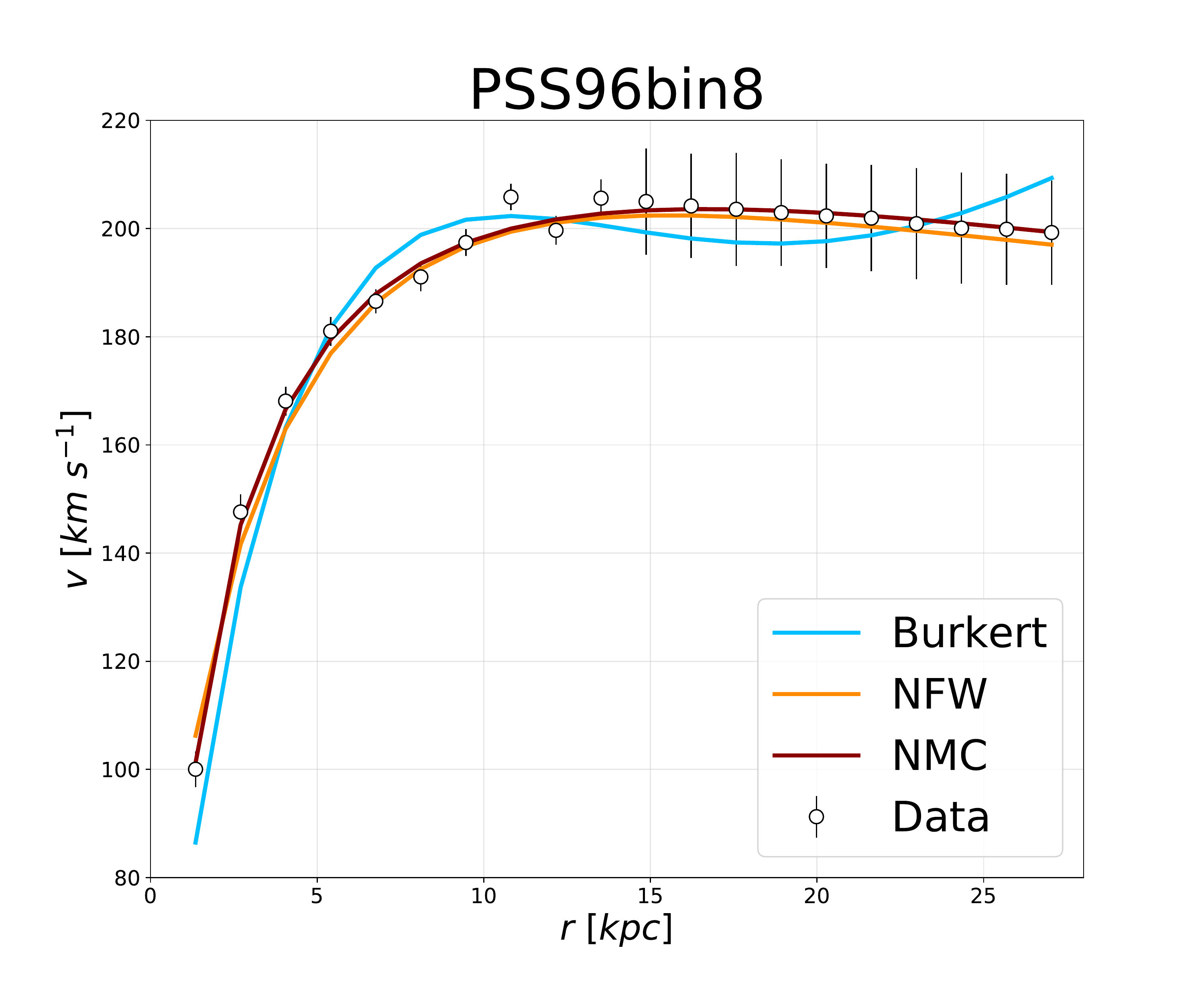}
    \includegraphics[width=.495\textwidth]{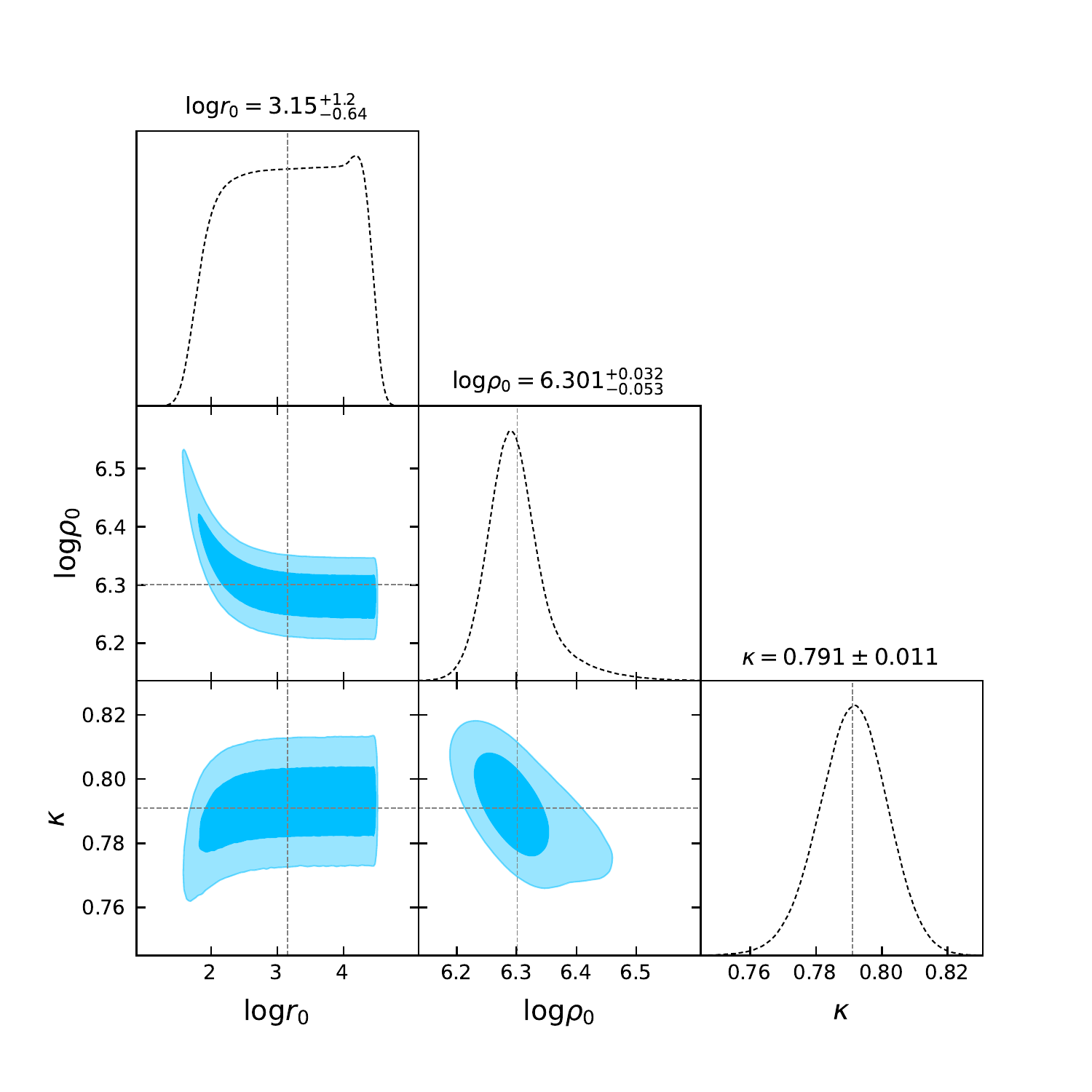}
    \includegraphics[width=.495\textwidth]{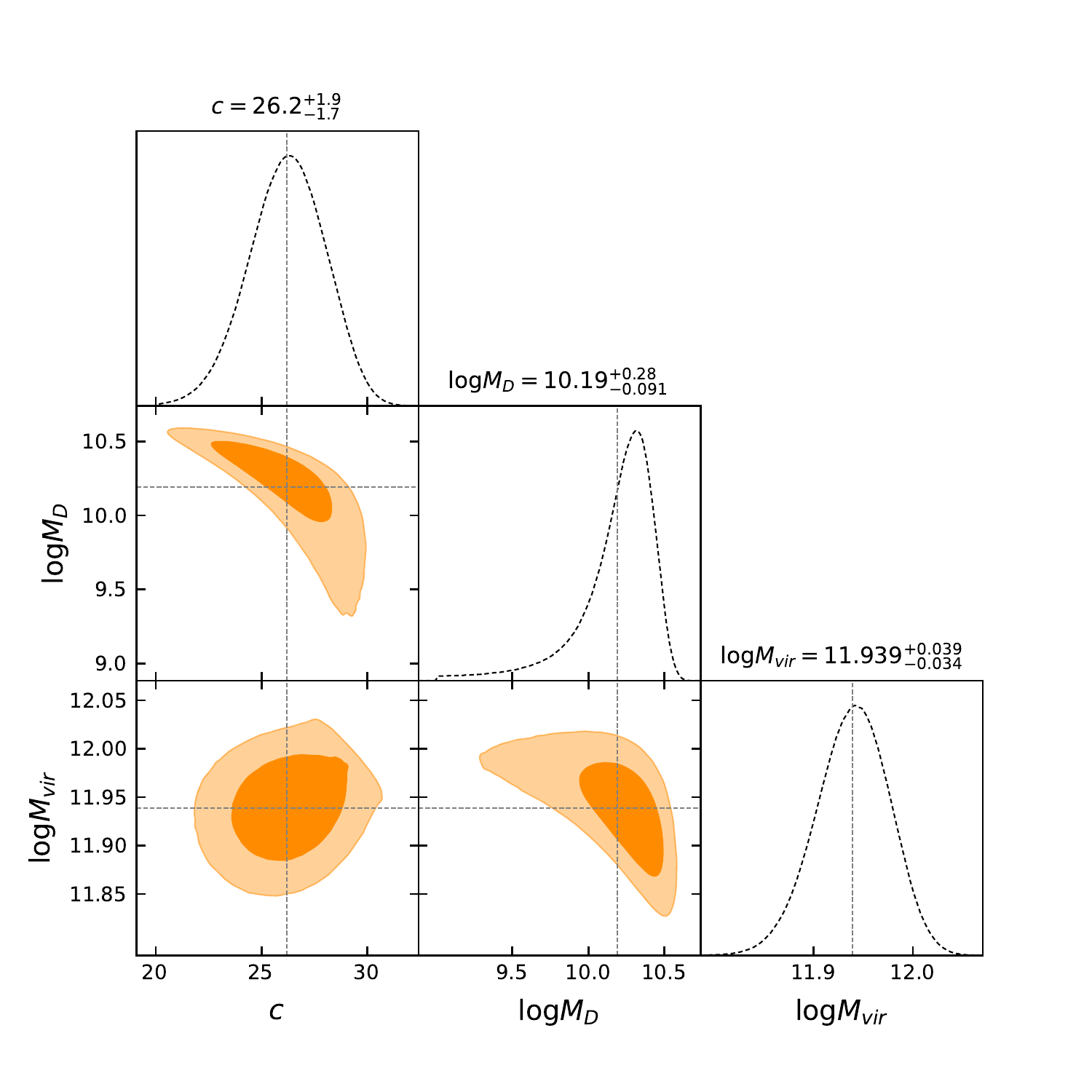}
    \includegraphics[width=.495\textwidth]{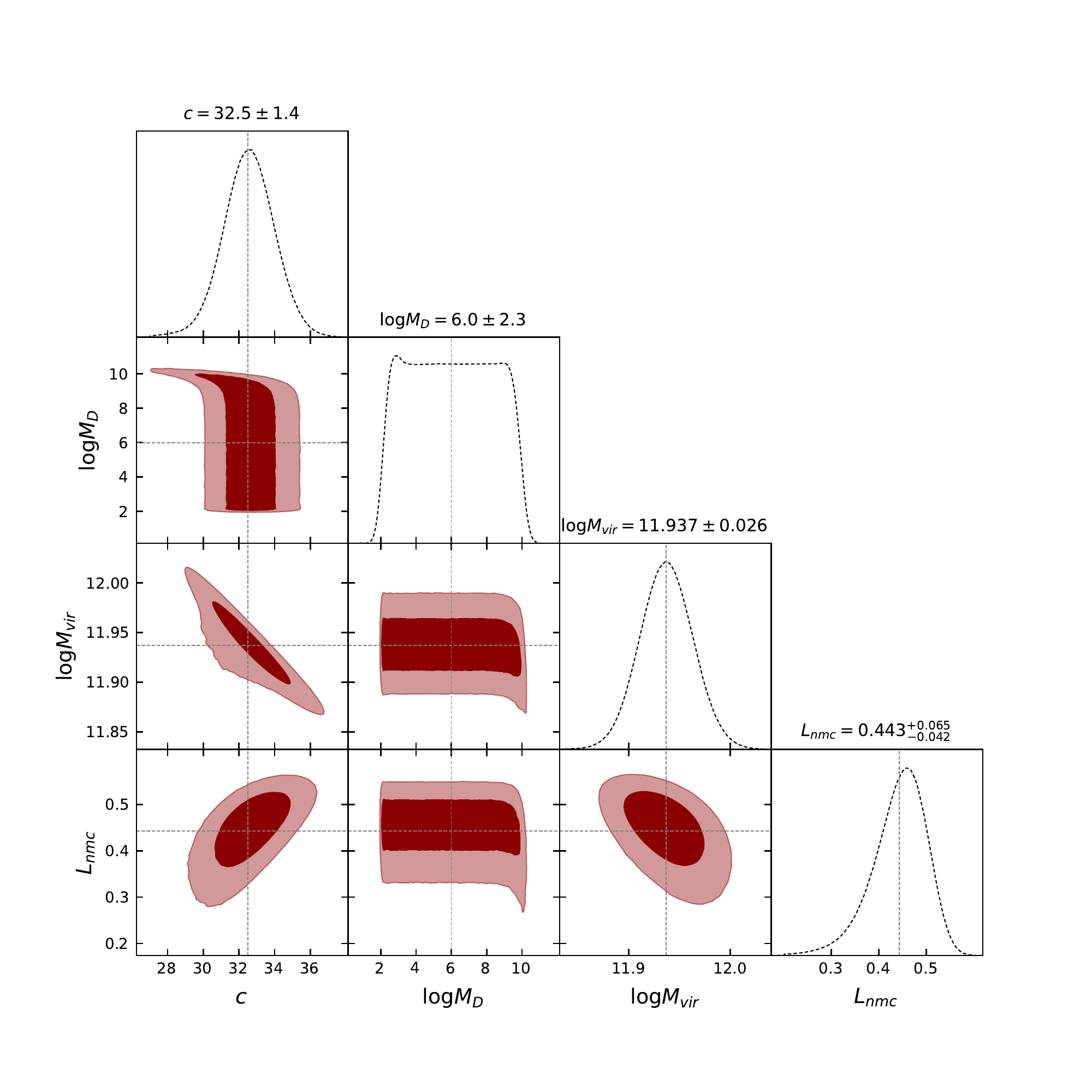}
    \end{center}
    \caption{Bin 8 of PSS96}
\end{figure}
\clearpage
\begin{figure}[ht]
    \begin{center}
    \includegraphics[width=.495\textwidth]{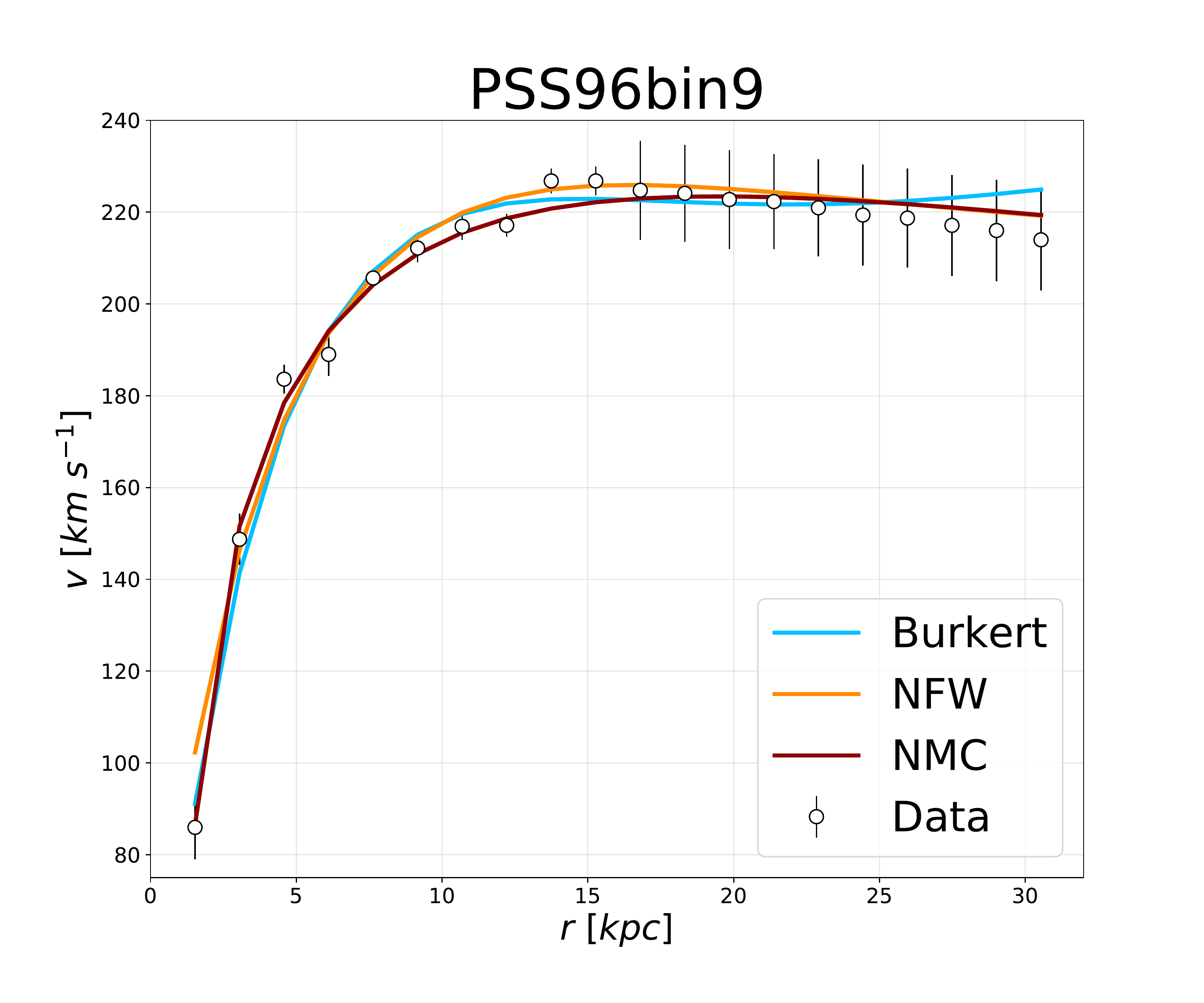}
    \includegraphics[width=.495\textwidth]{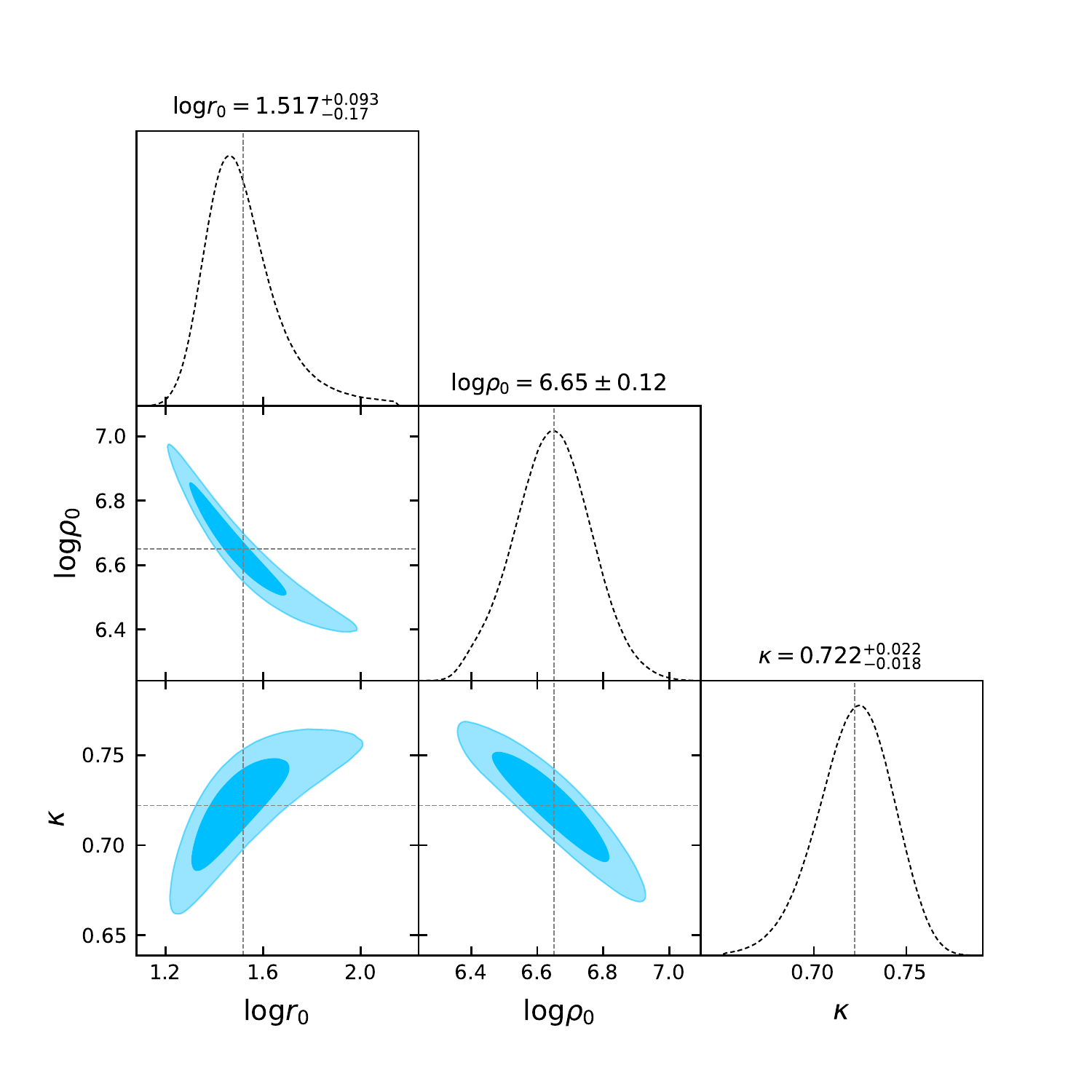}
    \includegraphics[width=.495\textwidth]{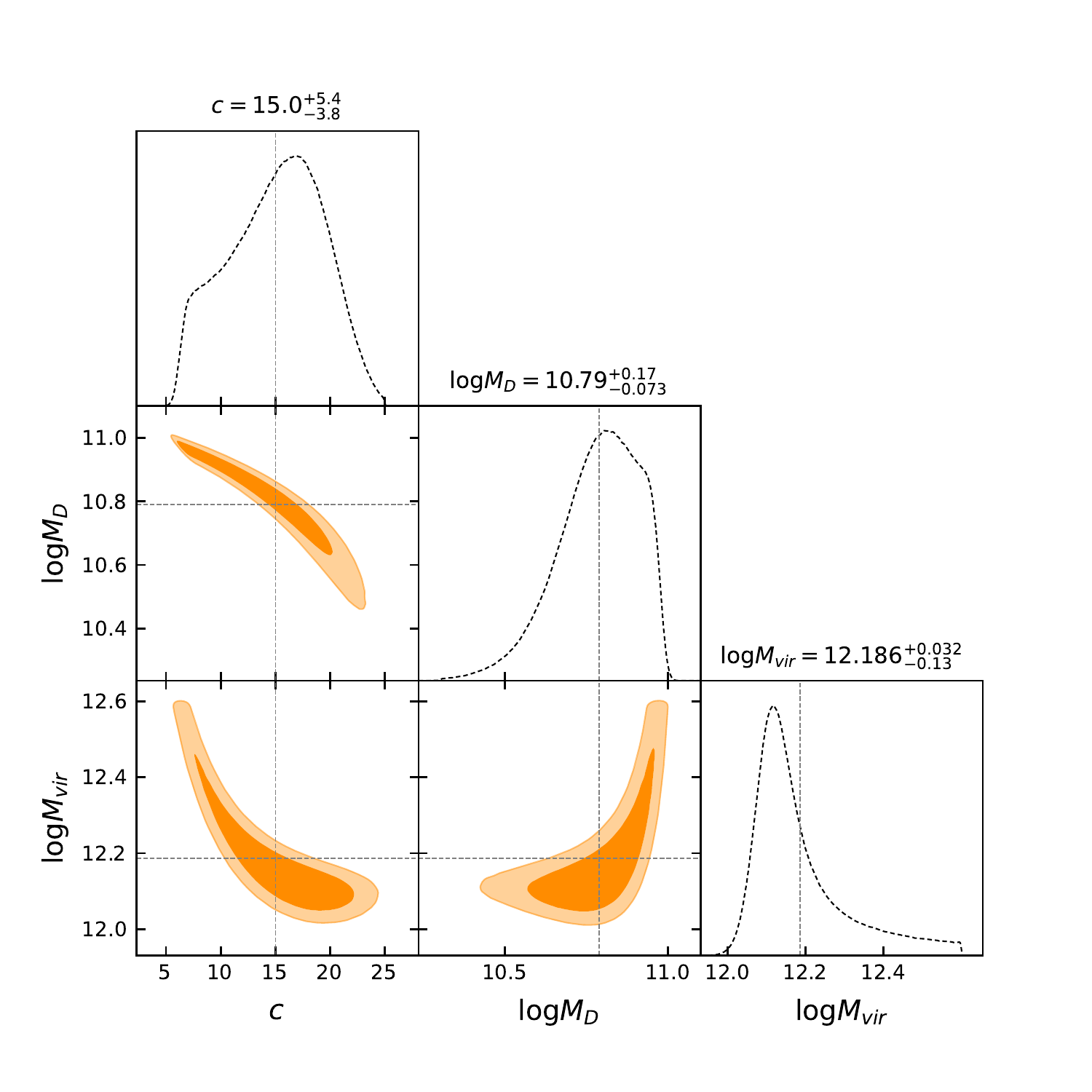}
    \includegraphics[width=.495\textwidth]{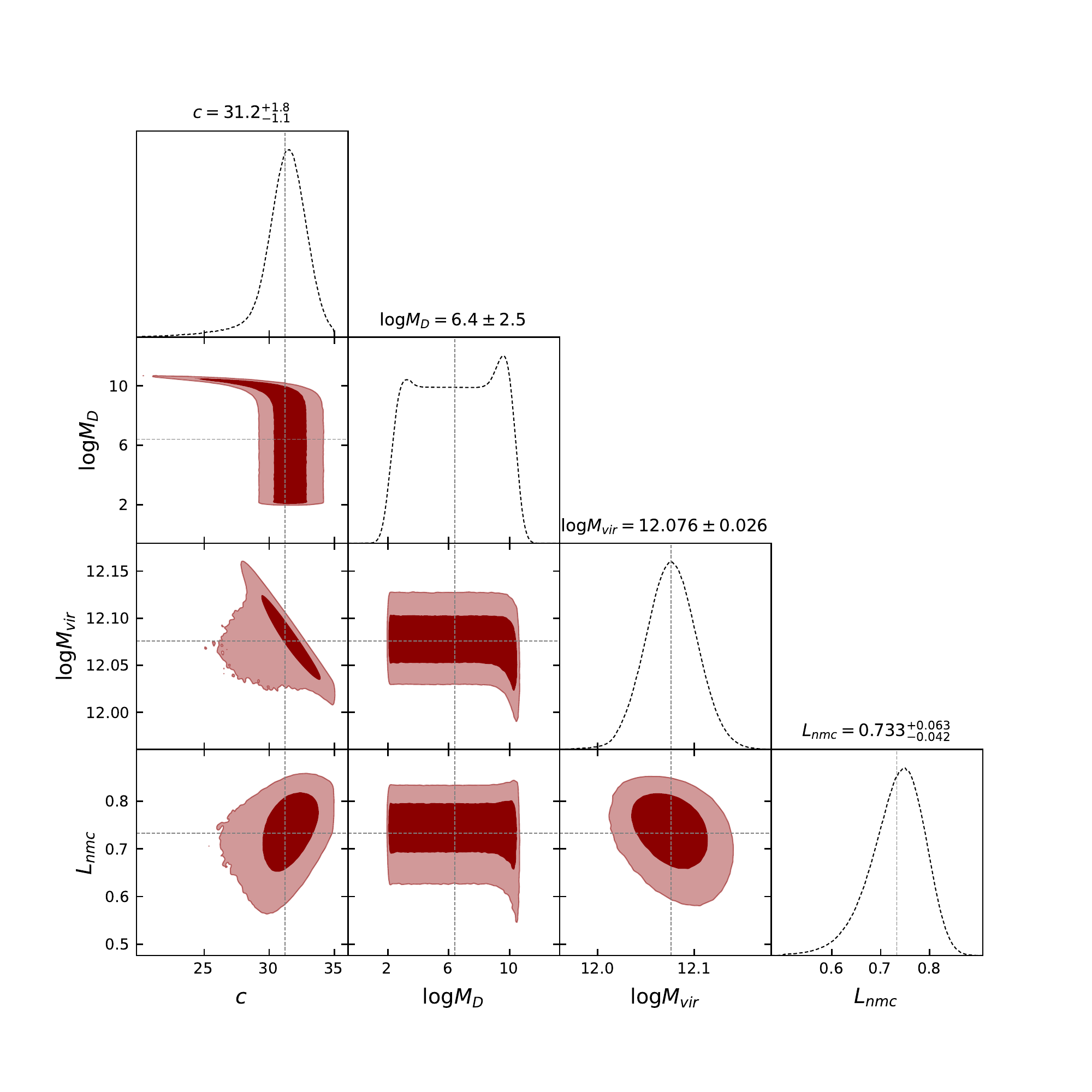}
    \end{center}
    \caption{Bin 9 of PSS96}
\end{figure}
\clearpage
\begin{figure}[ht]
    \begin{center}
    \includegraphics[width=.495\textwidth]{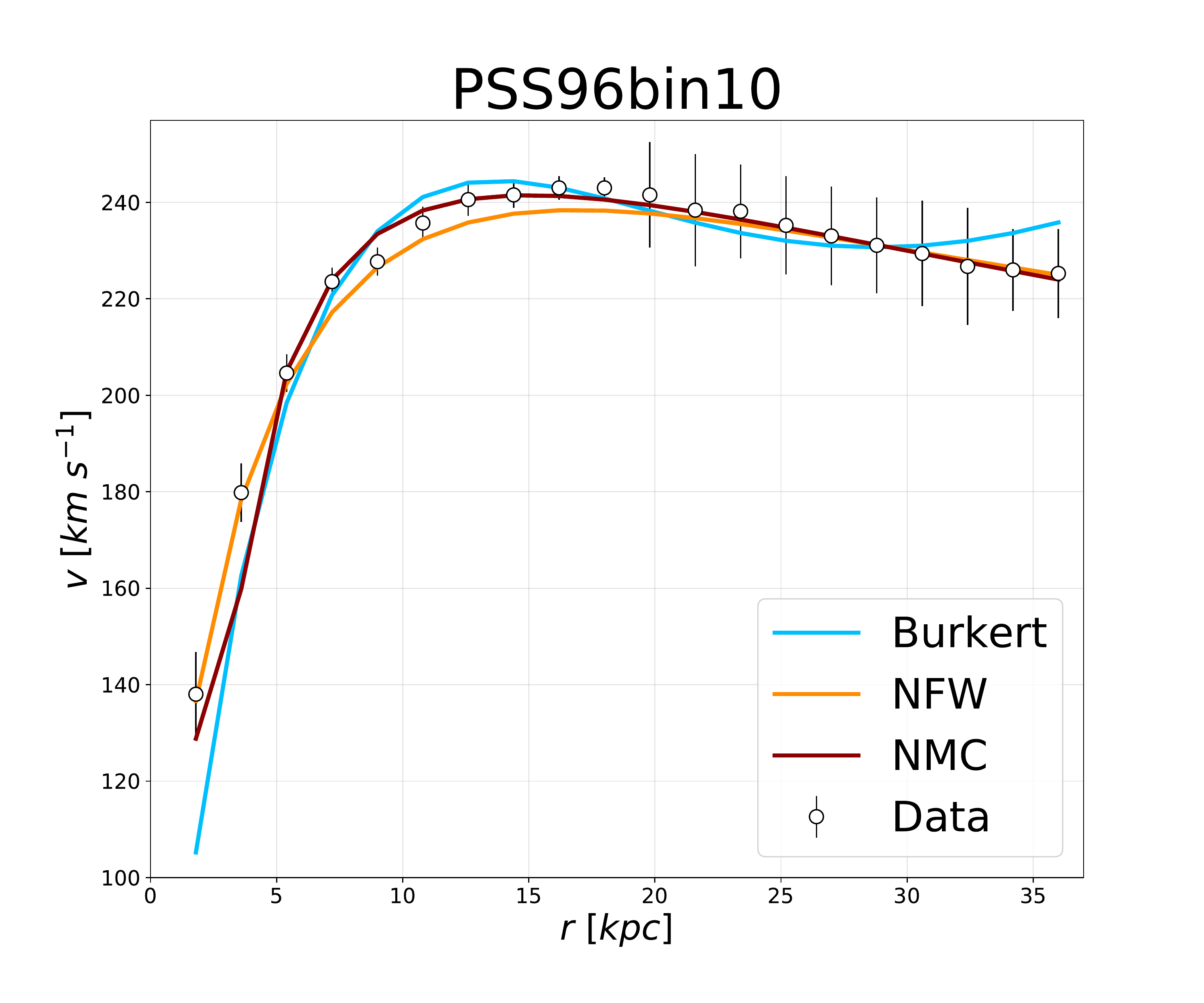}
    \includegraphics[width=.495\textwidth]{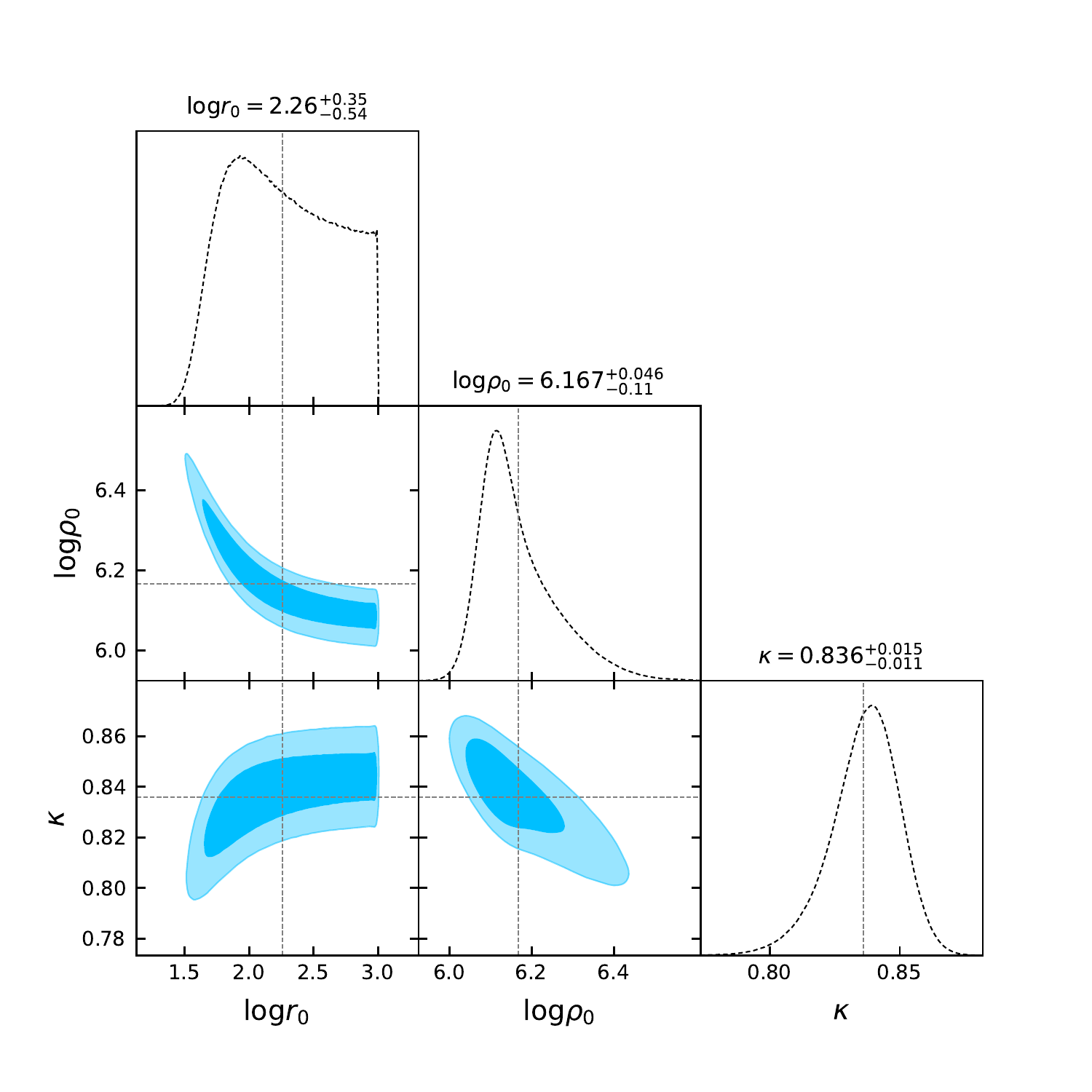}
    \includegraphics[width=.495\textwidth]{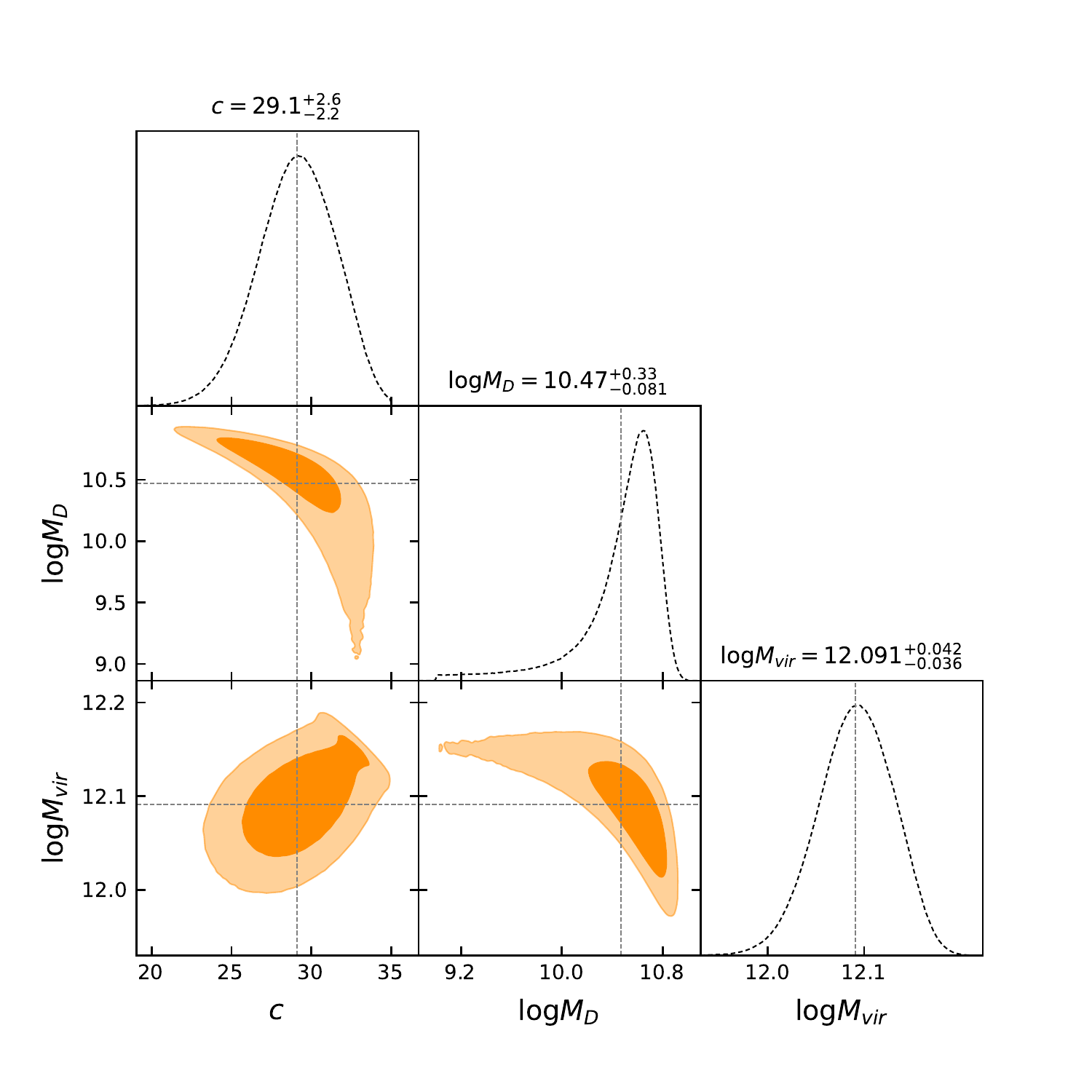}
    \includegraphics[width=.495\textwidth]{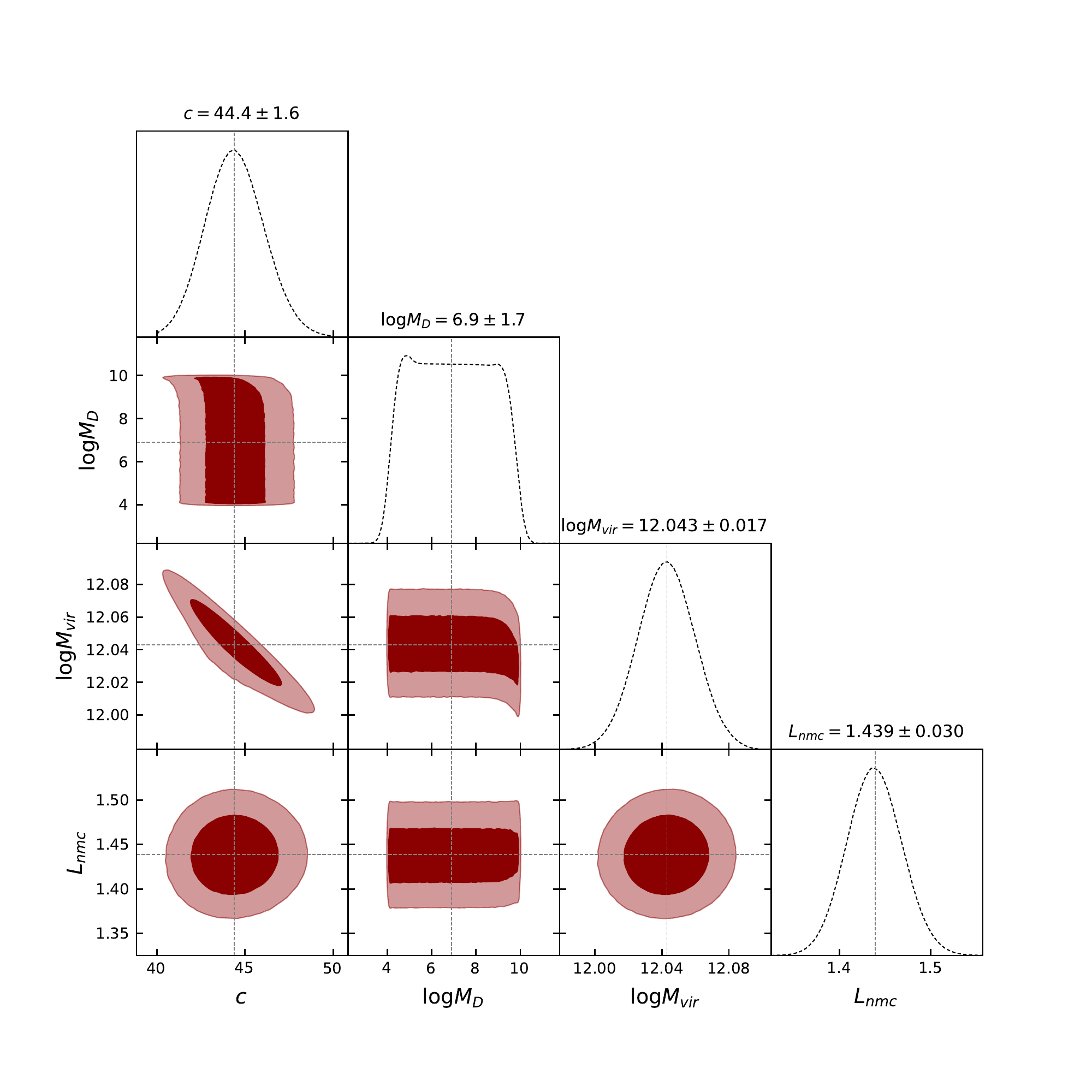}
    \end{center}
    \caption{Bin 10 of PSS96}
\end{figure}
\clearpage
\begin{figure}[ht]
    \begin{center}
    \includegraphics[width=.495\textwidth]{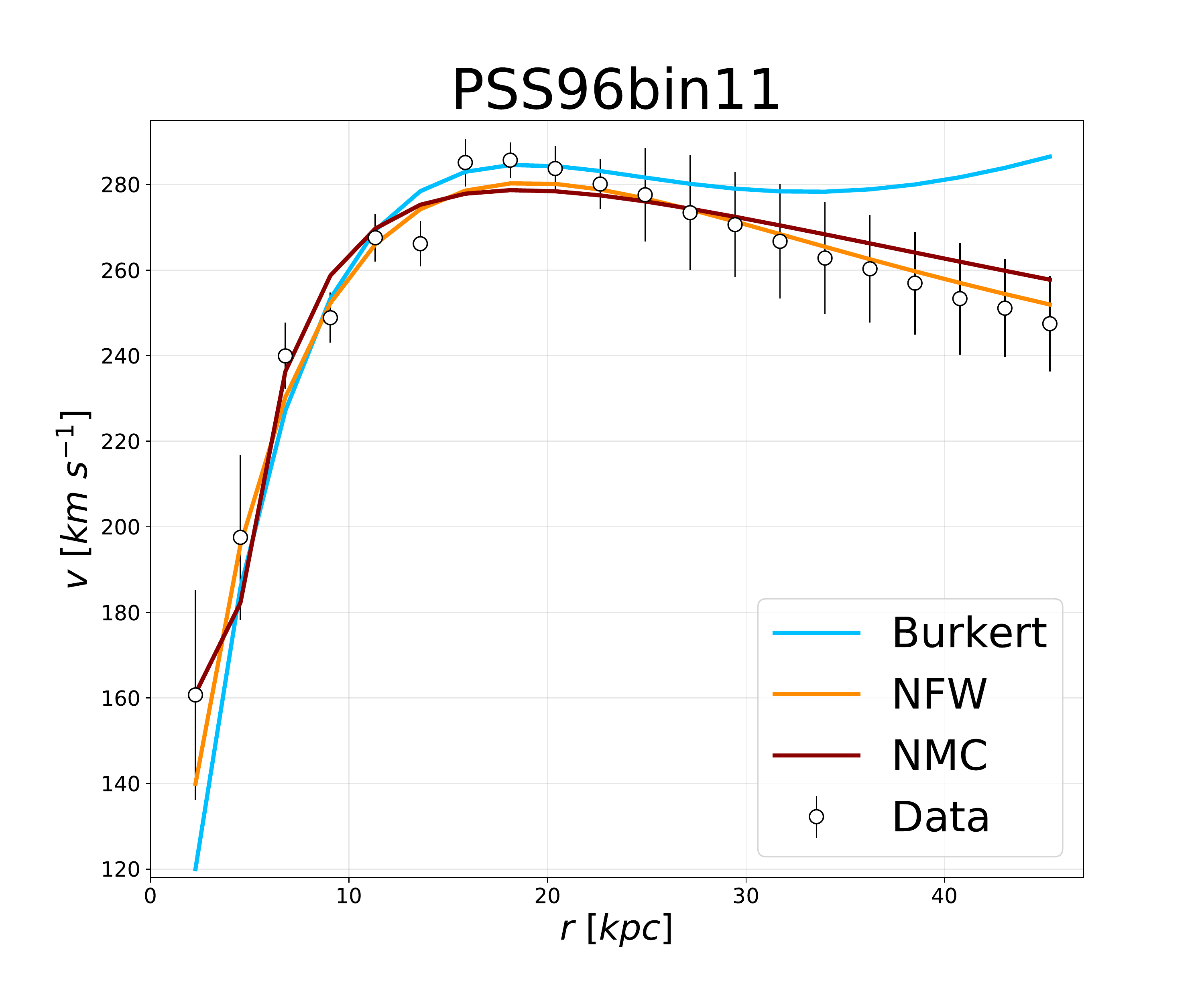}
    \includegraphics[width=.495\textwidth]{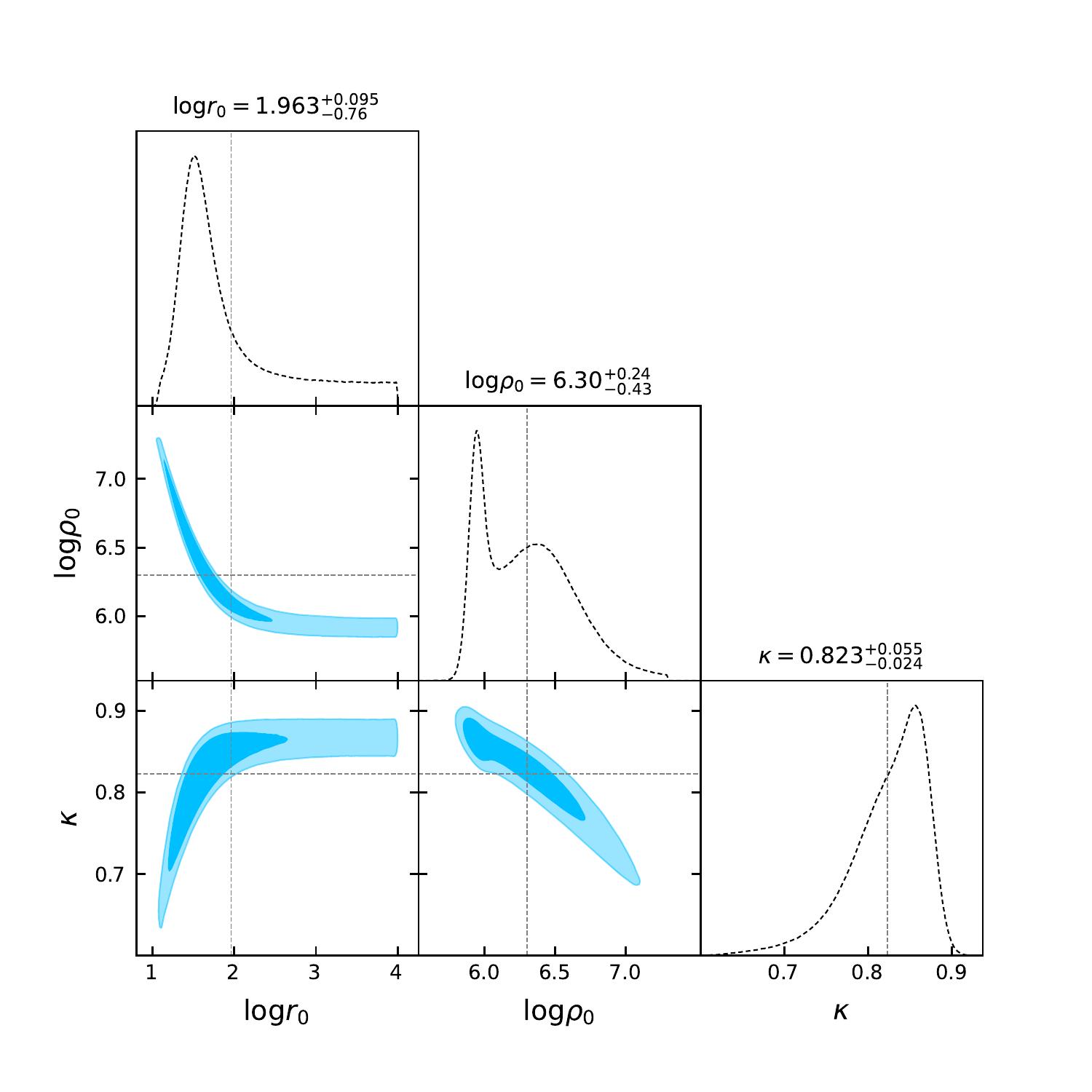}
    \includegraphics[width=.495\textwidth]{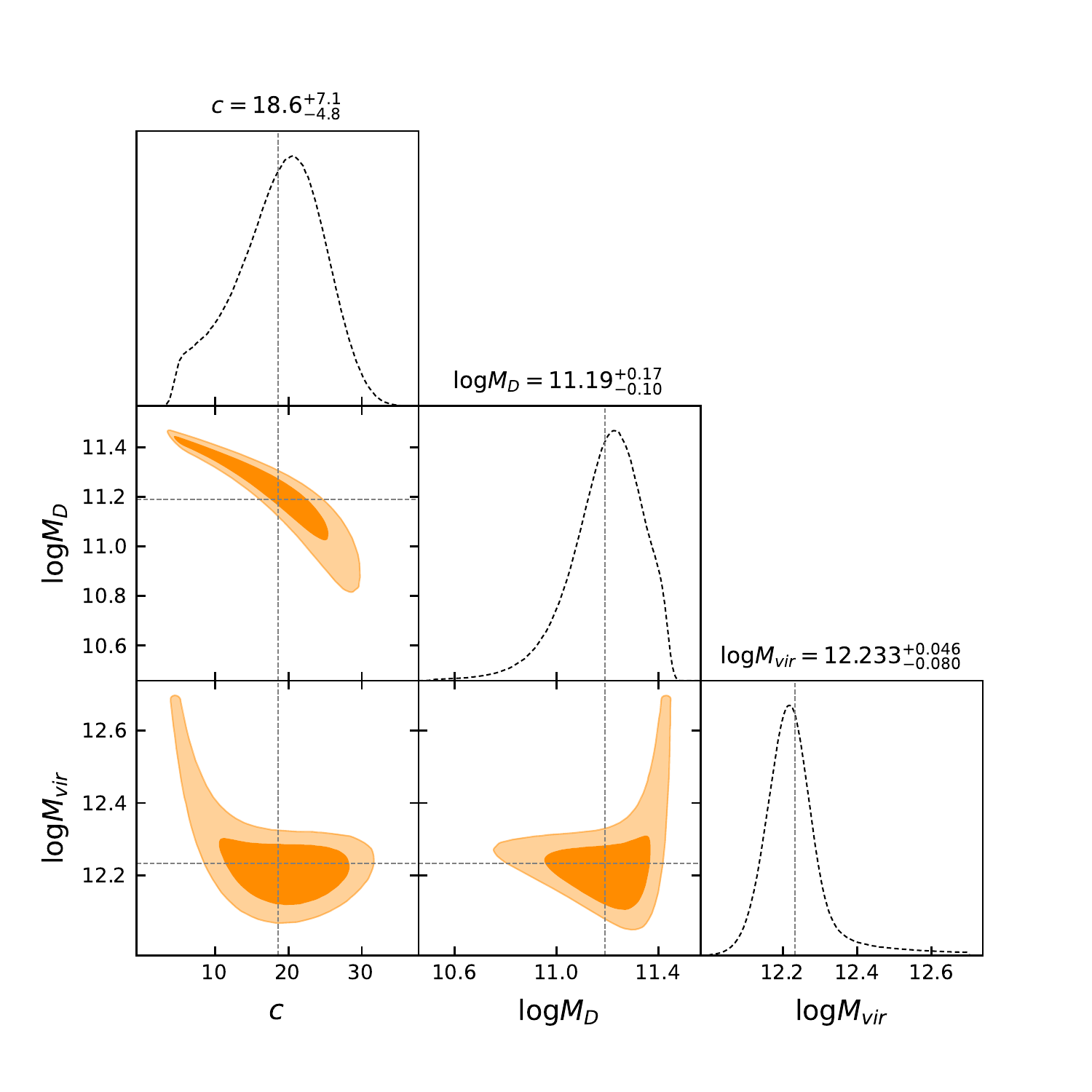}
    \includegraphics[width=.495\textwidth]{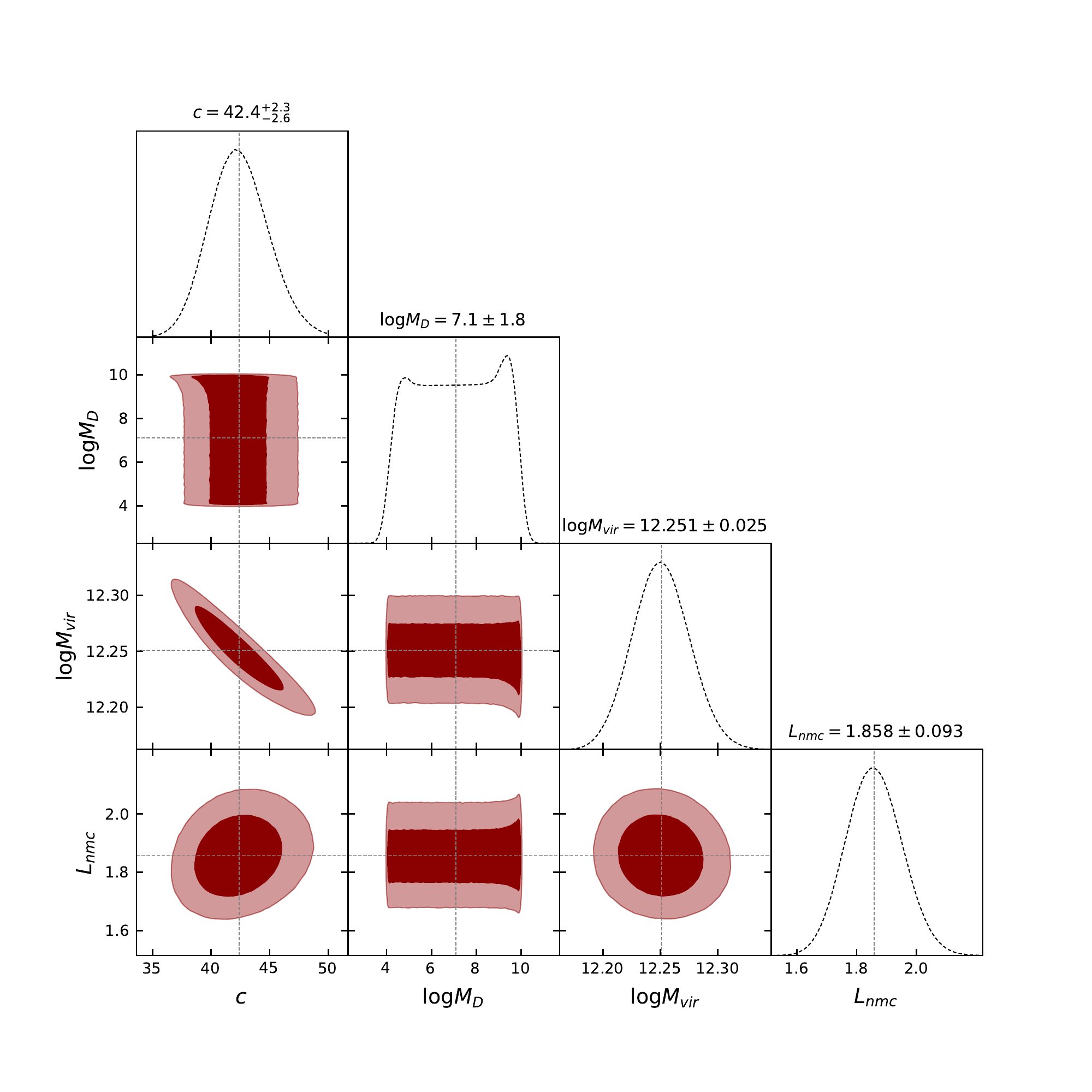}
    \end{center}
    \caption{Bin 11 of PSS96}
\end{figure}
\clearpage
\begin{figure}[ht]
    \begin{center}
    \includegraphics[width=.495\textwidth]{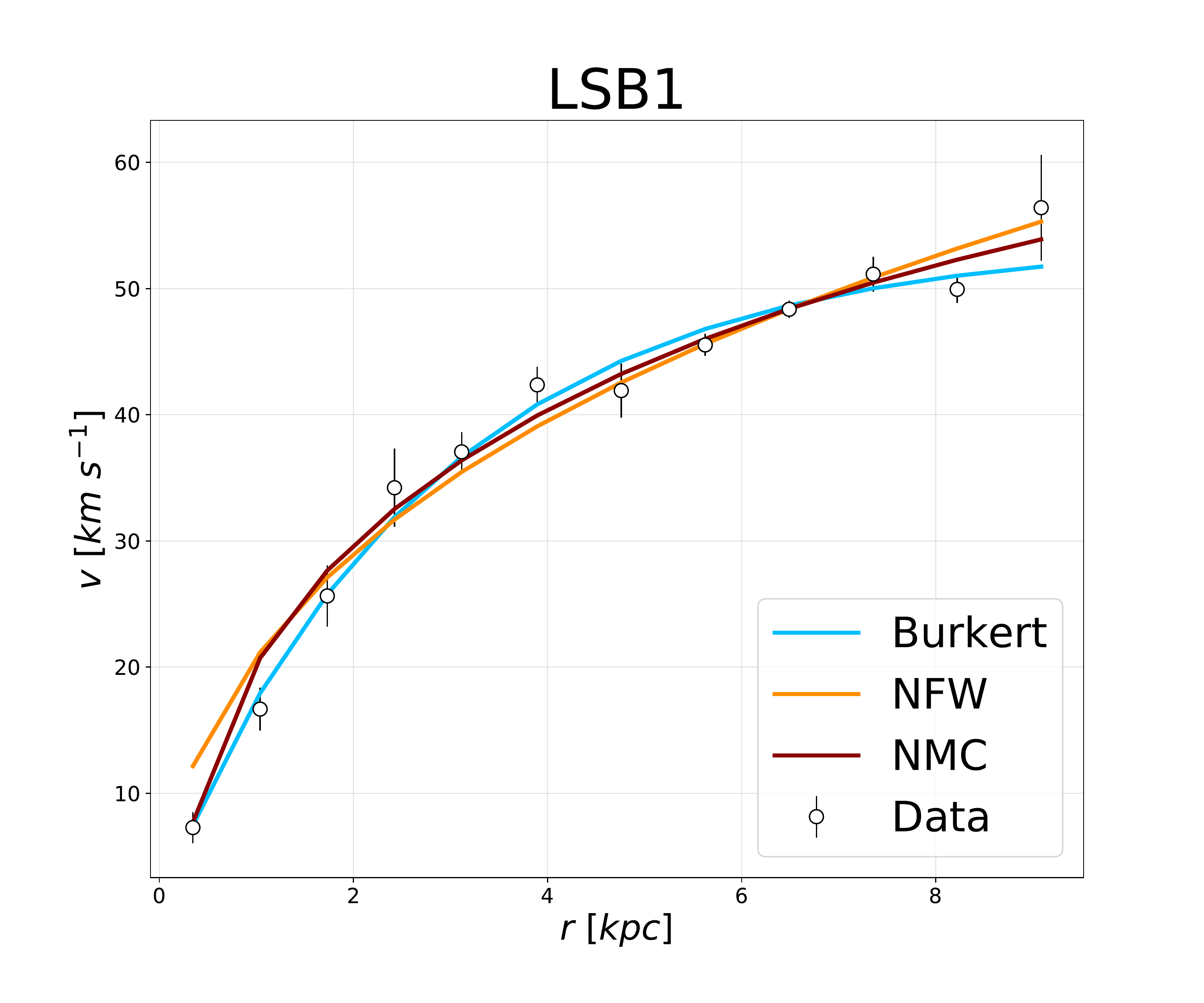}
    \includegraphics[width=.495\textwidth]{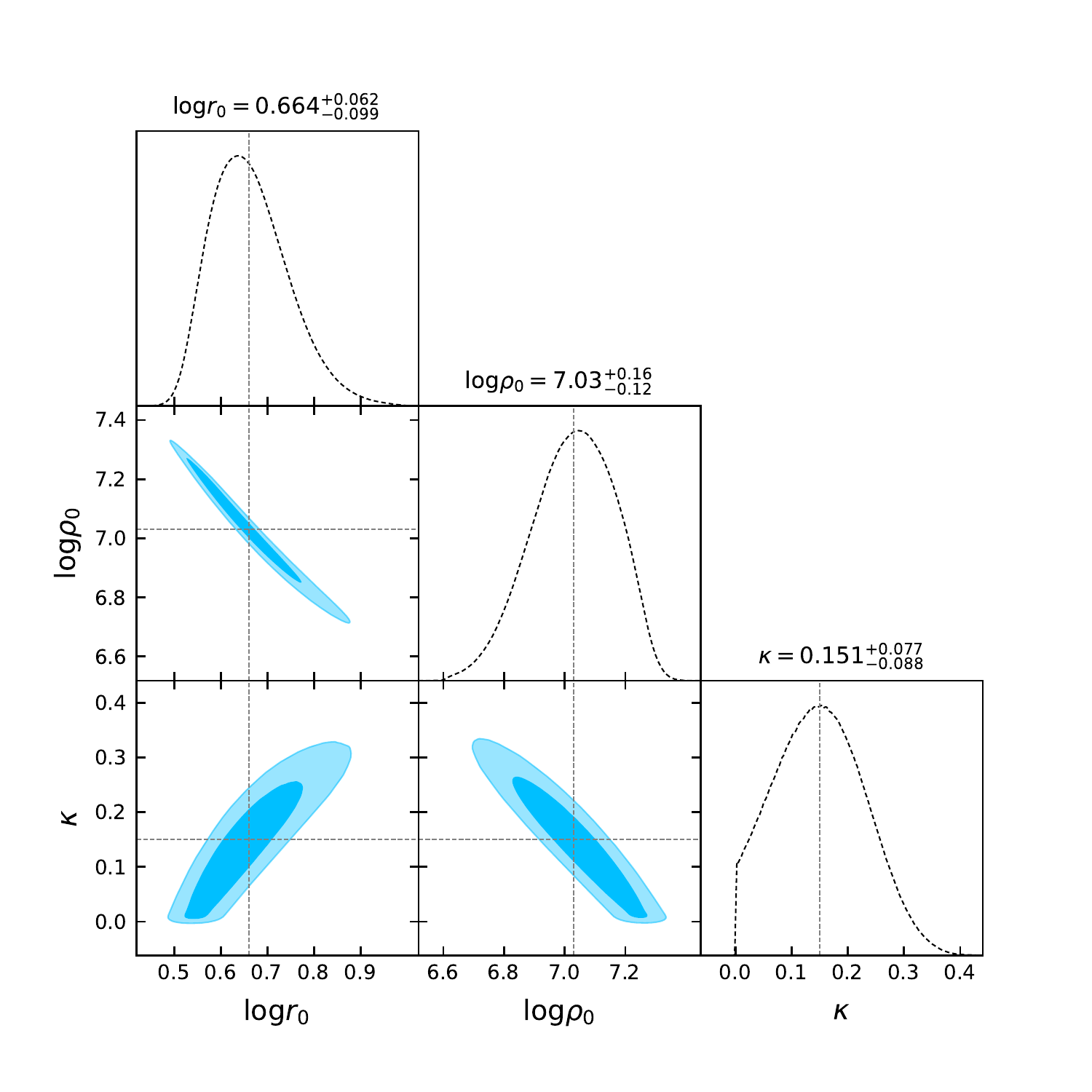}
    \includegraphics[width=.495\textwidth]{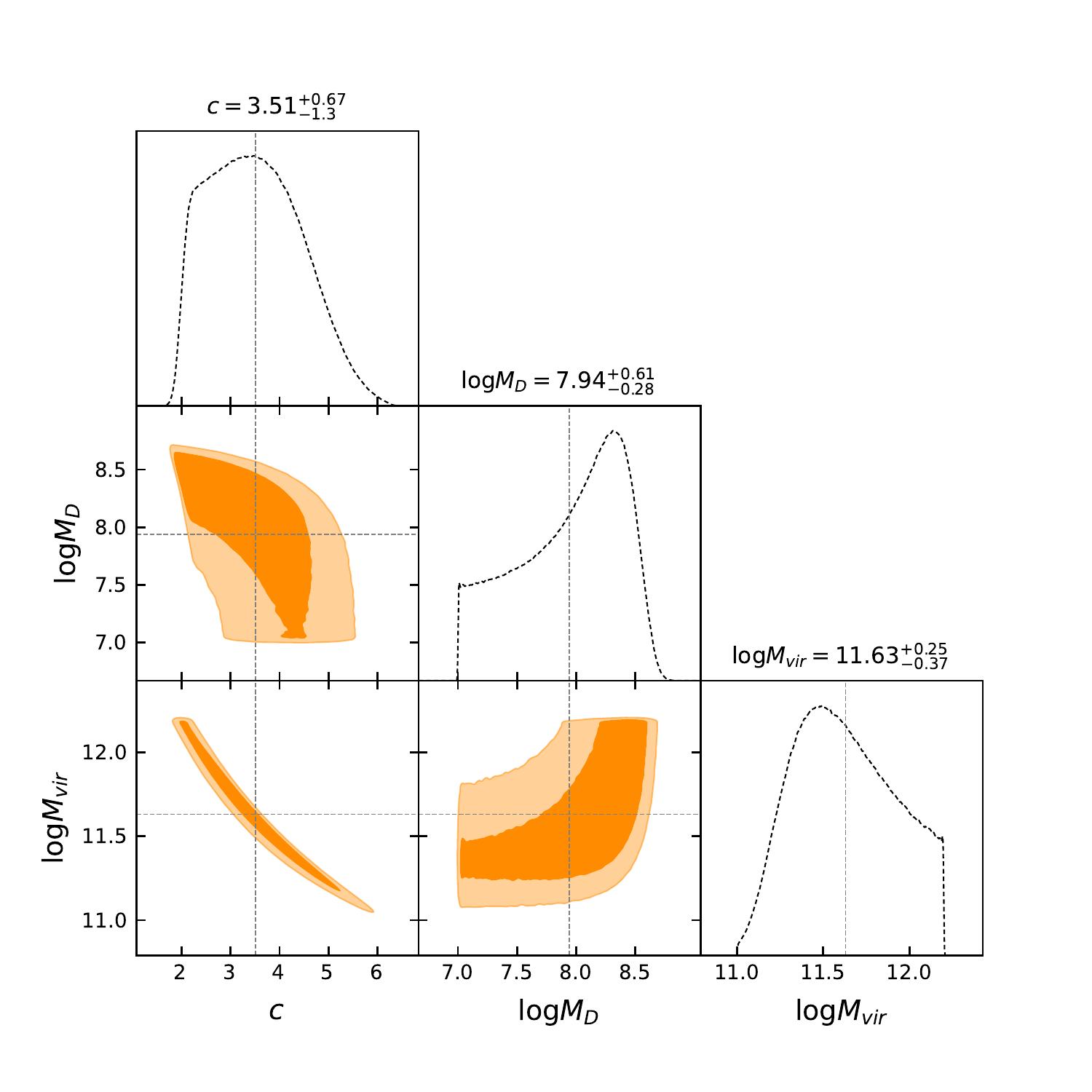}
    \includegraphics[width=.495\textwidth]{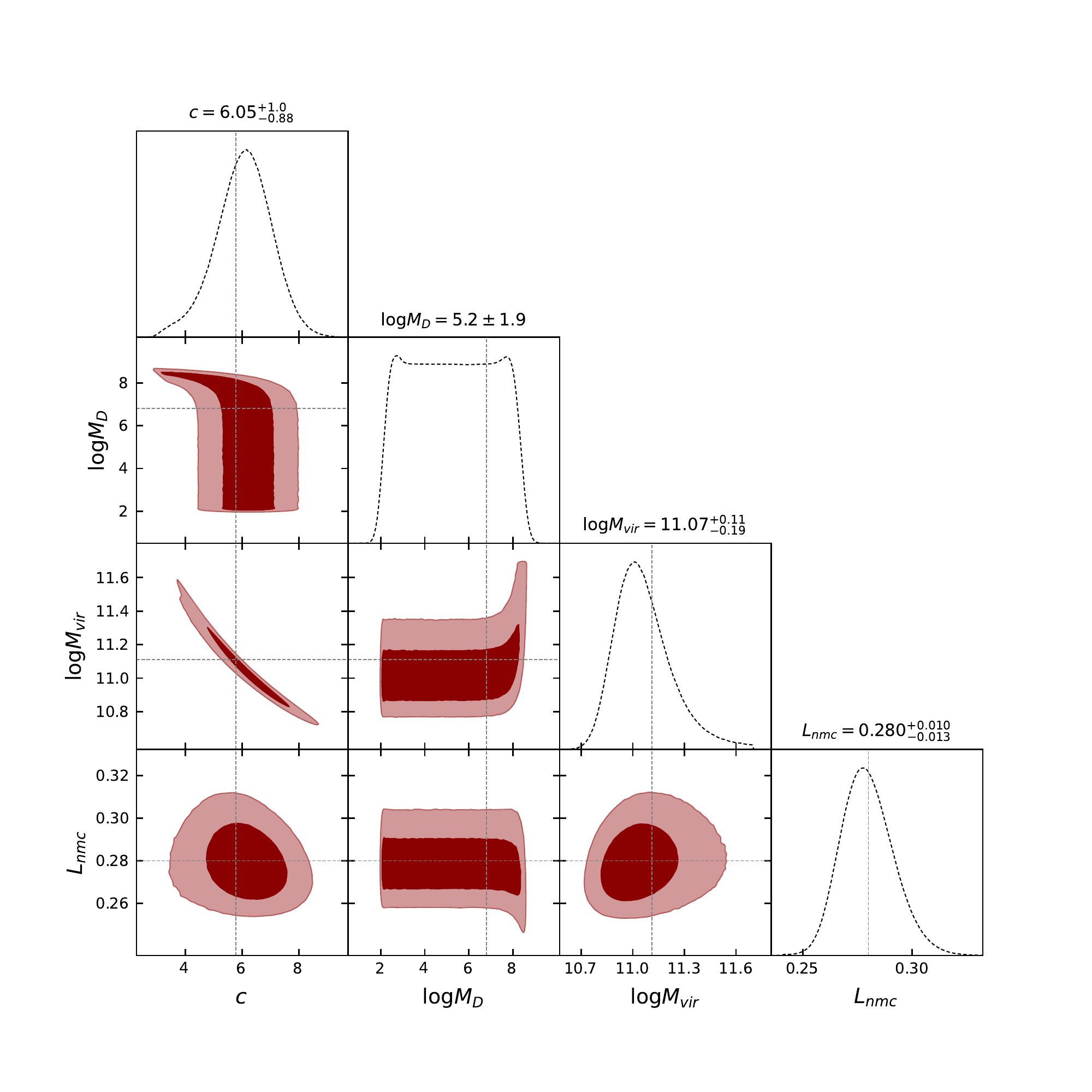}
    \end{center}
    \caption{Bin 1 of LSB galaxies}
\end{figure}
\clearpage
\begin{figure}[ht]
    \begin{center}
    \includegraphics[width=.495\textwidth]{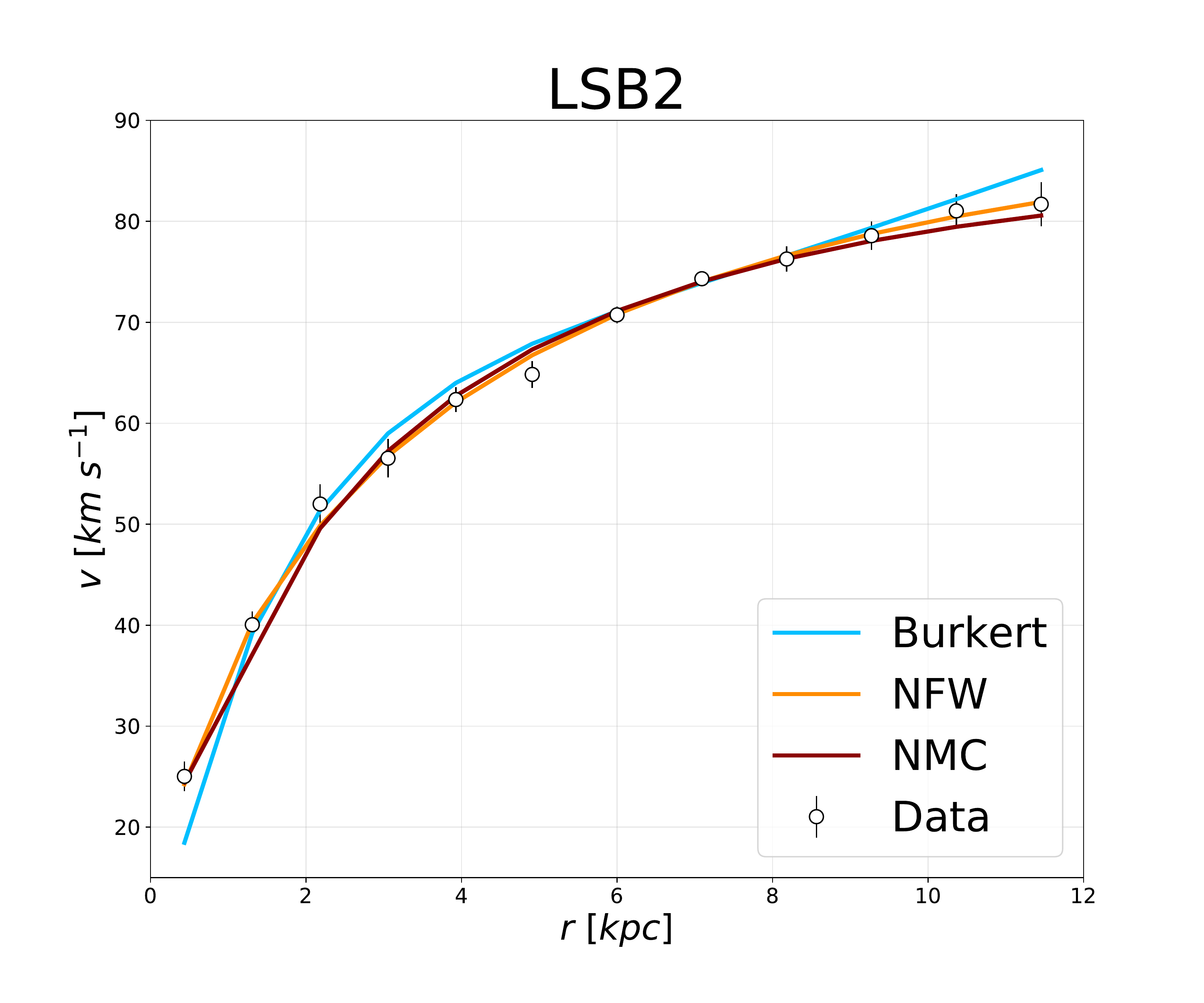}
    \includegraphics[width=.495\textwidth]{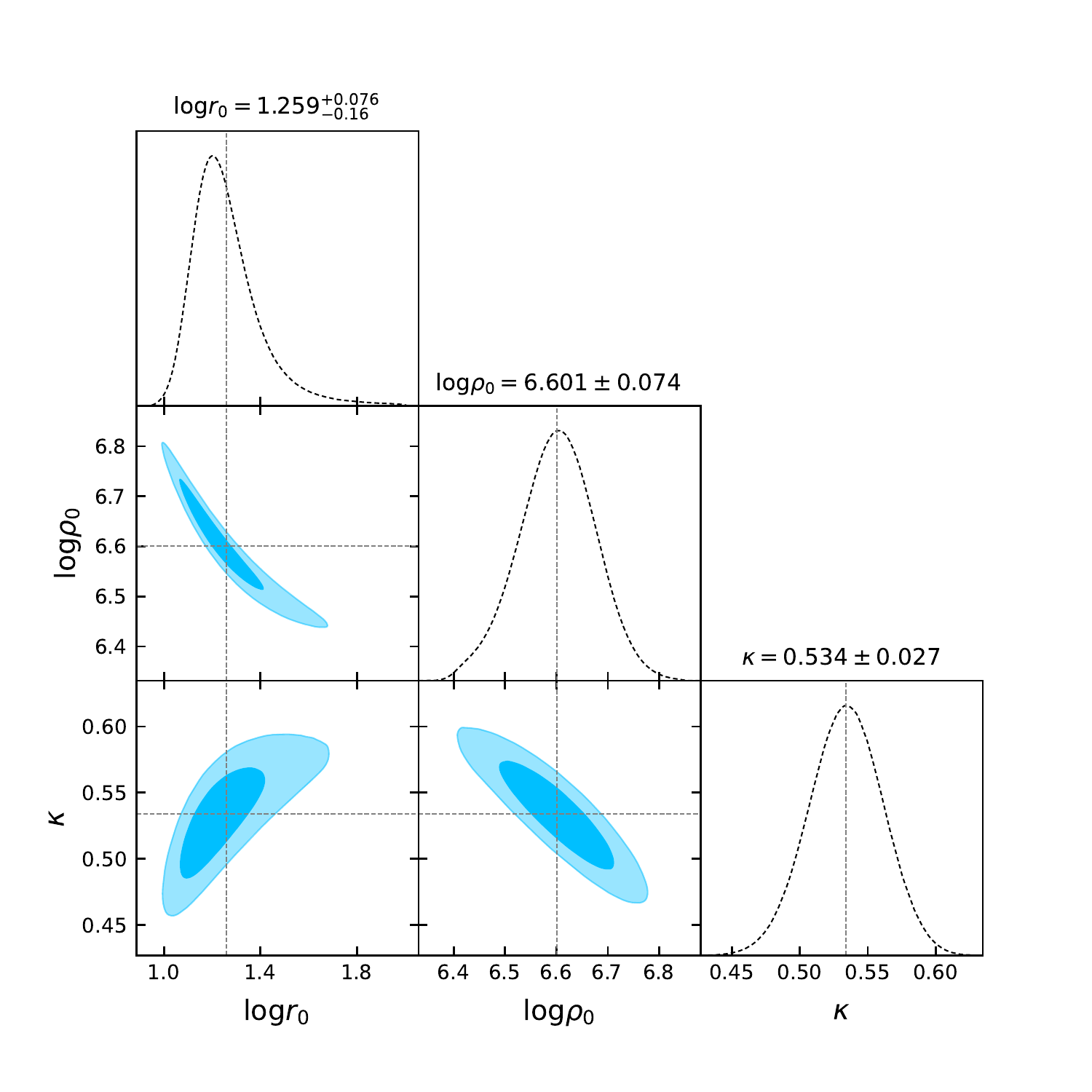}
    \includegraphics[width=.495\textwidth]{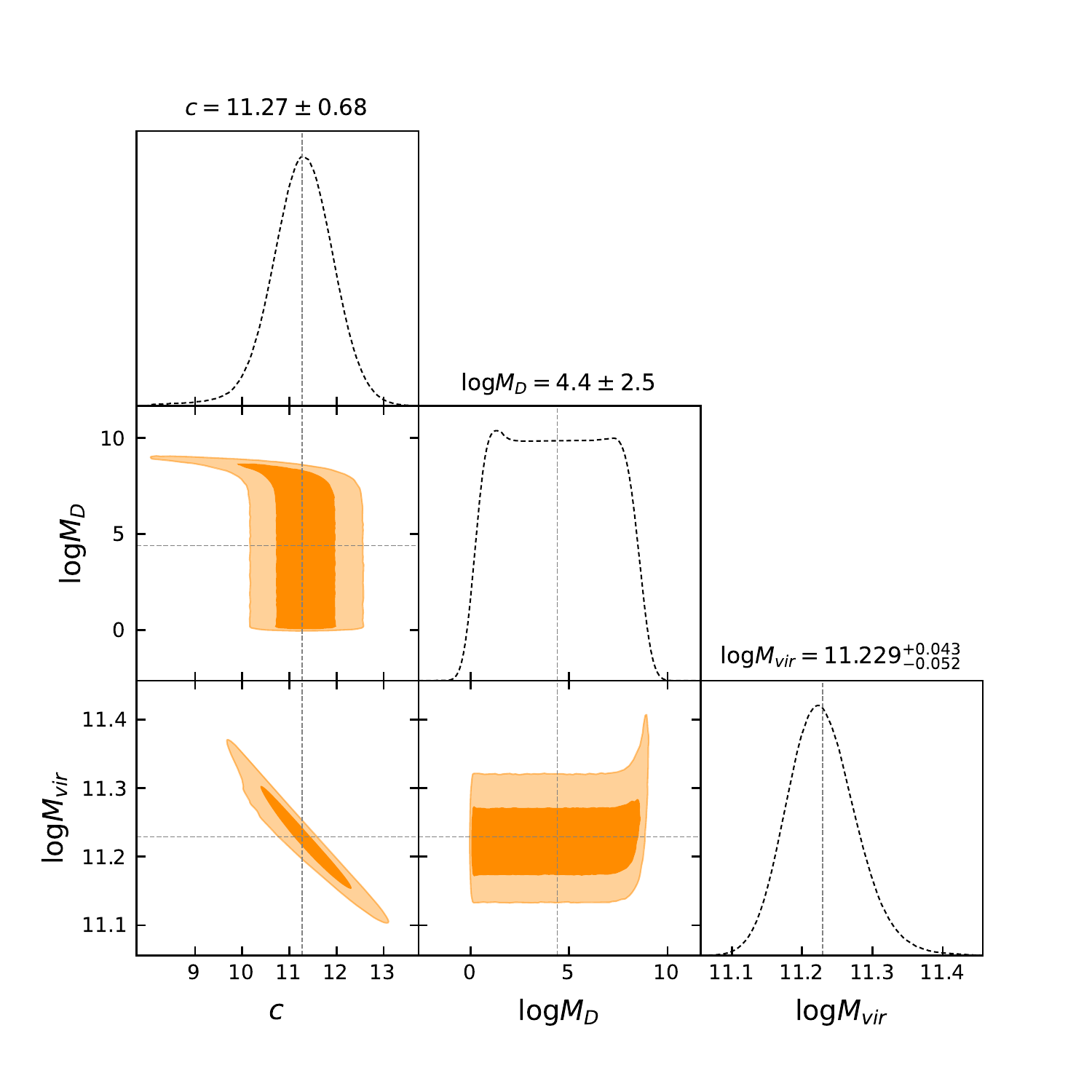}
    \includegraphics[width=.495\textwidth]{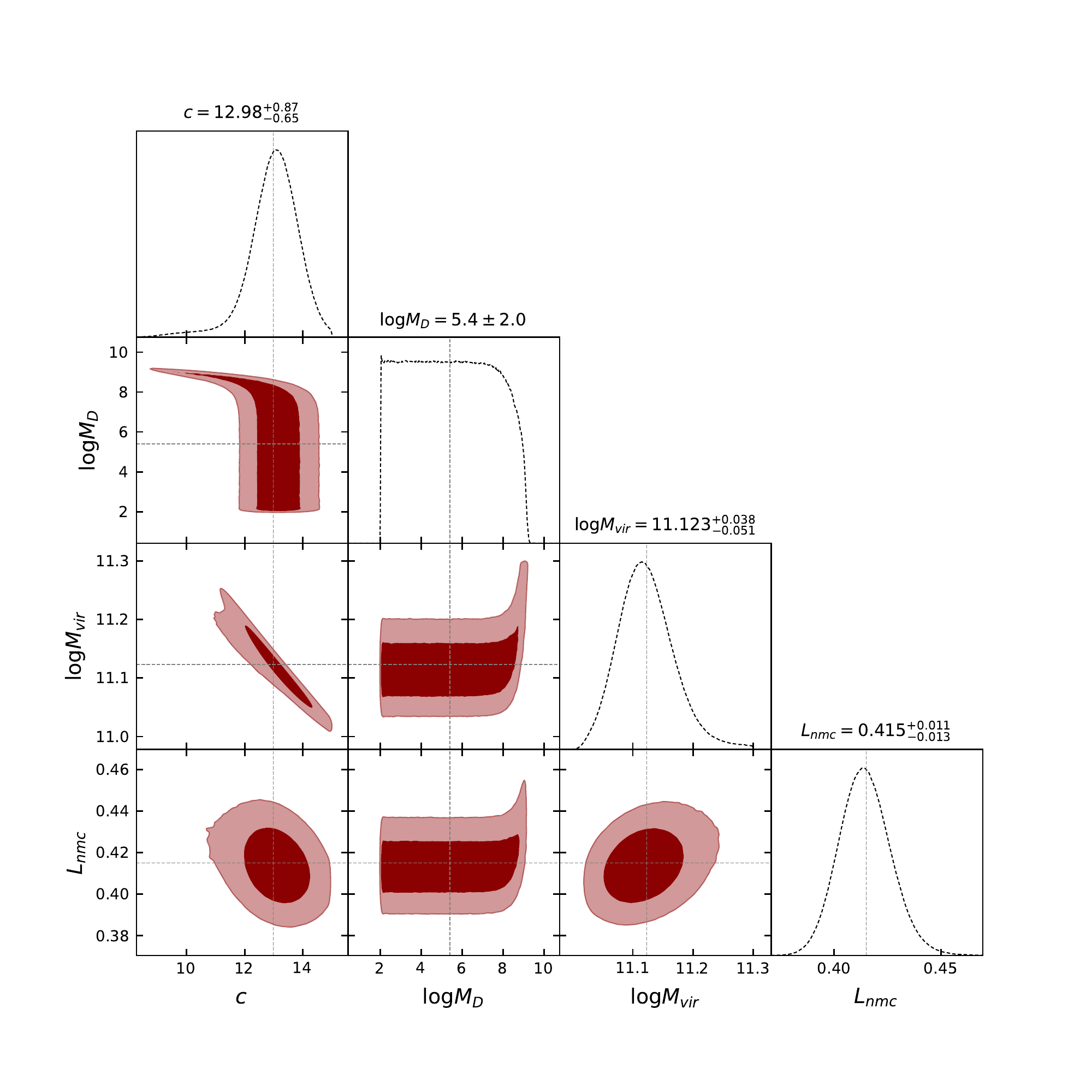}
    \end{center}
    \caption{Bin 2 of LSB galaxies}
\end{figure}
\clearpage
\begin{figure}[ht]
    \begin{center}
    \includegraphics[width=.495\textwidth]{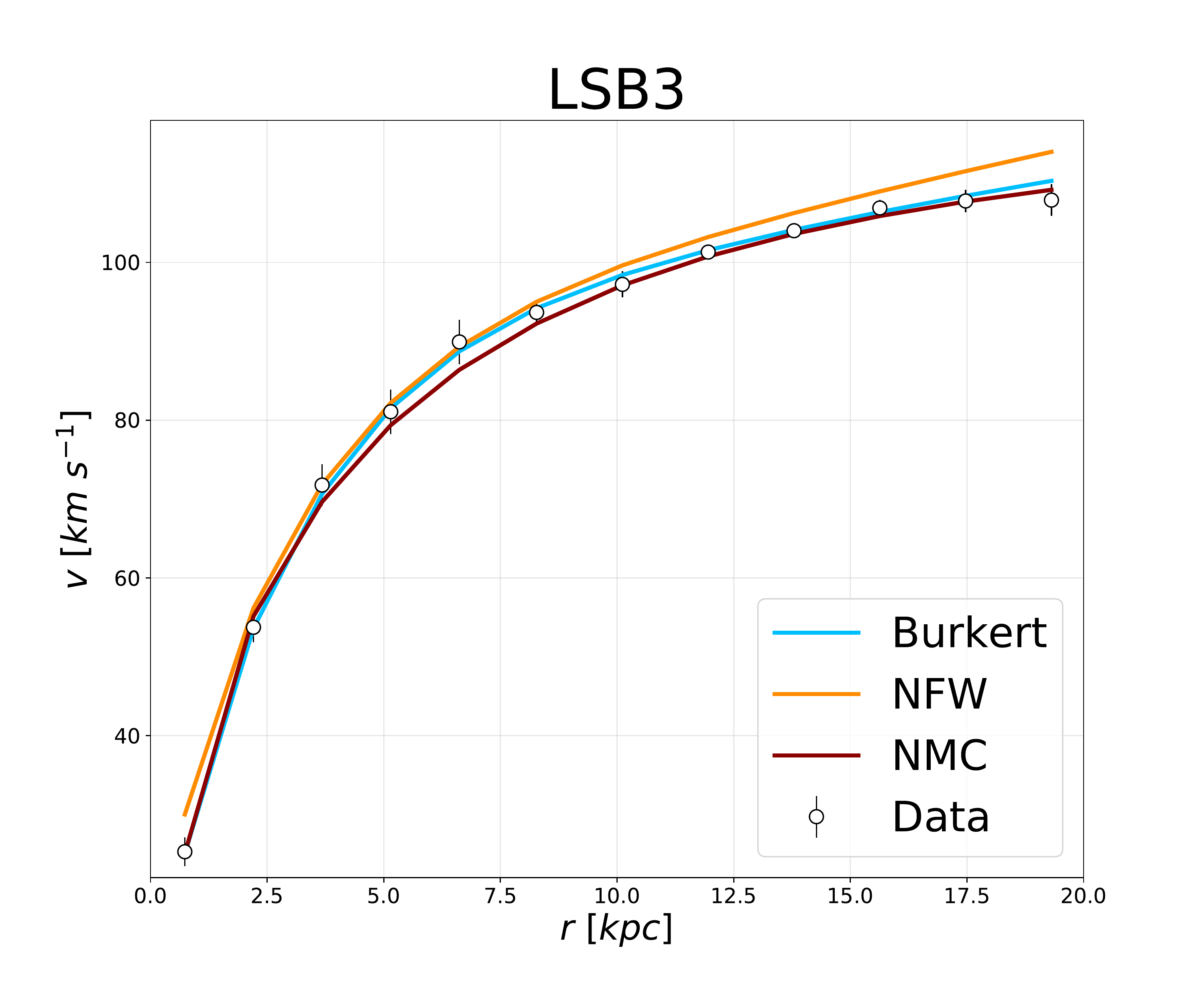}
    \includegraphics[width=.495\textwidth]{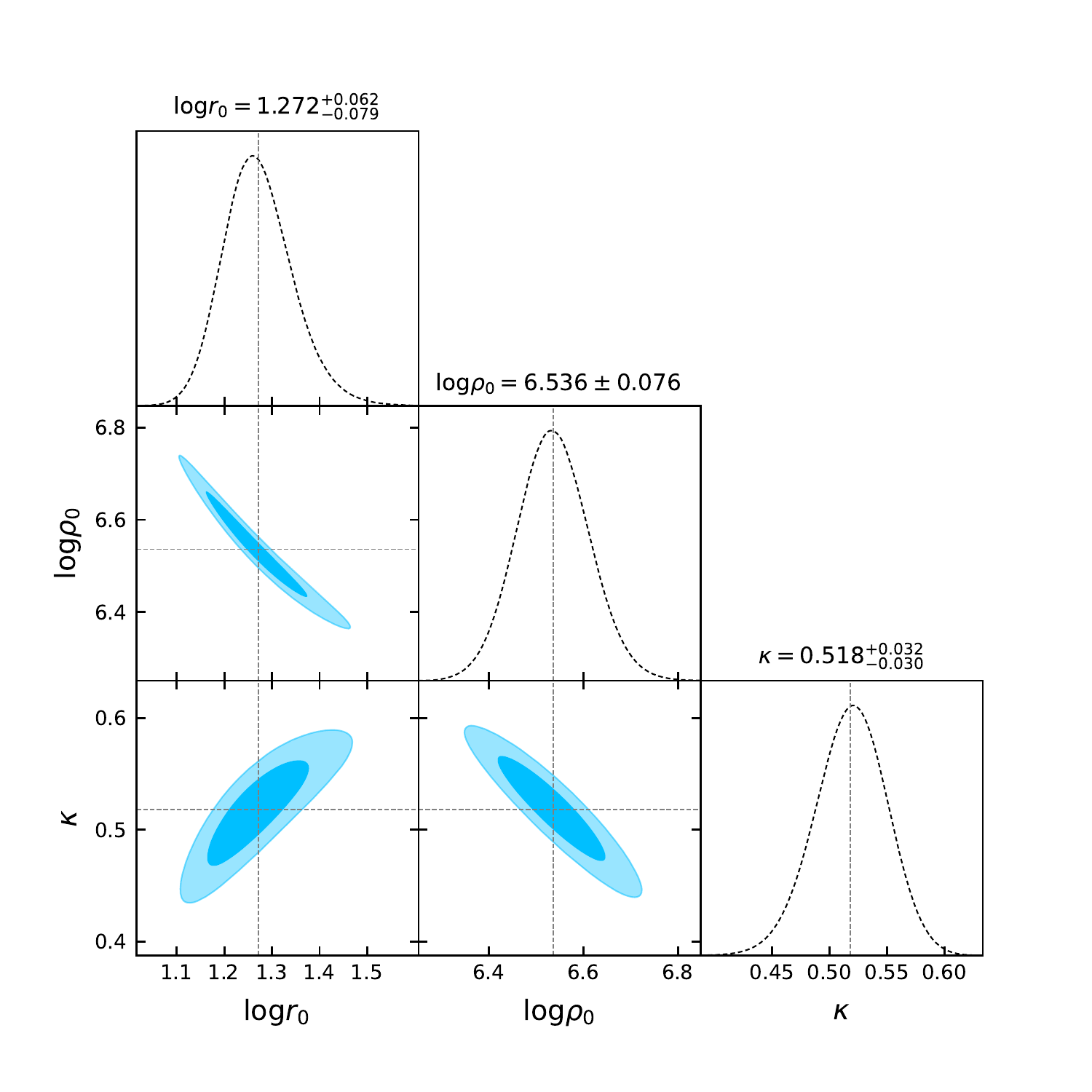}
    \includegraphics[width=.495\textwidth]{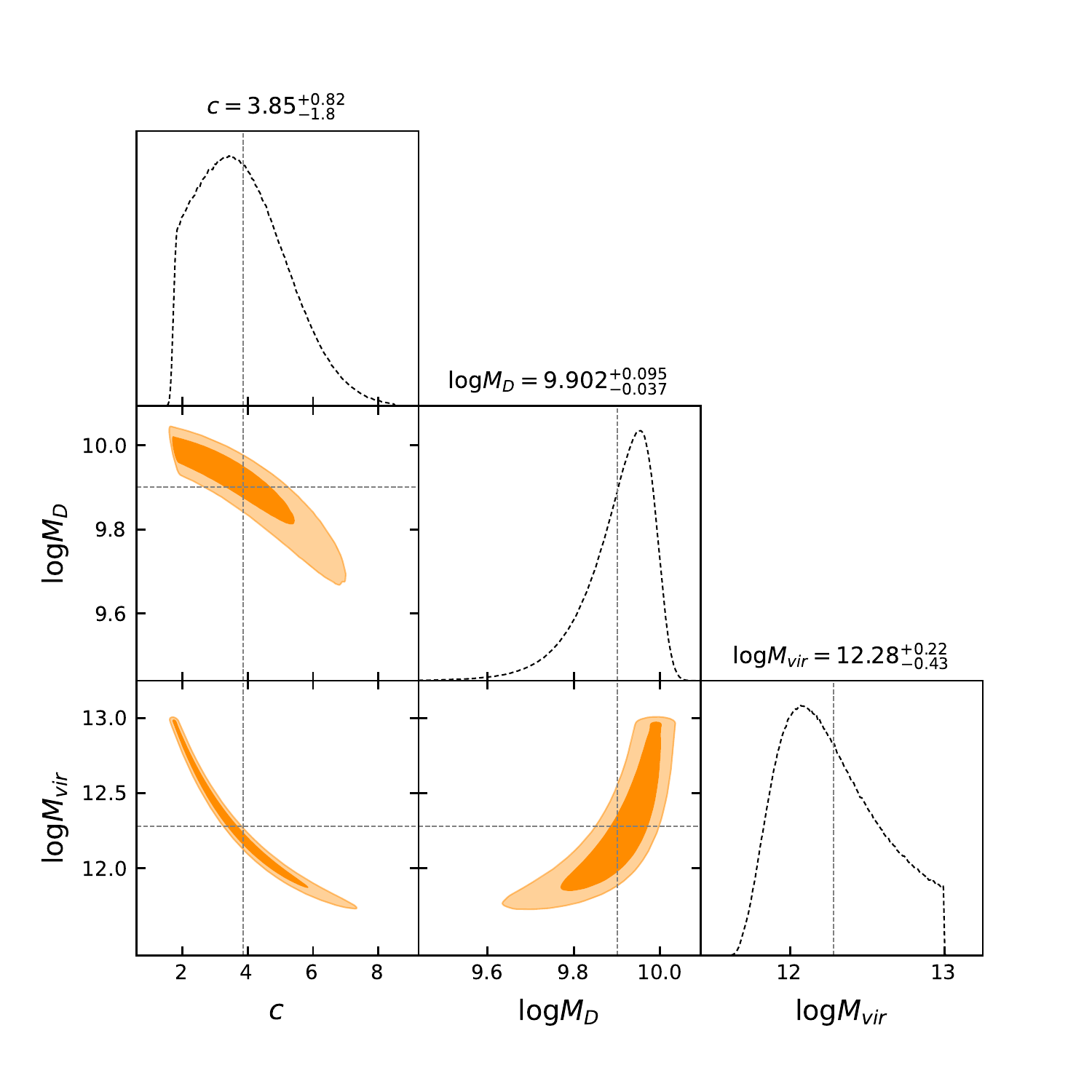}
    \includegraphics[width=.495\textwidth]{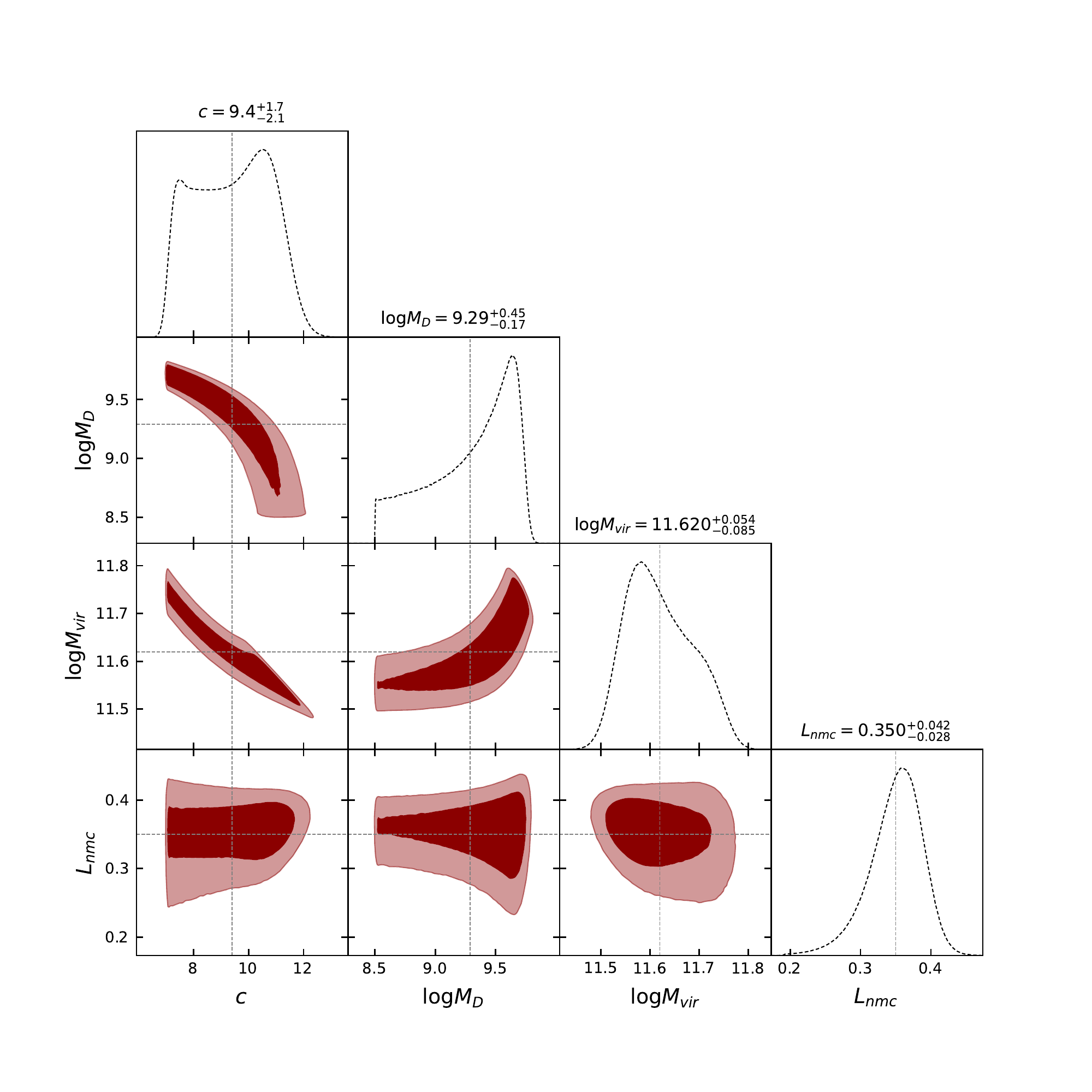}
    \end{center}
    \caption{Bin 3 of LSB galaxies}
\end{figure}
\clearpage
\begin{figure}[ht]
    \begin{center}
    \includegraphics[width=.495\textwidth]{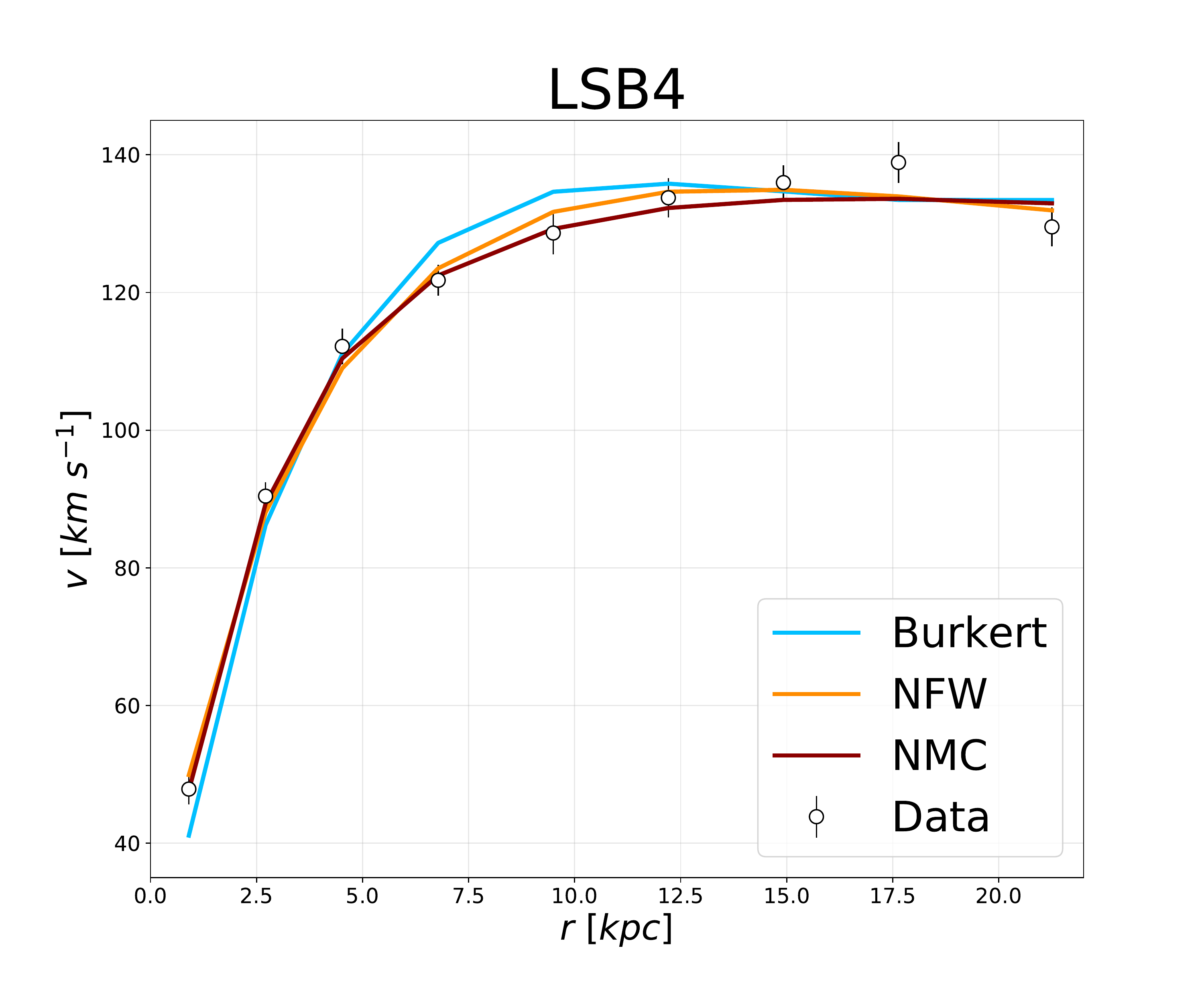}
    \includegraphics[width=.495\textwidth]{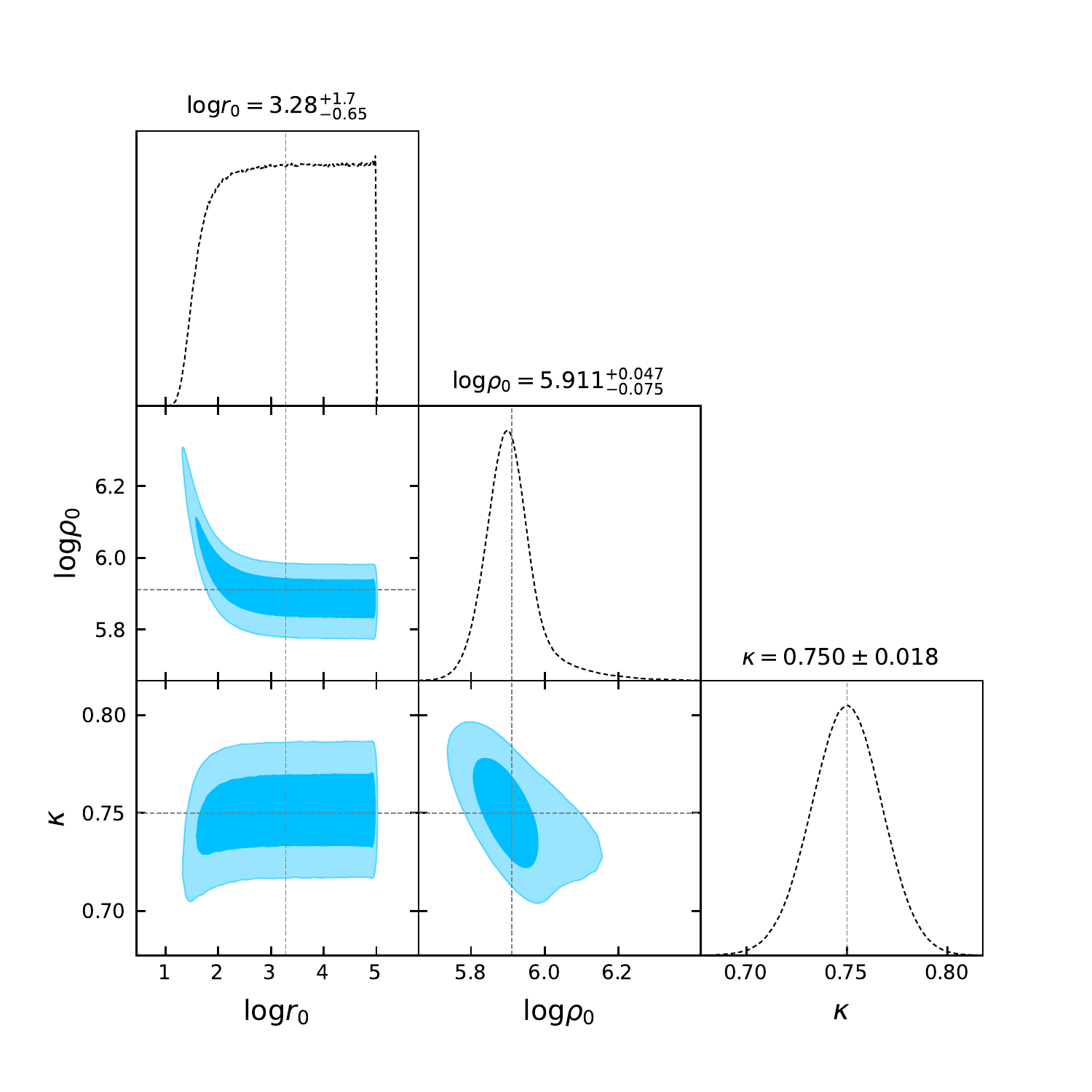}
    \includegraphics[width=.495\textwidth]{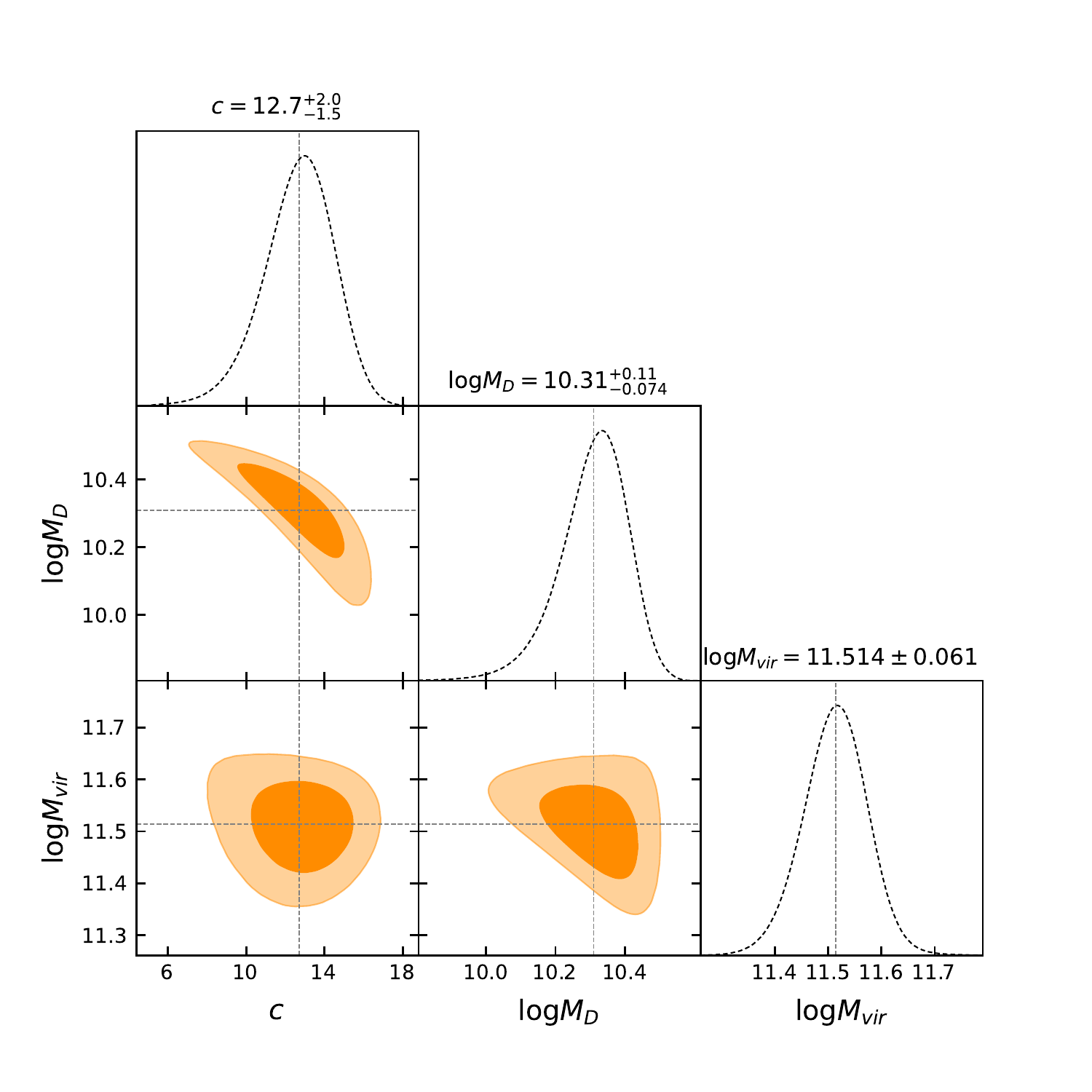}
    \includegraphics[width=.495\textwidth]{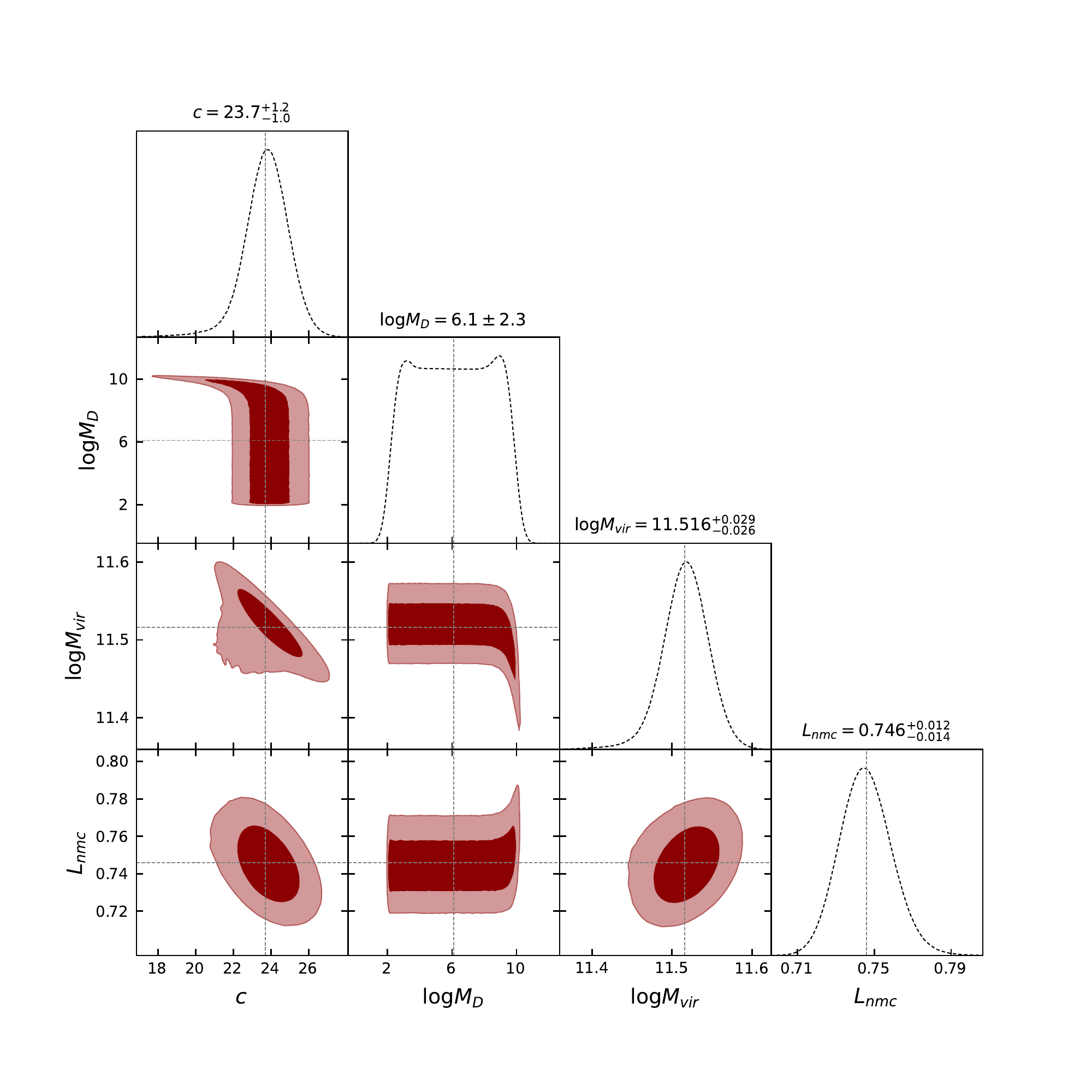}
    \end{center}
    \caption{Bin 4 of LSB galaxies}
\end{figure}

%



\end{document}